\documentclass[graybox]{svmult}

\usepackage{type1cm}    
\usepackage{makeidx}         
\usepackage{graphicx}        
\usepackage{multicol}        
\usepackage[bottom]{footmisc}

\usepackage{newtxtext}       %
\usepackage{newtxmath}       

\makeindex     

\begin{document}

\title*{Global polarization effect and spin-orbit coupling in strong interaction}
\author{Jian-Hua Gao, Zuo-Tang Liang, Qun Wang and Xin-Nian Wang}
\institute{Jian-Hua Gao \at Shandong Provincial Key Laboratory of Optical Astronomy and Solar-Terrestrial Environment,
Institute of Space Sciences, Shandong University, Weihai, Shandong 264209, China, \email{gaojh@sdu.edu.cn}
\and Zuo-tang Liang \at Key Laboratory of Particle Physics and Particle Irradiation (MOE), 
Institute of Frontier and Interdisciplinary Science, 
Shandong University, Qingdao, Shandong 266237, China, \email{liang@sdu.edu.cn}
\and Qun Wang \at Department of Modern Physics, University of Science and Technology
of China, Hefei, Anhui 230026, China, \email{qunwang@ustc.edu.cn}
\and Xin-Nian Wang \at Key Laboratory of Quark and Lepton Physics (MOE) and Institute of
Particle Physics, Central China Normal University, Wuhan, 430079, China; 
and Nuclear Science Division, MS 70R0319, 
Lawrence Berkeley National Laboratory, Berkeley, California 94720, \email{xnwang@lbl.gov}}
%
%
\maketitle

\abstract{In non-central high energy heavy ion collisions 
the colliding system posses a huge orbital angular momentum in the direction 
opposite to the normal of the reaction plane. 
Due to the spin-orbit coupling in strong interaction, such huge orbital angular momentum 
leads to the polarization of quarks and anti-quarks in the same direction. 
This effect, known as the global polarization effect, has been recently 
observed by STAR Collaboration at RHIC that confirms the theoretical prediction made more than ten years ago.  
The discovery has attracted much attention on the study of spin effects in heavy ion collision. 
It opens a new window to study properties of QGP and a new direction in high energy heavy ion physics 
--- Spin Physics in Heavy Ion Collisions. 
In this chapter, we review the original ideas and calculations that lead to the predictions. 
We emphasize the role played by spin-orbit coupling in high energy spin physics 
and discuss the new opportunities and challenges in this connection. }

\section{Introduction}
\label{secint}

Recently, the global polarization effect (GPE) of $\Lambda$ and $\bar\Lambda$ hyperons 
in heavy-ion collisions (HIC) has been observed~\cite{STAR:2017ckg} by the STAR
Collaboration at the Relativistic Heavy Ion Collider (RHIC) in Brookhaven National Laboratory (BNL). 
The discovery confirms the theoretical prediction~\cite{Liang:2004ph} made more than ten years 
ago and has attracted much attention on the study of spin effects in HIC. 
This opens a new window to study properties of QGP and a new direction
in high energy heavy ion physics --- Spin Physics in HIC. 
New experiments along this line are being carried out and/or planned. 
It is therefore timely to summarize the original ideas and theoretical calculations~\cite{Liang:2004ph,Liang:2004xn,Gao:2007bc} 
that lead to the predictions and discuss new opportunities and challenges.

Spin, as a fundamental degree of freedom of elementary particles, 
plays a very important role in modern physics and often brings us surprises.  
There are many well known examples in the field of particle and nuclear physics. 
The anomalous magnetic moments of nucleons are usually regarded as 
one of the first clear signatures for the existence of inner structure of nucleon. 
The explanation of these anomalous magnetic moments in 1960s was one of the great successes of the quark model 
that lead us to believe that it provides us the correct picture for hadron structure.  

High energy spin physics experiments started since 1970s. 
Soon after the beginning, a series of striking spin effects have been observed 
that were in strong contradiction to the theoretical expectations at that time and been pushing 
the studies move forward.  
The most famous ones might be classified as following.
 
(i) Proton's ``spin crisis'' : Measurements of spin dependent structure functions in deeply inelastic lepton-nucleon scatterings,  
started by E80 and E143 Collaborations at SLAC~\cite{Baum:1980mh,Baum:1983ha}
and later on by the European Muon Collaboration (EMC) at CERN~\cite{Ashman:1987hv,Ashman:1989ig}, 
seem to suggest that the contribution of the sum of spins of quarks and anti-quarks to proton spin is consistent with zero.  
This has triggered the so-called spin crisis of the proton and the intensive study on the spin structure of nucleon~\cite{Aidala:2012mv}.
 
(ii) Single spin left-right asymmetry (SSA): 
It has been observed~\cite{Klem:1976ui,Dragoset:1978gg,Adams:1991cs,Liang:2000gz} that 
in inclusive hadron-hadron collisions with singly transversely polarized beams or targets, 
the produced hadron has a large azimuthal angle dependence characterized by the left-right asymmetry. 
The observed asymmetry can be as large as $40\%$ 
but the theoretical expectation at the quark level using pQCD at the leading order was close to zero. 

(iii) Transverse hyperon polarization: 
It has been observed~\cite{Lesnik:1975my,Bunce:1976yb,Bensinger:1983vc,Gourlay:1986mf,Krisch:2007zza} 
that hyperons produced in unpolarized hadron-hadron and hadron-nucleus collisions are 
transversely polarized with respect to the production plane. 
The observed polarization can reach a magnitude as high as $40\%$ 
but the leading order pQCD expectation was again close to zero. 

(iv) Spin asymmetries in elastic $pp$-scattering: 
It has been observed~\cite{O'Fallon:1977cp,Crabb:1978km,Cameron:1985jy,Krisch:2007zza} that 
the azimuthal dependence, called the spin analyzing power, in scattering with single-transversely polarized proton 
and doubly polarized asymmetries are very significant, much larger than theoretical expectations available at that time.  

Such striking spin effects came out often as such a shock to the field of strong interaction physics 
that lead to the famous comment by Bjorken~\cite{Bjorken:1996dc} in a QCD workshop that 
``Polarization phenomena are often graveyards of fashionable theories. ...''.  
In last decades, the study on such spin effects lead to one of the most active fields in strong interaction or QCD physics. 

At the same time, high energy HIC physics has become the other active field in strong interaction physics 
in particular after the quark gluon plasma (QGP) has been discovered at RHIC~\cite{Gyulassy:2004zy,Adams:2005dq}.  
The study on properties of QGP in HIC is the core of high energy HIC physics currently. 

We recall that RHIC is not only the first relativistic heavy ion collider in the world 
but also the first polarized high energy proton-proton collider. 
It is therefore natural to ask whether we can do spin physics in HIC.  

Spin physics in HIC was however used to be regarded as difficult or impossible because 
the polarization of the nucleon in a heavy nucleus is very small even if the nucleus is completely polarized. 
The breakthrough came out in 2005 when it was realized that~\cite{Liang:2004ph} there is however 
a great advantage to study spin and/or angular momentum effects in HIC,  
i.e., the reaction plane in a HIC can be determined experimentally by measuring flows and/or spectator nucleons 
and there exist a huge orbital angular momentum for the participating system in a non-central HIC with respect to the reaction plane! 
It provides a unique place in high energy reactions to study the mutual exchange of orbital angular momentum and the spin polarization.  
The discovery of GPE leads to an active field of Spin Physics in HIC~\cite{Liang:2019clf}. 

In this chapter, we review the original ideas and calculations~\cite{Liang:2004ph,Liang:2004xn,Gao:2007bc} that lead to the prediction of GPE in HIC. 
We present also a rough comparison to data available and an outlook for future studies.  
The rest of the chapter is arranged as follows: 
In Sec. 2, we present the orbital angular momentum of the colliding system in non-central HIC and the resulting gradient in 
momentum or rapidity distribution. 
In Sec. 3, we recall the origin of spin-orbit coupling and famous example in electromagnetic and strong interaction systems. 
In Sec. 4, we present calculations at the quark level and results for the global quark polarization in HIC. 
In Sec. 5, we discuss the global hadron polarization and finally a short summary and outlook is presented in Sec. 6.

\section{{Orbital angular momenta of QGP in HIC}}
\label{secoam}

\subsection{The reaction plane in HIC}
\label{secreactionplane}

\begin{figure}[htbp]
\begin{center}
\resizebox{3.7in}{2.0in}{\includegraphics{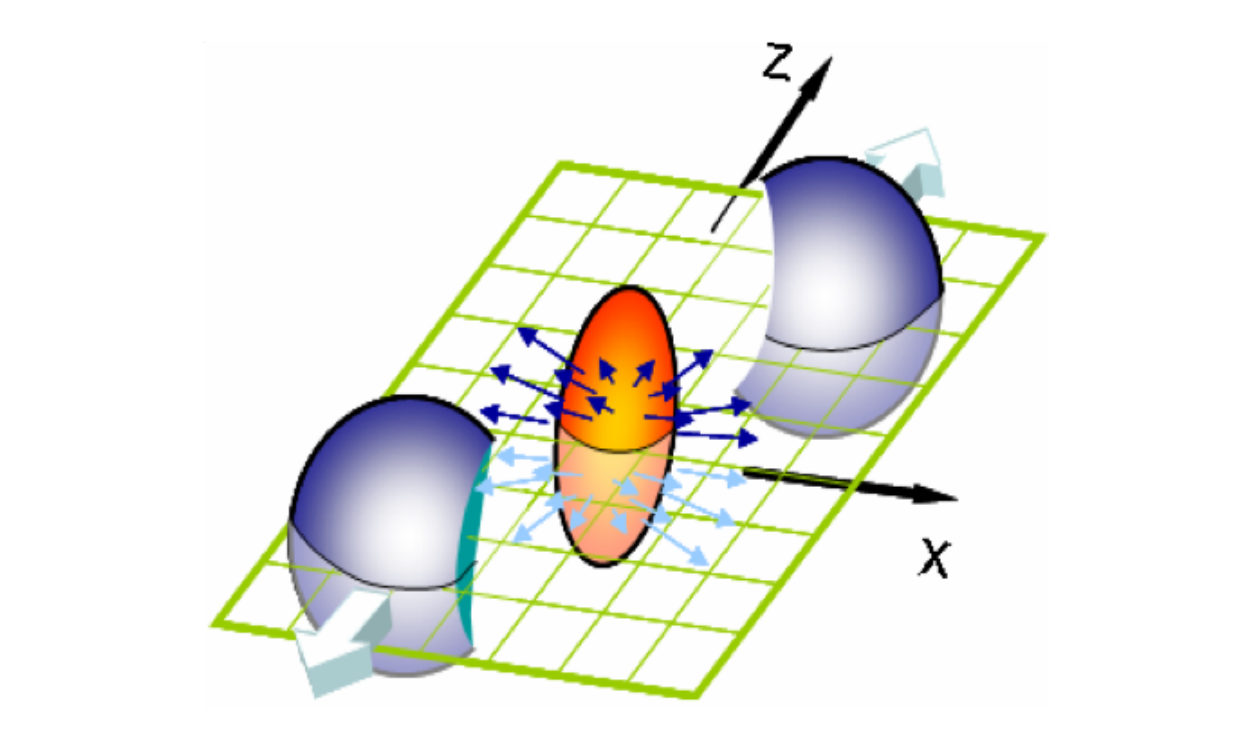}}
\end{center}
\caption{Illustration diagram for the reaction plane in a non-central heavy ion collision. 
In contrast to high energy $pp$ or $e^+e^-$ collisions, the reaction plane in 
a high energy heavy collision can be determined experimentally.}
\label{figreplane}
\end{figure}

We consider two colliding nuclei with the projectile of beam momentum per nucleon $\vec p_{in}$.
For a non-central collision, there is a transverse separation between the centers of the two colliding nuclei. 
The impact  parameter $\vec b$ is defined as the transverse vector pointing from the target to the projectile. 
The reaction plane of a HIC is usually defined by $\vec b$ and $\vec p_{in}$ and is illustrated in Fig.~\ref{figreplane}. 
The overlap parts, hereafter referred as the colliding system, 
interact with each other and form the system denote by the red core in the middle 
while the other parts, denoted by the blues parts in the figure, are just spectators 
and move apart in the original directions. 

\begin{figure}[htbp]
\begin{center}
\resizebox{2.5in}{2.1in}{\includegraphics{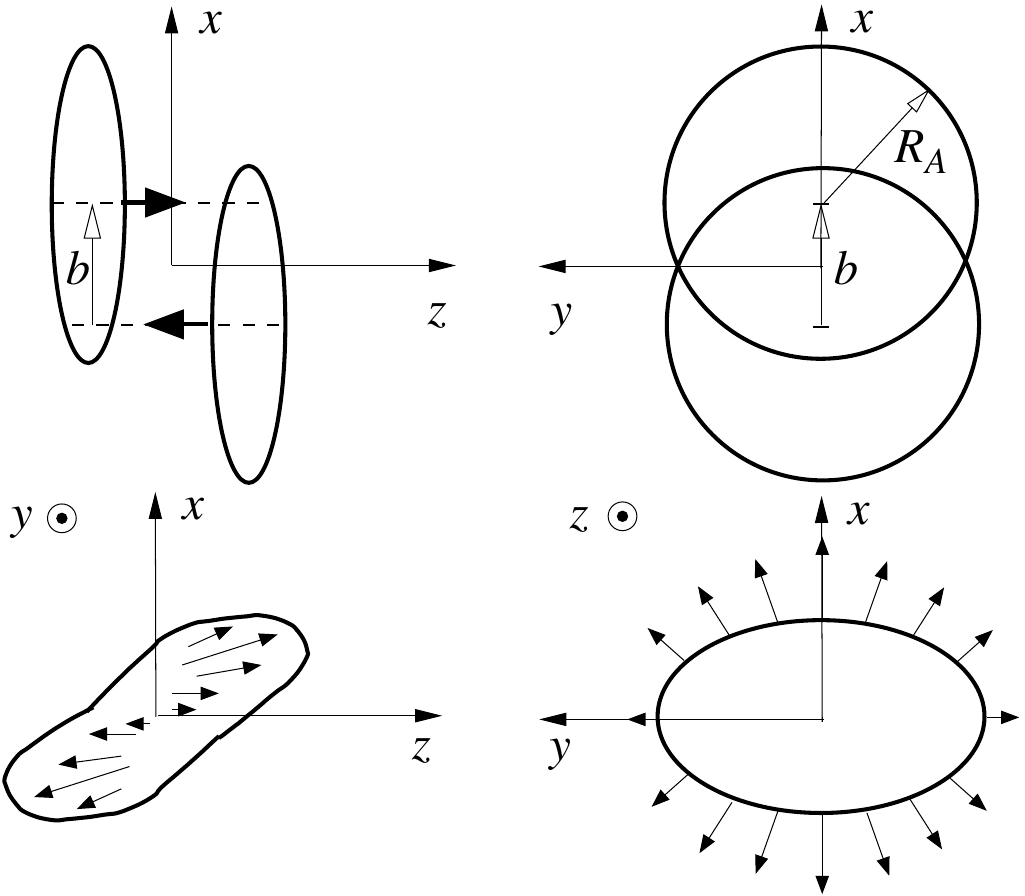}}
\end{center}
\caption{Illustration of the geometry and coordinate system 
for the non-central HIC with impact parameter $\vec{b}$. 
The global angular momentum of the produced matter is along the minus ${y}$ direction,
opposite to the reaction plane. 
This figure is taken from~\cite{Liang:2004ph}. }
\label{figgeo}
\end{figure}

The geometry and the coordinate system are further specified in Fig.~\ref{figgeo}. 
The beam direction of the colliding nuclei is taken as the $z$ axis, 
as illustrated in the upper-left panel in the figure.
The transverse separation is called the impact parameter $\vec{b}$  
defined as the transverse distance of the projectile 
from the target nucleus and is taken as in the ${x}$-direction.
The normal of the reaction plane is given by,
\begin{equation}
\label{eqvecb}
\vec{n}\equiv {\vec p}_{in}\times{\vec{b}}/|{\vec p}_{in}\times{\vec{b}}|,
\end{equation} 
and is taken as the ${y}$-direction, where $\vec p_{in}$ is the momentum per nucleon in  the incident nucleus $A$. 

Usually in a high energy reaction such as a hadron-hadron, or lepton-hadron or $e^+e^-$ annihilation, 
the size of the reaction region is typically less than 1fm. 
The reaction plane in such collisions can be defined theoretically but can not be determined experimentally. 
However, in a HIC, the reaction region is usually much larger and colliding parts  give rise to a quark matter system 
with very high temperature and high density and expand violently while the spectators just leave 
the region in the original directions. 
Since the colliding system is not isotropic, the pressures in different directions are also different in different directions 
thus lead to a system that expands non-isotropically.  
In the transverse directions they behave like an ellipse as illustrated in the lower-right panel in Fig.~\ref{figgeo}. 
Such a non-isotropy is described by the elliptic flow $v_2$ and the directed flow $v_1$ that can be 
measured experimentally (see e.g. \cite{star-v2,star-v1}). 
Clearly, by measuring $v_2$, one can determine the reaction plane 
and further determine the direction of the plane by measuring the directed flow $v_1$. 

In experiments, the reaction plane in a HIC can not only be determined by measuring $v_2$ and $v_1$ 
but also determined by measuring the sidewards deflection of the forward- and backward-going 
fragments and particles in the beam–beam counter detectors~\cite{STAR:2017ckg}.  
This is quite unique in different high energy reactions.

\subsection{The global orbital angular momentum}

Just as illustrated in Figs.~\ref{figreplane} and \ref{figgeo}, 
in a non-central HIC, there is a transverse separation between the overlapping parts 
of the two colliding nuclei in the same direction as the impact parameter $\vec b$.
Hence the whole system that takes part in the reaction, i.e. the colliding system 
carries a finite orbital angular momentum $L_y$ along the direction orthogonal to the reaction plane.
We call $L_y$ the global orbital angular momentum. 
The magnitude of this global orbital angular momentum $L_y$ can be calculated by, 
\begin{equation}
\label{eqgly}
L_y=-p_{in}\int x\ dx \left(\frac{dN_{\rm{part}}^P}{dx}-\frac{dN_{\rm{part}}^T}{dx}\right), 
\end{equation}
where ${dN_{\rm part}^{P,T}}/{dx}$ is the transverse distributions (integrated over $y$ and $z$) of participant
nucleons in each nucleus $A$ along the $x$-direction, 
the superscript $P$ or $T$ denotes projectile or target respectively.  
These transverse distributions are given by,
\begin{equation}
\label{eqdNdx}
\frac{dN_{\rm part}^{P,T}}{dx}=\int dydz~\rho_A^{P,T}(x,y,z,b),
\end{equation}
where $\rho_A^{P,T}(x,y,z,b)$ is the number density of participant nucleons
in nucleus $A$ in the coordinate system defined in Fig.~\ref{figgeo}.

The number density $\rho_A^{P,T}(x,y,z,b)$ of participant nucleons in nucleus $A$ 
can easily be calculated if we take a hard sphere distribution of nucleons in the nucleus $A$.
In this model, the overlapping area has a clear boundary 
and the participant nucleon density is given by the overlapping area of two hard spheres, 
as illustrated in the upper-right panel of Fig.~\ref{figgeo}, i.e., 
\begin{eqnarray}
\rho _{A,HS}^{P,T}(x,y,z,b)
&=&f_{A,HS}^{P,T}(x,y,z,b)~\theta\left(R_A-\sqrt{(x\pm b/2)^2+y^2+z^2}\right), \label{eqHSrho}
\end{eqnarray}
where $f_{A,HS}^{P,T}(x,y,z,b)$ is the hard sphere nuclear distribution in $A$ that is given by,
\begin{equation}
\label{eqHSf}
f_{A,HS}^{P,T}(x,y,z,b)
=\frac{3A}{4\pi R_A^3}\theta\left(R_A-\sqrt{(x\mp b/2)^2+y^2+z^2}\right),
\end{equation}
where $R_A=1.12A^{1/3}$ fm is the nuclear radius and $A$ the atomic number.

If we take the Woods-Saxon nuclear distribution, i.e., 
\begin{equation}
\label{eqWSf}
f_{A,WS}^{P,T}(x,y,z,b)=C_0\left(1+\exp {\frac{\sqrt{(x\mp b/2)^2+y^2+z^2}-R_A}{a}}\right)^{-1} \,, 
\end{equation}
there is no clear boundary of the overlapping region and 
the participant nucleon number density is calculated using the Glauber model and is given by,
\begin{eqnarray}
\label{eqWSrho}
&&\rho_{A,WS}^{P,T}(x,y,z,b)=f_{WS}^{P,T}(x,y,z,b) 
\left\{ 1-\exp\Bigl[ -\sigma_{NN}\int dz f_{WS}^{T,P}(x,y,z,b)\Bigr]\right\} \, , ~~~
\end{eqnarray}
where $\sigma_{NN}$ is the total cross section of nucleon-nucleon scatterings, 
$C_0$ is the normalization constant,
\begin{equation}
C_0=A/4\pi \int {r^2dr} \left(1+e^{(r-R_A)/a}\right)^{-1}~, 
\end{equation}
 and $a$ is the width parameter set to $a=0.54$ fm.

\begin{figure}[htbp]
\begin{center}
\includegraphics[width=7cm]{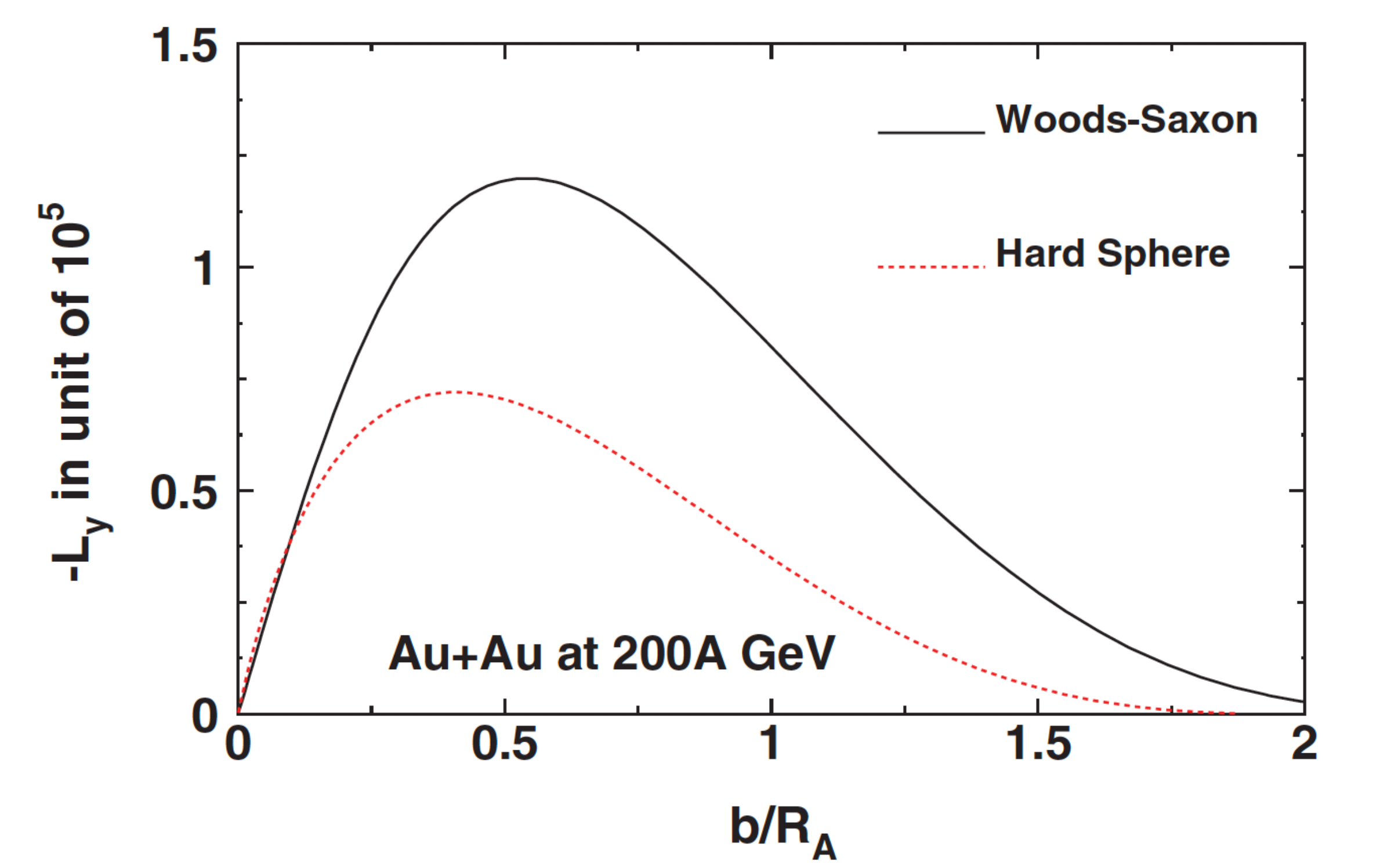}
\end{center}
\caption{Global orbital angular momentum of the colliding system 
in the non-central HIC as a function of the impact parameter obtained from 
the Woods-Saxon and hard-sphere distributions, respectively.
This figure is taken from \cite{Gao:2007bc}.}
\label{figgly}
\end{figure}

The calculations have been carried out in~\cite{Liang:2004ph} and \cite{Gao:2007bc}.  
The obtained results are shown in Fig.~\ref{figgly}.
From the results shown in Fig.~\ref{figgly}, we see that though there are significant differences 
between two nuclear geometry models 
the global orbital angular momentum $L_y$ of the overlapped parts of two colliding nuclei  
is huge and is of the order of $10^4$ at most impact parameters.

\subsection{The transverse gradient of the momentum distribution and the local orbital angular momentum}

How the global orbital angular momentum discussed above is transferred to 
the final state particles depends on the equation of state (EOS) of the dense matter.
At low energies, the final state is expected to be the normal nuclear matter with an EOS of rigid nuclei.
In such cases, a rotating compound nucleus can be formed when the colliding energy
is comparable or smaller than the nuclear binding energy. 
The finite value of the global orbital angular momentum of the non-central collision at such low energies
provides a useful tool for the study of the properties of
super-deformed nuclei under such rotation~\cite{Cederwall:1994gz}.

At high colliding energies such as those at RHIC, 
the dense matter is expected to be partonic with an EOS of QGP. 
Given such a soft EOS, the global orbital angular momentum would probably not 
lead to the global rotation of the dense matter system. 
Instead, the global angular momentum could be distributed across the overlapped
region of nuclear scattering and is manifested in the shear of
the longitudinal flow leading to a finite value of local vorticity density. 
Under such longitudinal fluid shear, a pair of scattering partons will on average carry a finite value
of relative orbital angular momentum that will be referred to as the local orbital angular momentum 
in the opposite direction to the reaction plane as defined in Eq.~(\ref{eqvecb}).

By momentum conservation, the average initial collective longitudinal momentum at any given transverse position can be calculated
as the total momentum difference between participating projectile and target nucleons. 
Since the total multiplicity in HIC is proportional to the number of participant nucleons~\cite{phobos2003},
we can make the same assumption for the produced partons with a proportionality constant fixed at a given
center of mass energy $\sqrt{s}$.
How the global angular momentum is distributed to the longitudinal flow shear and 
the magnitude of the local relative orbital angular momentum depends on the parton production mechanism 
and their longitudinal momentum distributions. 
We consider two different scenarios: the Landau fireball and the Bjorken scaling model.

\subsubsection{Results from the Landau fireball model}

In the Landau fireball model, we assume that the produced partons thermalize quickly 
and have a common longitudinal flow velocity at a given transverse position in the overlapped region. 
The average collective longitudinal momentum per parton can be written as
\begin{equation}
p_z(x,b,\sqrt{s})=p_0 R_N(x,b,\sqrt{s}),  \label{eqpzxb}
\end{equation}
where $p_0=\sqrt{s}/2c(s)$ is an energy dependent constant, 
$\sqrt{s}$ is the center of mass energy of a colliding nucleon pair, 
$c(s)$ is the average number of partons produced per participating nucleon; 
and $R_N(x,b,\sqrt{s})$ is the ratio defined as, 
\begin{equation}
R_N(x,b,\sqrt{s})=\left(\frac{dN_{\rm part}^P}{dx} - \frac{dN_{\rm part}^T}{dx}\right)\Big/\left(\frac{dN_{\rm part}^P}{dx} + \frac{dN_{\rm part}^T}{dx}\right)
\label{eqRNxb}
\end{equation}

\begin{figure}[ht]
\begin{center}
\includegraphics[width=7cm]{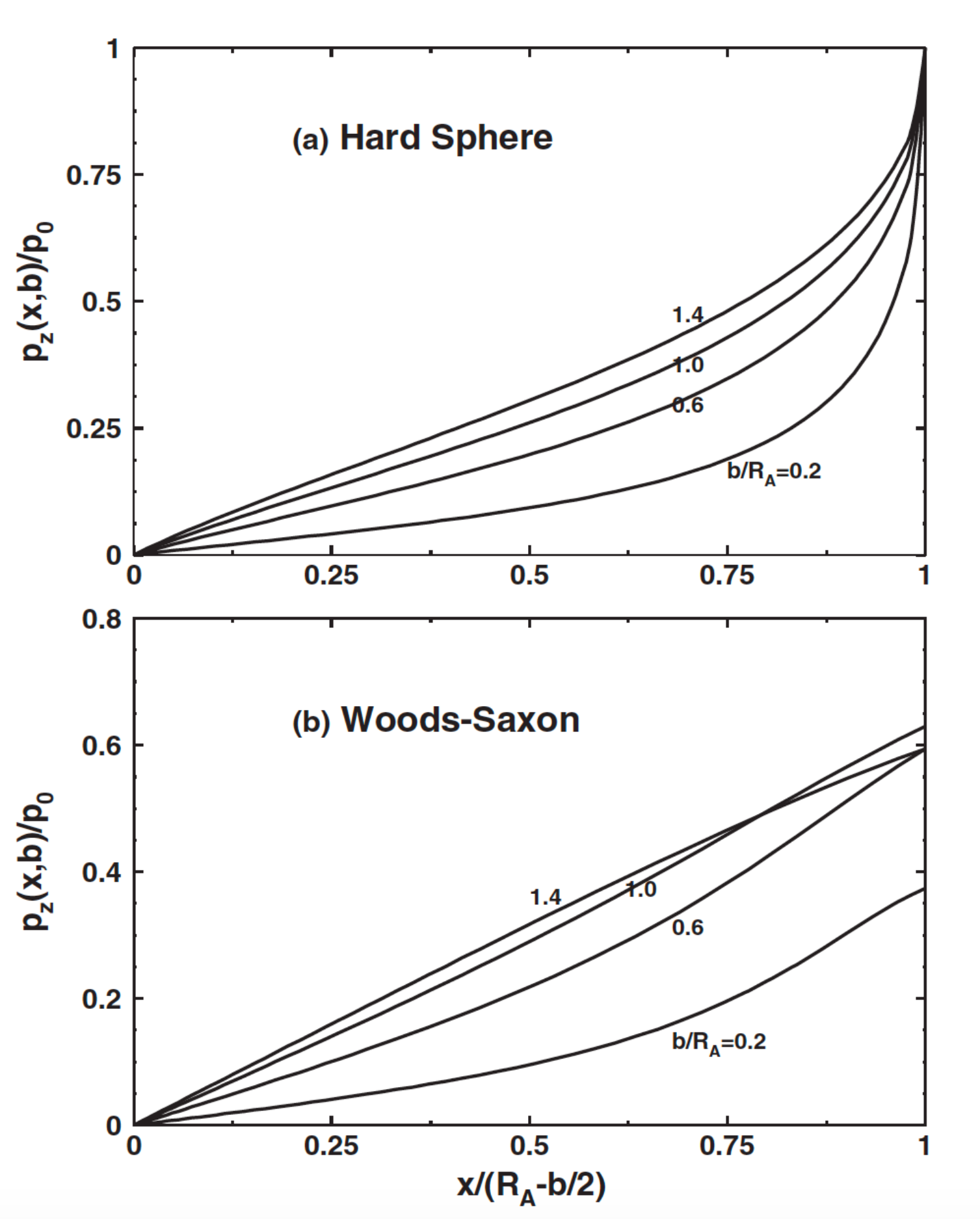}
\end{center}
\caption{The average longitudinal momentum distribution $p_z(x,b,\sqrt{s})$ in
unit of $p_0=\sqrt{s}/[2c(s)]$ as a function of $x/(R_A-b/2)$ for
different values of $b/R_A$ with the hard sphere (upper panel)
and Woods-Saxon (lower panel) nuclear distributions.
This figure is taken from \cite{Gao:2007bc}.}
\label{figpxb}
\end{figure}

It is clear that in the symmetric $AA$ collision (where the beam and target nuclei are the same),   
the ratio $R_N(x,b,\sqrt{s})$ thus the distribution $p_z(x,b,\sqrt{s})$ 
is an odd function in both $x$ and $b$ and therefore vanishes at $x=0$ or $b=0$. 
In Fig.~\ref{figpxb}, $p_z(x,b,\sqrt{s})$ is plotted as a function of $x$ at different impact parameters $b$. 
We see clearly that $p_z(x,b,\sqrt{s})$ is a monotonically increasing function of $x$ until the edge of the overlapped region
$|x\pm b/2|=R_A$ beyond which it drops to zero (gradually for Woods-Saxon geometry).

From $p_z(x,b,\sqrt{s})$ one can compute the
transverse gradient of the average longitudinal collective
momentum per parton $dp_z/dx$ which is an even function of $x$ and vanishes at $b=0$. 
One can then estimate the longitudinal momentum difference $\Delta p_z$ 
between two neighboring partons in QGP. 
On average, the relative orbital angular momentum for two colliding partons separated 
by $\Delta x$ in the transverse direction is 
\begin{equation}
\label{eqlly}
l_y \equiv -(\Delta x)^2\frac{dp_z}{dx}.
\end{equation} 
With the hard sphere nuclear distribution, 
$l_y$ is proportional to
\begin{equation}
\label{eqdp0dx}
\frac{dp_0}{dx}\equiv \frac{p_0}{R_A}=\frac{\sqrt{s}}{2c(s)R_A}.
\end{equation}
\begin{figure}[ht]
\begin{center}
\includegraphics[width=7cm]{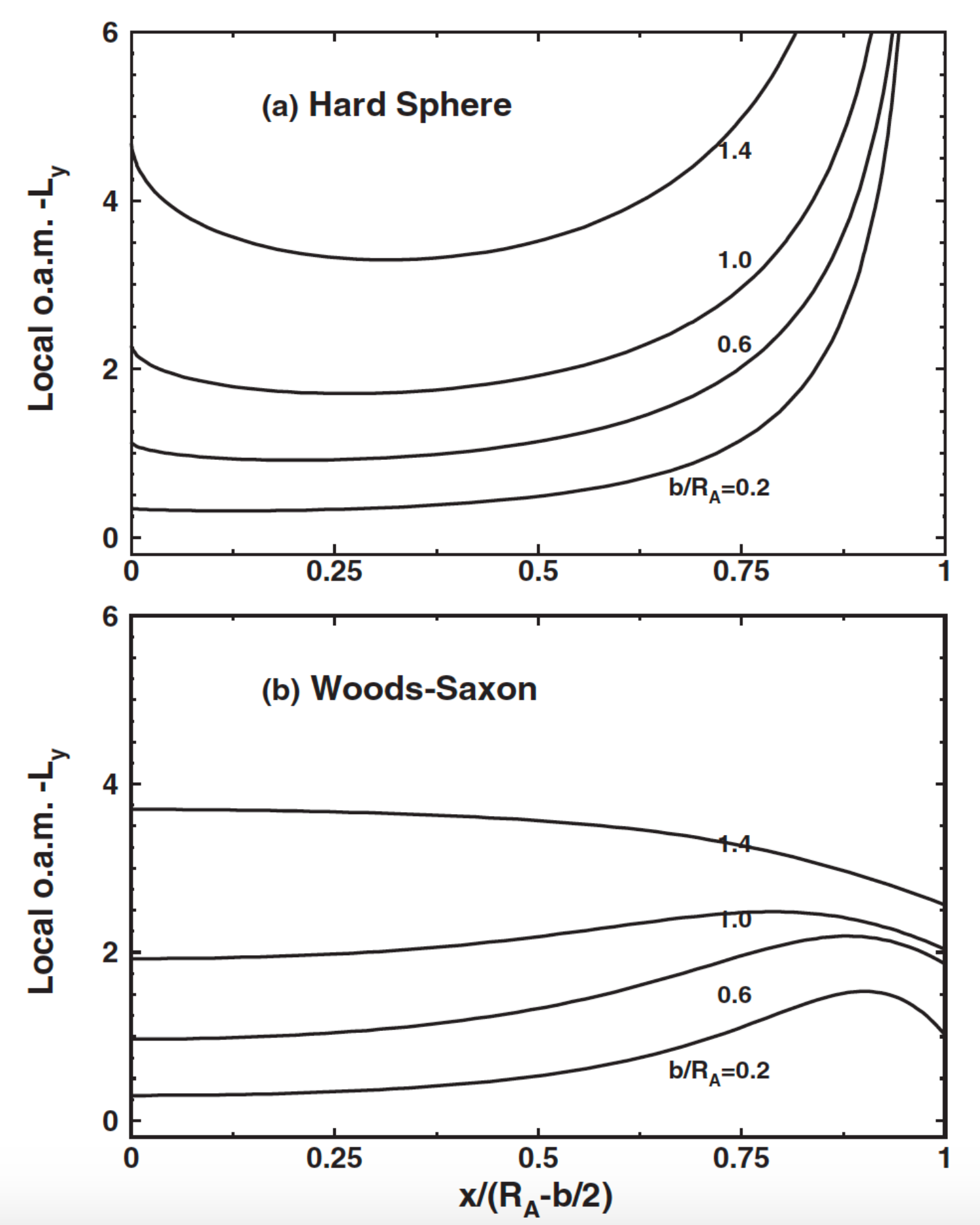}
\end{center}
\caption{The average orbital angular momentum $l_y \equiv -(\Delta x)^2dp_z/dx$
of two neighboring partons separated by $\Delta x=1$ fm as a
function of the scaled transverse coordinate $x/(R_A-b/2)$ for
different values of the impact parameter $b/R_A$ with the hard-sphere (upper panel)
and Woods-Saxon (lower panel) nuclear distributions.
This figure is taken from \cite{Gao:2007bc}.}
\label{figlly}
\end{figure}
This provides a measure of order of magnitude of $dp_z/dx$. 
In $Au+Au$ collisions at $\sqrt{s}=200$ GeV, the number of charged
hadrons per participating nucleon is about 15~\cite{phobos2003}.
Assuming the number of partons per (meson dominated) hadron is about 2, 
we have $c(s)\simeq 45$ (including neutral hadrons).
Given $R_A=6.5$ fm, $dp_0/dx \simeq 0.34$ GeV/fm 
and we obtain o value of $l_0\equiv -(\Delta x)^2 dp_0/dx \simeq -1.7$ for $\Delta x=1$ fm.

In Fig.~\ref{figlly}, we show the average local orbital angular momentum
$l_y$ given by Eq.~(\ref{eqlly}) for two neighboring partons separated by $\Delta x=1$ fm
as a function of $x$ for different impact parameter $b$ for both
Woods-Saxon and hard-sphere nuclear distributions.
We see that $l_y$ is in general of the order of 1 and is comparable or larger than the spin of a quark.
It is expected that $c(s)$ should depend logarithmically on the colliding energy $\sqrt{s}$,
therefore $l_y$ should increases with growing $\sqrt{s}$.

\subsubsection{Results from the Bjorken scaling model}

In a three dimensional expanding system, there could be strong correlation
between longitudinal flow velocity and spatial coordinate of the fluid cell. 
The most simplified picture is the Bjorken scaling scenario~\cite{Bjorken:1982qr} 
in which the longitudinal flow velocity is identical to the spatial velocity $\eta=\log[(t+z)/(t-z)]$. 
With such correlation, the local interaction and thermalization require that 
a parton only interacts with other partons in the same region of longitudinal momentum or rapidity $Y$. 
The width of such region in rapidity is determined by the half-width of the thermal distribution
$f_{th}(Y,p_T)=\exp[-p_T\cosh(Y-\eta)/T]$~\cite{Levai:1994dx}, 
which is approximately $\Delta_Y\approx 1.5$ (with $\langle p_T\rangle\approx 2T$ and $T$ is the local temperature).
The relevant measure of the local relative orbital angular momentum between two interacting partons is, 
therefore, the difference in parton rapidity distributions at transverse distance of  the
order of the average interaction range.

The variation of the rapidity distributions with respect to the transverse coordinate 
can be described by the normalized rapidity distribution $f_p(Y,x)$ at given $x$, 
\begin{equation}
f_p(Y,x,b,\sqrt{s}~)=\frac{d^2N}{dxdY}\Big/\frac{dN}{dx}, 
\label{eqfpYx}
\end{equation}
where $d^2N/dxdY$ denotes the number density of particles produced with respect to $x$ and $Y$ 
and $dN/dx\equiv\int dY d^2N/dxdY$ is the distribution of particles with respect to $x$.  
At a given x, the overall average value of the rapidity is given by,
\begin{equation}
\langle Y(x,b,\sqrt{s}~)\rangle =\int YdY f_p(Y,x,b,\sqrt{s}~) .
\label{eqaYx}
\end{equation}
$\langle Y(x,b,\sqrt{s}~)\rangle$ just corresponds to $p_z(x,b,\sqrt{s}~)$ given by Eq.~(\ref{eqpzxb}) discussed in the Landau fireball model. 
It measures the overall behavior of the rapidity distribution of partons at given transverse coordinate $x$. 
To further quantify such longitudinal fluid shear, one can calculate the average rapidity 
within an interval $\Delta_Y$ at a given rapidity $Y$, i.e., 
\begin{equation}
\langle Y_l(Y,x,b,\sqrt{s}~)\rangle \approx Y +\frac{\Delta_Y^2}{12}\frac{1}{f_p}\frac{\partial f_p}{\partial Y} 
=Y +\frac{\Delta_Y^2}{12} \frac{\partial\ln f_p}{\partial Y} .
\label{eqaYlx}
\end{equation}
Here, we use the subscript $l$ to denote that this is the average of $Y$ in a localized interval 
$[Y-\Delta_Y/2, Y+\Delta_Y/2]$ to differentiate it from the overall average $\langle Y(x,b,\sqrt{s})\rangle$ given by Eq.~(\ref{eqaYx}). 
The average rapidity shear or the difference in average rapidity for
two partons separated  by a unit of transverse distance $\Delta x$ is then given by,
\begin{equation}
\frac{\partial}{\partial x} \langle Y_l(Y,x,b,\sqrt{s}~)\rangle \approx
\frac{\Delta_Y^2}{12} \frac{\partial^2 \ln f_p}{\partial Y\partial x} .
\label{eqdaYdx}
\end{equation}
The averaged longitudinal momentum is,
\begin{equation}
\langle p_z \rangle \approx p_T \sinh\langle Y_l\rangle 
\approx p_T\left( \sinh Y + \cosh Y~ \frac{\Delta_Y^2}{12} \frac{\partial \ln f_p}{\partial Y}\right) .
\label{eqapz}
\end{equation}
The corresponding local relative longitudinal momentum shear is given by,
\begin{equation}
\frac{\partial \langle p_z\rangle}{\partial x}\approx p_T\cosh Y ~
\frac{\partial \langle Y_l\rangle}{\partial x} \approx p_T\cosh Y ~ \frac{\Delta_Y^2}{12} \frac{\partial^2 \ln f_p}{\partial Y\partial x} \,\, .
\label{eqdapzdx}
\end{equation}

The corresponding local orbital angular momentum $l_y$ for two partons separated by 
a transverse separation $\Delta x$ at a given rapidity $Y$ is  
$\langle l_y(Y)\rangle=- \Delta x\Delta \langle p_z\rangle=  -(\Delta x)^2 \partial\langle p_z\rangle/\partial x $. 
We transform it into the co-moving frame or the center of mass frame of the two partons and obtain,
\begin{equation}
\langle l^*_y(Y,x,b,\sqrt{s}~)\rangle =-\Delta x~\langle p_z^*\rangle \approx -(\Delta x)^2p_T \frac{\Delta_Y^2}{24} \frac{\partial^2 \ln f_p}{\partial Y\partial x} .
\label{eqllyY}
\end{equation}
We see that they are all determined by a key quantity 
\begin{equation}
\xi_p(Y,x,b,\sqrt{s}~)\equiv\frac{\partial^2 \ln f_p(Y,x,b,\sqrt{s})}{\partial Y\partial x},
\label{eqxip}
\end{equation}
that is determined by $d^2N/dxdY$. 
In terms of $\xi_p(Y,x,b,\sqrt{s}~)$, we have, 
\begin{eqnarray}
&&\frac{\partial \langle Y_l\rangle}{\partial x}\approx  \frac{\Delta_Y^2}{12}~\xi_p , \label{eqdaYldxxip}\\
&&\frac{\partial \langle p_z\rangle}{\partial x}\approx  \frac{\Delta_Y^2}{12}~\xi_p~p_T\cosh Y , \label{eqdapzdxxip}\\
&&\langle l^*_y(Y,x,b,\sqrt{s}~)\rangle  \approx - \frac{\Delta_Y^2}{24}~\xi_p~(\Delta x)^2p_T .~~~~~~~~~
\label{eqllyYxip}
\end{eqnarray}

The $Y$-dependence averaged over the transverse separation $x$ is 
determined by the average value of $\xi_p(Y,x,b,\sqrt{s})$ defined by,
\begin{equation}
\langle\xi_p\rangle
=\int dx 
~\xi_p(Y,x,b,\sqrt{s}~) \frac{d^2N}{dxdY}\Big/\frac{dN}{dY},
\label{eqadfpdYx}
\end{equation}
where $dN/dY=\int dx (d^2N/dxdY)$ is the rapidity distribution of partons produced in a $AA$ collision at 
the given impact parameter $b$. In the binary approximation,
\begin{equation}
\frac{dN}{dY}=N_{\rm part} \frac{dN_{pp}}{dY}. 
\label{eqdNdY}
\end{equation}

To proceed with numerical calculations, 
one needs a dynamical model to estimate the local rapidity distribution $d^2N/dxdY$ of produced partons. 
For this purpose, two models, the HIJING Monte-Carlo model~\cite{Wang:1991ht,Wang:1996yf} 
and the model proposed by Brodsky, Gunion and Kuhn (denoted as BGK model)~\cite{Brodsky:1977de}, 
have been used~\cite{Gao:2007bc,Liang2019}. 
We present the results~\cite{Gao:2007bc,Liang2019} obtained in the following respectively.

(i) {\it Results obtained using HIJING}
 
In~\cite{Gao:2007bc}, the HIJING Monte Carlo model~\cite{Wang:1991ht,Wang:1996yf} 
was used to calculate the hadron rapidity distributions at different transverse coordinate $x$ 
and assume that parton distributions of the dense matter are proportional to the final hadron spectra. 
We show the results obtained in this way in~\cite{Gao:2007bc} in the following.

\begin{figure}[ht]
\begin{center}
\includegraphics[width=7cm]{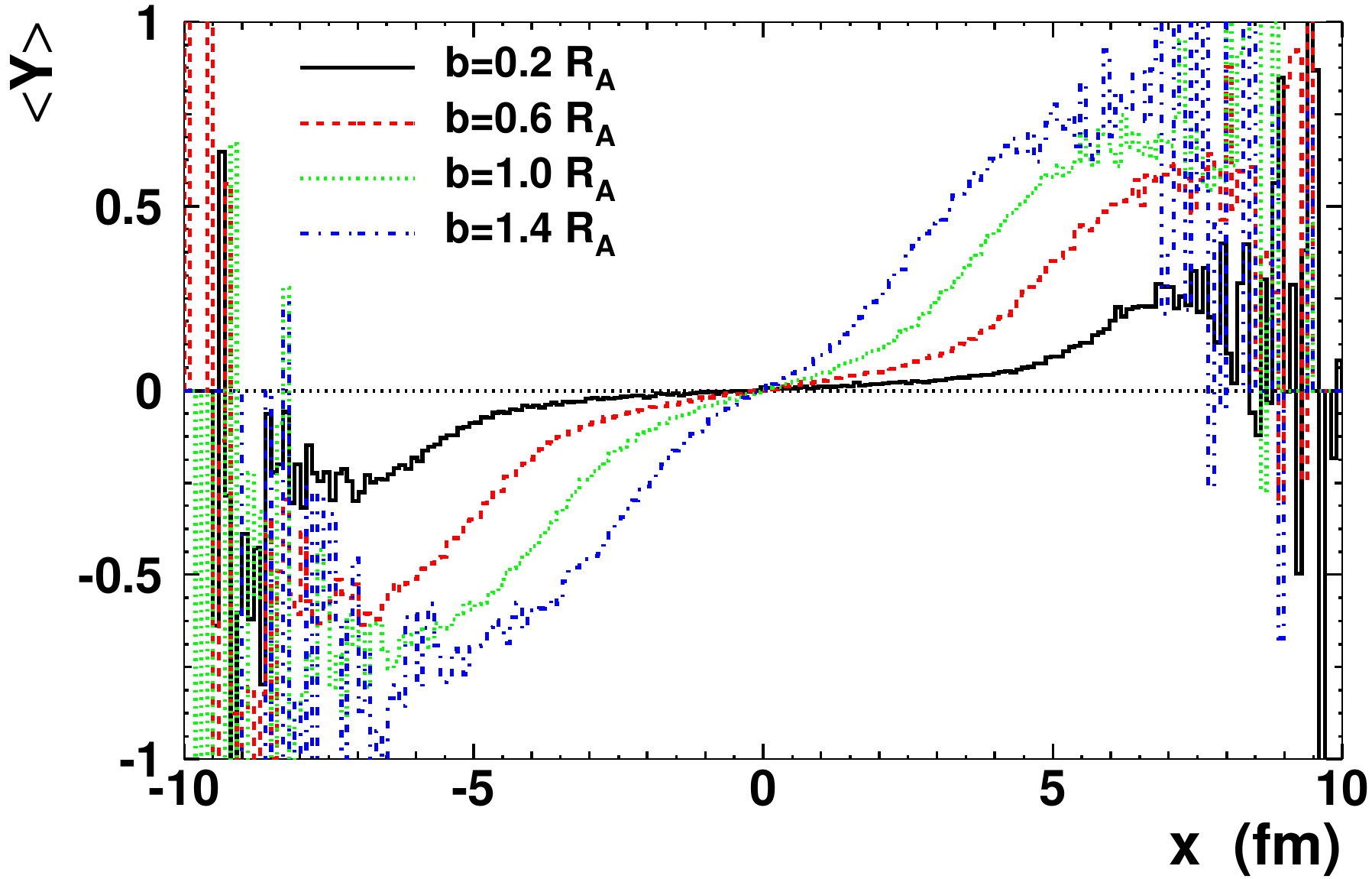}
\end{center}
\caption{The average rapidity $\langle Y\rangle$ of the
final state particles as a function of the transverse coordinate $x$ from HIJING
Mont Carlo simulations~\cite{Wang:1991ht,Wang:1996yf} of non-central
$Au+Au$ collisions at $\sqrt{s}=200$ GeV.
This figure is taken from \cite{Gao:2007bc}.}
\label{figyvsxHIJING}
\end{figure}

Shown in Fig.~\ref{figyvsxHIJING} is the average rapidity of particles in final state
as a function of the transverse coordinate $x$ for different values of the impact parameter $b$. 
We see that, besides the edge effects, the distributions have exactly the same qualitative 
features as given by the wounded nucleon model in Fig.~\ref{figpxb}.

\begin{figure}[ht]
\begin{center}
\includegraphics[width=7cm]{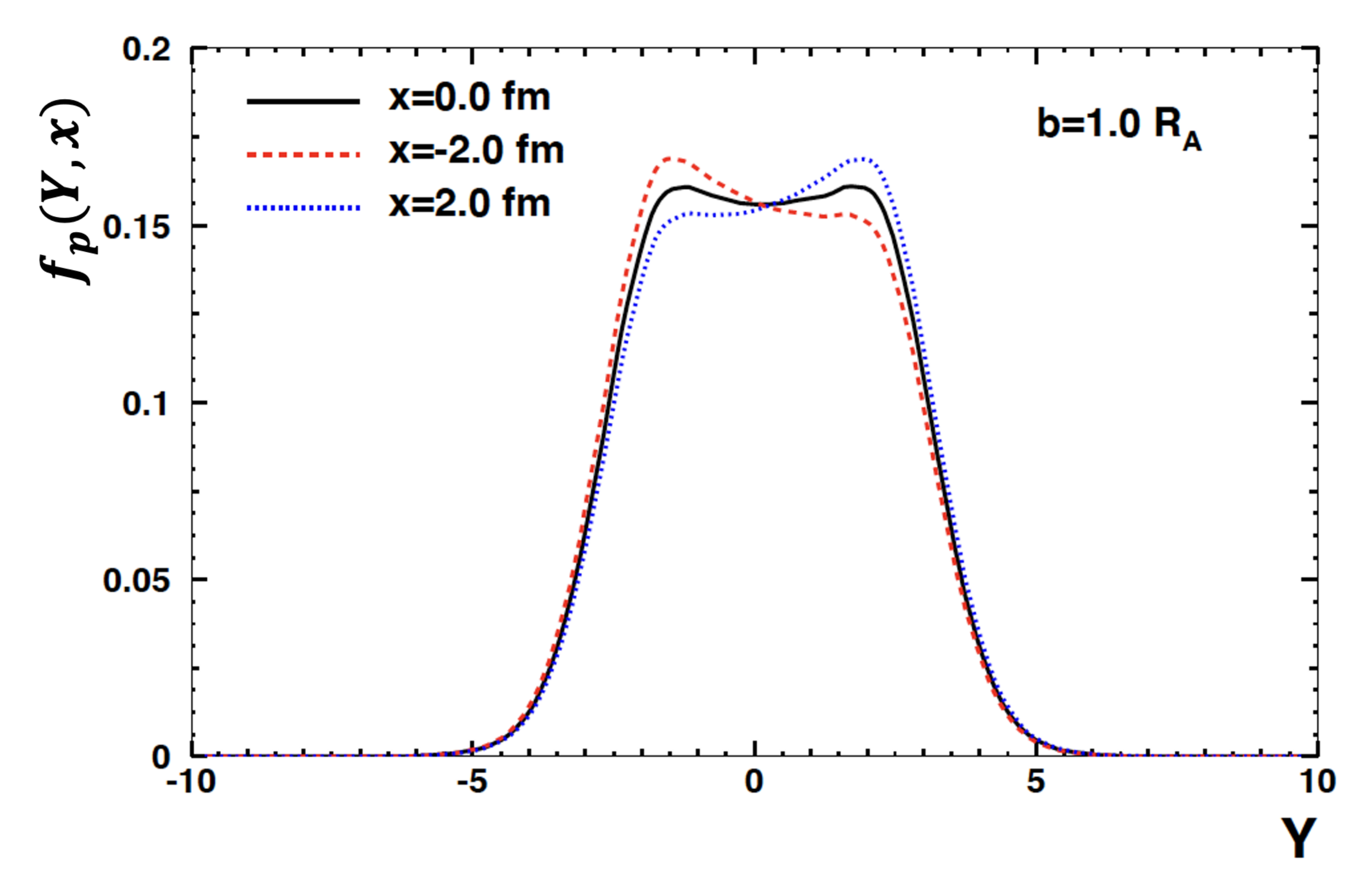}
\end{center}
\caption{The normalized rapidity distribution $f_p(Y,x,b,\sqrt{s})$ (in unit of 1/fm) of particles 
at different transverse position $x$ from HIJING simulations of non-central $Au+Au$ collisions
at $\sqrt{s}=200$ GeV.
This figure is taken from \cite{Gao:2007bc}.}
\label{figfpYxHIJING}
\end{figure}

In Fig.~\ref{figfpYxHIJING}, 
we see the results of normalized rapidity distributions $f_p(Y,x,b,\sqrt{s})$ 
at different values of the transverse coordinate $x$. 
We see that at finite values of $x$, 
$f_p(Y,x,b,\sqrt{s})$ evidently peak at larger values of rapidity $|Y|$. 
The shift in the shape of the rapidity distributions will provide the local longitudinal
fluid shear or finite relative orbital angular momentum for
two interacting partons in the local co-moving frame at any given rapidity $Y$. 
The fluid shear in the local co-moving frame at given rapidity $Y$ is finite
and peaks at large value of rapidity $|Y|\approx 2$. 
It is also generally smaller than the averaged fluid shear in the center of mass
frame of two colliding nuclei in the Landau fireball model.

\begin{figure}[ht]
\begin{center}
\includegraphics[width=7cm]{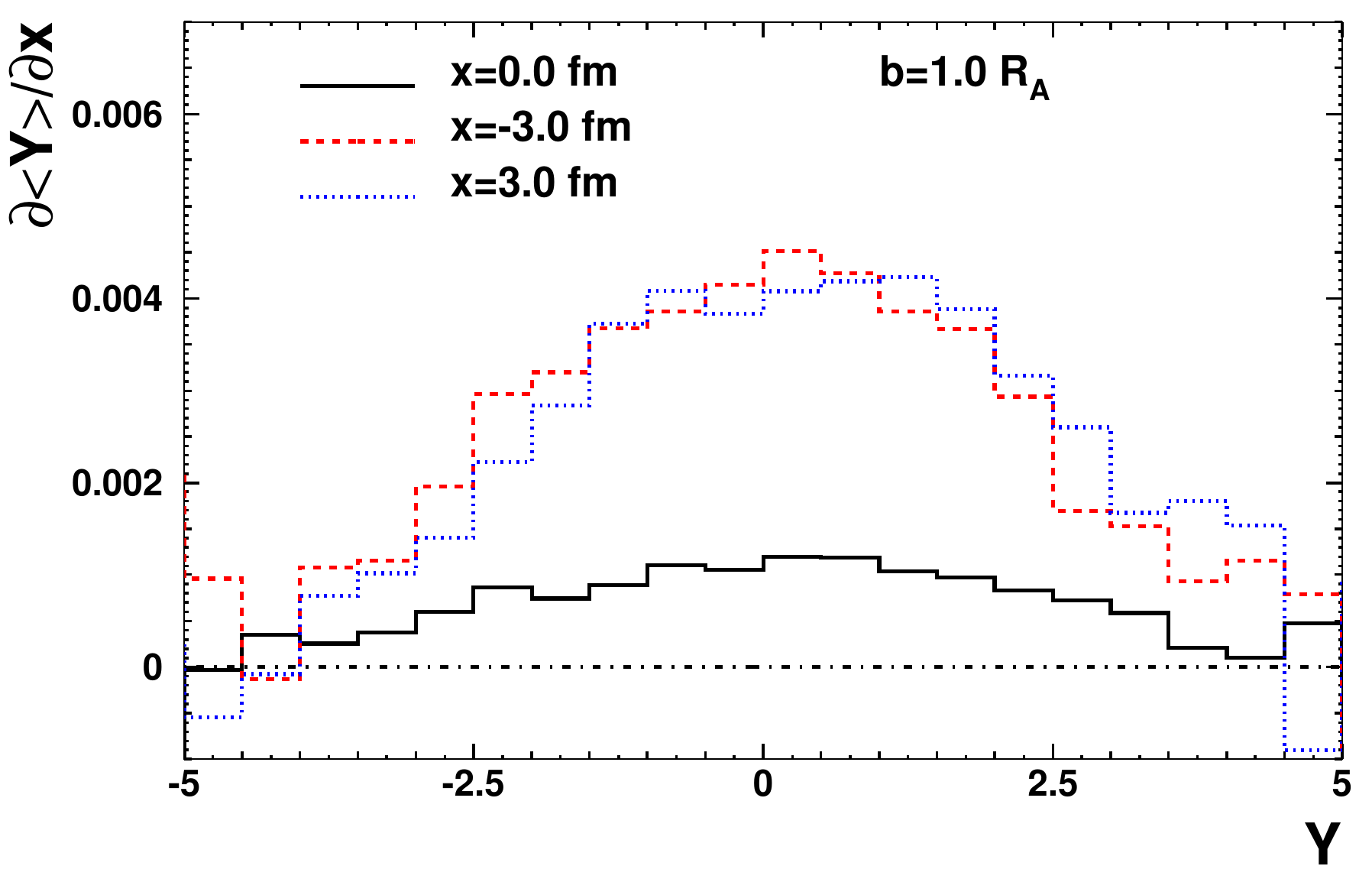}
\end{center}
\caption{(Color online) The average rapidity shear
$\partial \langle Y_l\rangle/\partial x$ within
a window $\Delta_Y=1$ as a function of the rapidity $Y$ at
different transverse position $x$ from HIJING calculation of
non-central $Au+Au$ collisions at $\sqrt{s}=200$ GeV.
This figure is taken from~\cite{Gao:2007bc}.}
\label{figdndydxHIJING}
\end{figure}

Shown in Fig.~\ref{figdndydxHIJING} is the average rapidity shear $\partial\langle Y_l\rangle/\partial x$ 
as a function of the rapidity $Y$ at different values of the transverse coordinate $x$ for $\Delta_Y=1$.
As we can see, the average rapidity shear has a positive and finite
value in the central rapidity region. 
As given by Eq.~(\ref{eqdapzdx}), the corresponding local relative longitudinal momentum shear 
$\partial \langle p_z\rangle/{\partial x}$ is determined by this rapidity shear multiplied by $p_T\cosh Y$.  
With $\langle p_T\rangle \approx 2T\sim 0.8$ GeV, we have
$\partial \langle p_z\rangle/\partial x\sim 0.003$
GeV/fm in the central rapidity region of a non-central $Au+Au$ collision
at the RHIC energy given by the HIJING simulations,
which is smaller than that from a Landau fireball model estimate.\\[-0.3cm]

(ii) {\it Results obtained using the BGK model}

In a recent paper~\cite{Liang2019}, a simple model~\cite{Brodsky:1977de} instead of HIJING~\cite{Wang:1991ht,Wang:1996yf} 
was used to repeat these calculations.   
Here, in this simple BGK model~\cite{Brodsky:1977de}, 
the rapidity distribution of produced hadrons is given by that in $pp$-collision, $dN_{pp}/dY$, 
multiplied by the following $Y$ linearly dependent factor, i.e., 
\begin{equation}
\frac{d^3N}{dxdydY}=\frac{dN_{pp}}{dY}\left[T_A^P(x,y,b)\frac{Y_L+Y}{2Y_L}+T_A^T(x,y,b)\frac{Y_L-Y}{2Y_L}\right], ~
\label{eqBGK}
\end{equation}
where $T_A^{P/T}$ is the thickness function for the projectile or target nucleus given by,
\begin{equation}
T_A^{P,T}(x,y,b)=\int dz~ \rho_A^{P,T}(x,y,z,b),
\end{equation}
$Y_L\approx \ln(\sqrt{s}/2m_N)$ is the maximum of the rapidity of the produced hadron;  
${dN_{pp}}/{dY}$ of hadrons produced in a $pp$-collision is taken 
as a modified Gaussian, 
\begin{equation}
\frac{dN_{pp}}{dY}=a_1\exp(-Y^2/a_2)/\sqrt{1+a_3\cosh^4 Y},
\label{eqdNppdY}
\end{equation}
where $a_1$, $a_2$ and $a_3$ are parameters depending on the collision energy. 
They are determined by fitting the results obtained 
from PYTHIA8.2~\cite{Sjostrand:2014zea} for $pp$ collisions.
A few examples obtained in~\cite{Liang2019} is given in Table~\ref{tabadNppdY}.

\begin{table}[h]
\caption{The parameters $a_1$, $a_2$ and $a_3$ for the rapidity distribution $dN_{pp}/dY$ 
given by Eq.~(\ref{eqdNppdY}) determined from PYTHIA8.2~\cite{Sjostrand:2014zea}. 
These numbers are taken from~\cite{Liang2019}.} \label{tabadNppdY}
\begin{center}
\begin{tabular}{cccc}\hline
~~~$\sqrt{s}$ (GeV)~~ & $a_1$ & $a_2$ & $a_3$ \\ \hline
200 & ~~4.584~~ & ~~~26.112~~~ & ~~ $9.70\times 10^{-8}$ \\
130 & 4.096 & 25.896 & ~~ $5.61\times 10^{-7}$ \\
62.4 & 3.862 & 18.911 &  ~~ $9.75\times 10^{-6}$ \\ 
39 & 3.420 & 18.779   & ~~ $6.61\times 10^{-5}$ \\ 
27 & 3.421 & 13.555   & ~~ $2.50\times 10^{-4}$ \\
11.5 & 2.784 & 10.488 & ~~ $5.90\times 10^{-3}$ \\ \hline
\end{tabular}
\end{center}
\end{table}

One great advantage to take this simple model~\cite{Brodsky:1977de} 
is that we have analytical expressions for all the quantities need so the calculations are quite simplified 
so that the physical significance can be easily demonstrated.   
In Ref.~\cite{Liang2019}, different results obtained using a hard sphere or Woods-Saxon nuclear distribution 
are given. In the following, we show those obtained using a hard sphere distribution as an example. 
Those obtained using Woods-Saxon are similar. 

\begin{figure}[ht]
\begin{center}
\includegraphics[width=8.7cm]{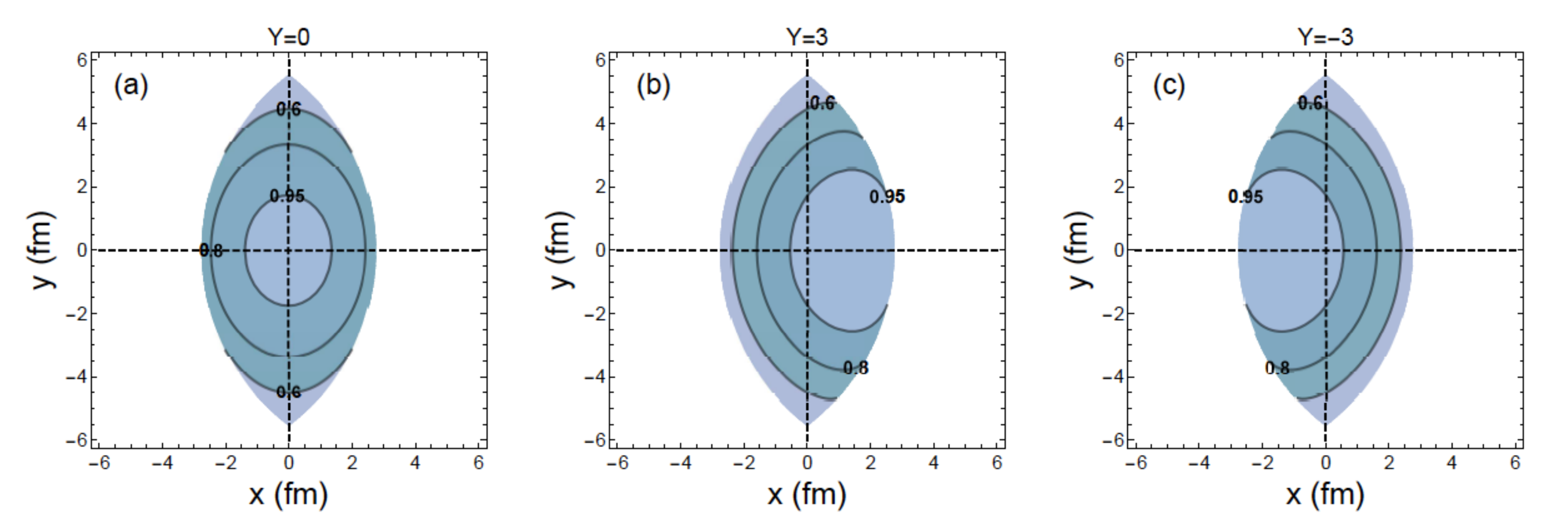}
\end{center}
\caption{Contour plots for distributions of hadrons obtained in BGK model~\cite{Brodsky:1977de} 
with a hard sphere 
nuclear distribution in the transverse plane for non-central $Au+Au$ collisions at $\sqrt{s}=200$ GeV 
at $b =1.2 R_A$ and different rapidities. 
The number on the contour line denotes the value on the line normalized by that at the origin. 
This figure is taken from \cite{Liang2019}.}
\label{figxyplotBGK}
\end{figure}

Shown in Fig.~\ref{figxyplotBGK} are the contour plots  for distributions of hadrons  in the transverse plane with different rapidities. 
This provides us a very intuitive picture how particles are distributed in the transverse plane at different rapidities. 
We see that at $Y=0$, the distributions are symmetric with respect to $x$ 
while at $Y=-3$ the center shifts to positive $x$ and at $Y=-3$ shifts to negative $x$. 
But they are all symmetric or even function of $y$.  

We integrate over the transverse coordinates and obtain,
\begin{eqnarray}
\frac{d^2N}{dxdY}&=&\frac{dN_{pp}}{dY}\left(\frac{dN_{\rm part}^P}{dx} \frac{Y_L+Y}{2Y_L}+\frac{dN_{\rm part}^T}{dx}\frac{Y_L-Y}{2Y_L}\right), ~ 
\label{eqdndxyBGK} \\
\frac{dN}{dx}&=&\frac{1}{2}\langle N_{pp}\rangle\left(\frac{dN_{\rm part}^P}{dx} +\frac{dN_{\rm part}^T}{dx}\right), ~
\label{eqdndxBGK}
\end{eqnarray}
where $\langle N_{pp}\rangle=\int dY (dN_{pp}/dY)$ is the average total number of particles produced in the $pp$ collision.
The normalized rapidity distribution at given $x$ is given by,
\begin{equation}
f_p(Y,x,b,\sqrt{s}~)=\frac{dN_{pp}}{\langle N_{pp}\rangle dY} 
\left[1+\frac{Y}{Y_L} R_N(x,b,\sqrt{s}~) \right], 
\label{eqfpYxBGK}
\end{equation}
where the ratio $R_N(x,b,\sqrt{s})$ is defined by Eq.~(\ref{eqRNxb}).  

The overall average value of $Y$ at a given $x$ is given by,
\begin{eqnarray}
\langle Y(x,b,\sqrt{s}~)\rangle
=\frac{\langle Y^2\rangle}{Y_L} R_N(x,b,\sqrt{s}~), 
\label{eqaYxBGK}
\end{eqnarray}
where $\langle Y^2\rangle=\int Y^2dY (dN_{pp}/dY)/\langle N_{pp}\rangle$ is the average value of $Y^2$ in $pp$ collision.

Compare Eq.~(\ref{eqdndxBGK}) with Eq.~(\ref{eqpzxb}), we see that $\langle Y(x,b,\sqrt{s}~)\rangle$ in this model has 
exactly the same behavior as $p_z(x,b,\sqrt{s})$ in the Landau fireball model. 

\begin{figure}[ht]
\begin{center}
\includegraphics[width=5.6cm]{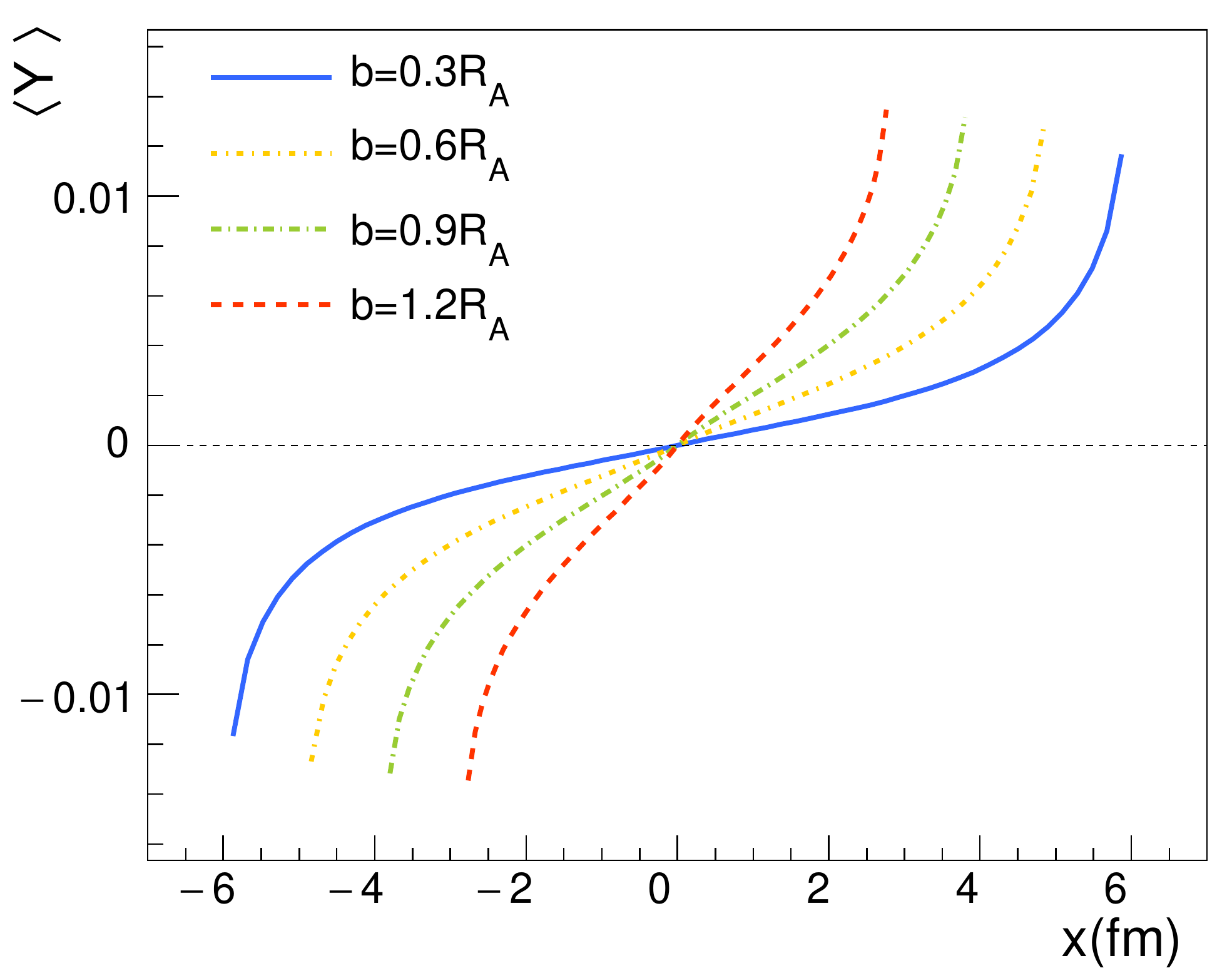}
\end{center}
\caption{The average rapidity $\langle Y\rangle$ of the final state particles 
as a function of the transverse coordinate $x$ from BGK~\cite{Brodsky:1977de} with a hard sphere 
nuclear distribution in non-central $Au+Au$ collisions at $\sqrt{s}=200$ GeV.
This figure is taken from \cite{Liang2019}.}
\label{figyvsxBGK}
\end{figure}

\begin{figure}[ht]
\begin{center}
\includegraphics[width=7.7cm]{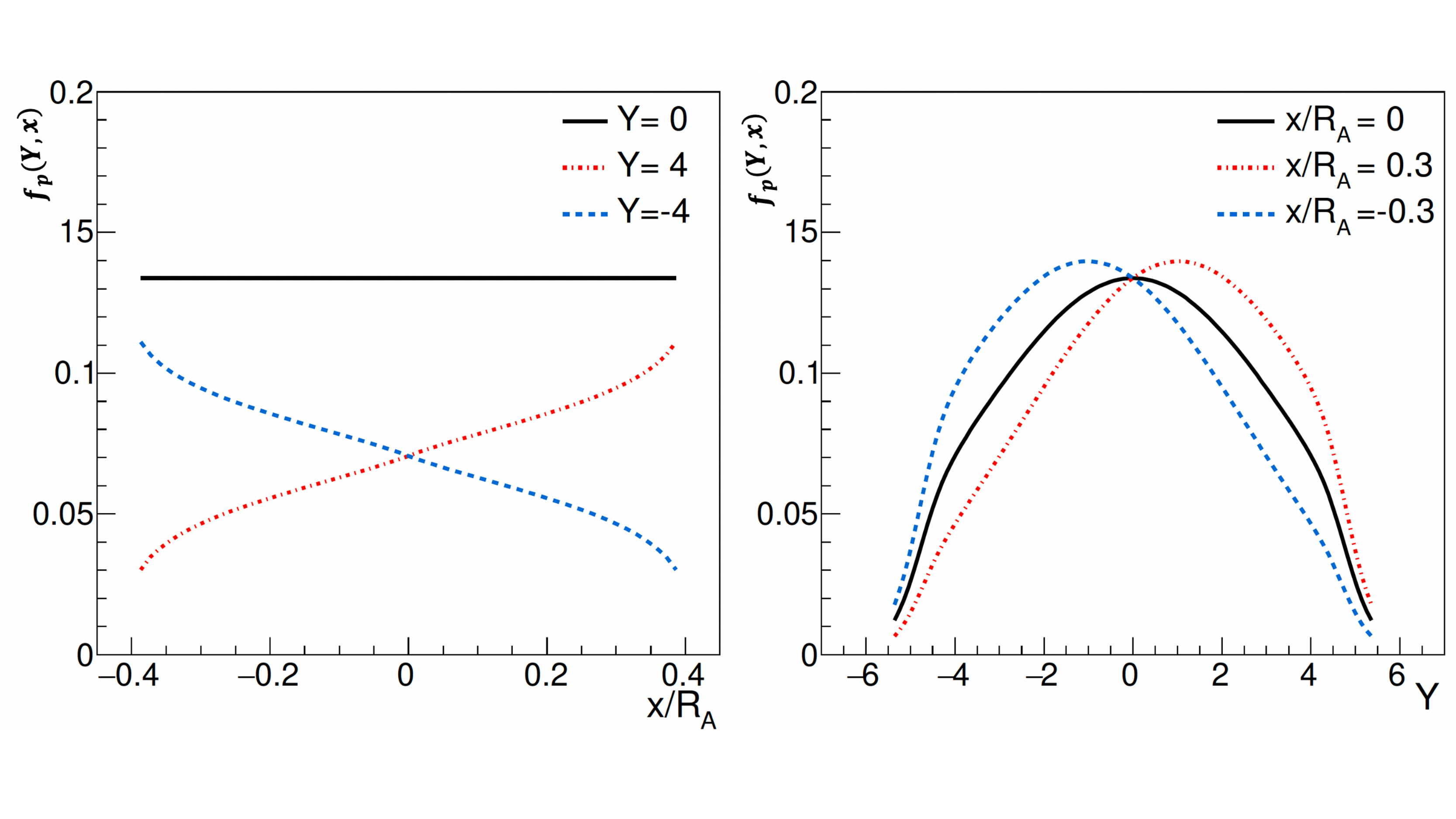}
\end{center}
\caption{
The normalized distribution $f_p(Y, x)$ of hadrons in BGK model~\cite{Brodsky:1977de} with 
a hard sphere 
nuclear distribution in the transverse plane for 
non-central $Au+Au$ collisions at $\sqrt{s}=200$ GeV and $b = 1.2R_A$
as a function of $x$ at different rapidity $Y$ (left panel),  
and as a function of $Y$ at different $x$ (right panel). 
This figure is taken from \cite{Liang2019}.}
\label{figfYxBGK}
\end{figure}

Fig.~\ref{figyvsxBGK} shows the average values of $Y$ as functions of $x$ plotted in the same format 
as that in Fig.~\ref{figyvsxHIJING}. 
We see that, besides those in the edge regions where the calculations need to be modified,
the results exhibit the same qualitative features as those in Fig.~\ref{figyvsxHIJING}, 
though the quantitative results show slight differences. 
Fig.~\ref{figfYxBGK} shows the corresponding normalized distributions $f_p(Y,x,b,\sqrt{s})$.
The right panel is to compare with Fig.~\ref{figfpYxHIJING} where HIJING monte-Carlo model was used. 
We see in particular a clear shift of the peak to positive $Y$ for $x>0$ and to negative $Y$ for $x<0$.

\begin{figure}[ht]
\begin{center}
\includegraphics[width=5.7cm]{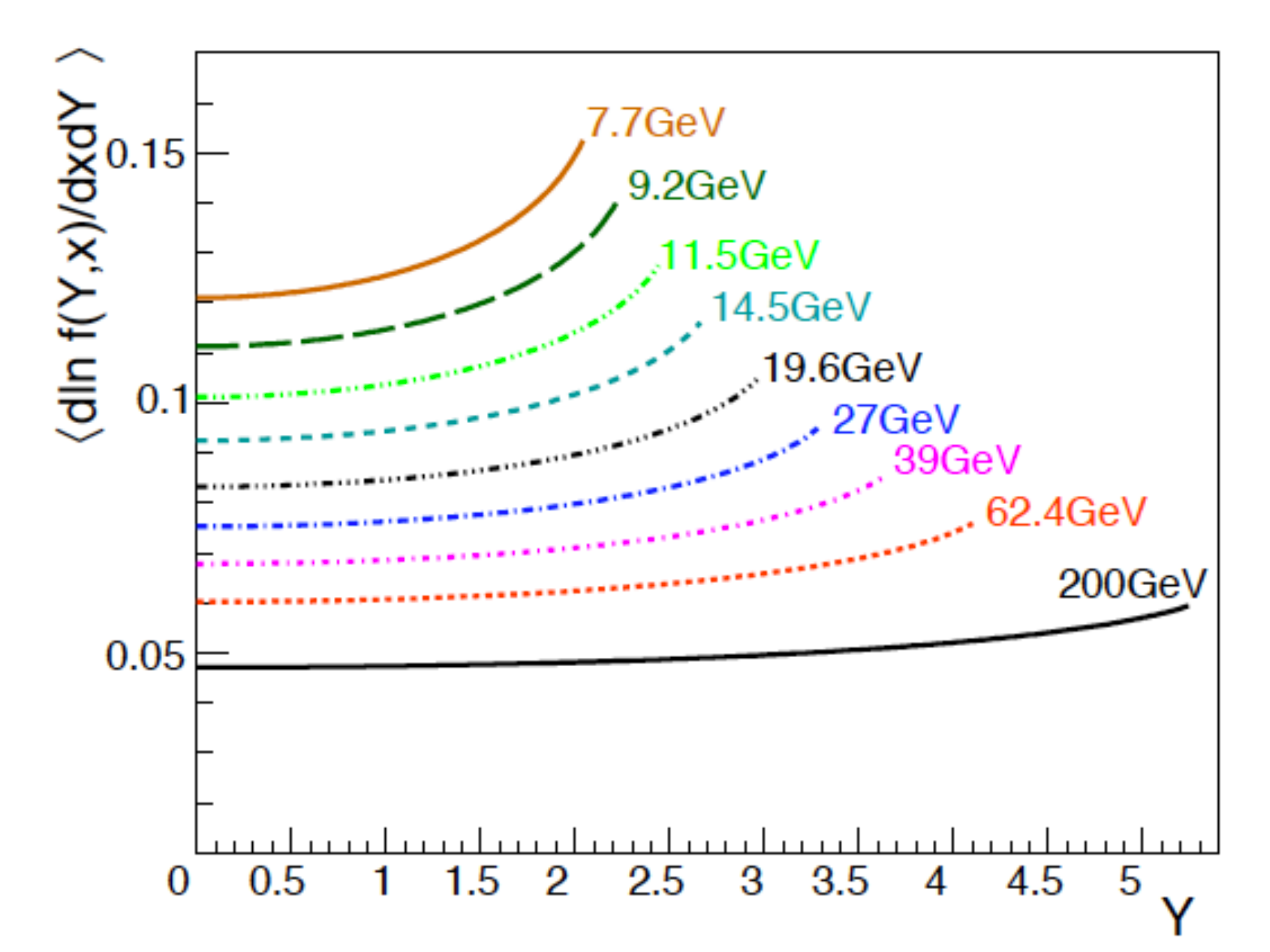}
\end{center}
\caption{
The averaged $\langle \xi_p\rangle=\langle\partial^2{\rm ln}f_p/\partial Y\partial x\rangle$ 
as a function of rapidity $Y$ of final state particles in BGK model~\cite{Brodsky:1977de} 
with a hard sphere  
nuclear distribution for non-central $Au+Au$ collisions at different energies and impact parameter $b = 1.2R_A$. 
This figure is taken from \cite{Liang2019}.}
\label{figllyBGK}
\end{figure}

To show the rapidity dependence of the local orbital angular momentum or momentum shear, 
Ref.~\cite{Liang2019} also calculated $\langle \xi_p\rangle $ 
defined in Eq.~(\ref{eqxip}) as a function of $Y$ at different energies. 
The obtained results are shown in Fig.~\ref{figllyBGK}. 
From this figure, we see that the rapidity dependence of $\langle \xi_p\rangle$ is quite weak 
except at the limiting region when $Y$ reaches its maximum. 
This represents the characteristics of the rapidity dependence of the microscopic local momentum shear and may also 
reflect the rapidity dependence of the corresponding macroscopic observable effects.  

\section{Spin-orbit coupling in a relativistic quantum system}
\label{secls}

The spin-orbit coupling is a well known effect in a quantum system. 
Here, we present a short discussion of the origin and a brief review of related phenomena.  

\subsection{Dirac equation and spin-orbit coupling}
\label{seclsDirac}

The spin-orbit coupling is an intrinsic property for a relativistic fermionic quantum system. 
This is derived explicitly from Dirac equation. 
A number of characteristics of Dirac equation show that it describes particles of spin-1/2, 
and the spin and orbital angular momentum couple to each other intrinsically even for free particles.  
Here, we recall a few of such characteristics in the following. 

First of all, it is well known that, even for a free Dirac particle, the Hamiltonian $\hat H$ 
does not commute with the orbital angular momentum $\hat{\vec L}$ 
and the spin ${\vec\Sigma}$ separately, but commutes with the total angular momentum $\hat{\vec J}=\hat{\vec L}+{\vec\Sigma}/2$,
i.e., 
$[\hat H, \hat{\vec L}~]=-i\vec\alpha\times\hat{\vec p}$, $[\hat H, \vec\Sigma]=2i\vec\alpha\times\hat{\vec p}$, but
$[\hat H, \hat{\vec J}~]=0$.
This shows clearly that spin and orbital angular momentum couple to each other 
and transform from one to another in a relativistic fermionic quantum system,
though the strength of the spin-orbit coupling can be different for an electromagnetic  or a strongly interacting system.  

Second, the magnetic momentum of a Dirac particle with electric charge $e$ is obtained simply 
by replacing the classical expression $\vec M=e\vec r\times\vec v/2$ with operators, i.e., 
$\hat{\vec M}={e}\vec r\times\vec\alpha/2$. 
In an eigenstate $|\psi\rangle$ 
of $\hat H$, if we take the non-relativistic approximation $E\approx m$, we obtain immediately that~\cite{Liang:1992hw}, 
\begin{equation}
\langle \vec M\rangle \approx \frac{e}{2m}\langle\varphi|(\hat{\vec L}+\vec\sigma)|\varphi\rangle,  \label{eqMnr}
\end{equation}
where $\varphi$ is the upper component of $\psi$.
This is just the well known result for point-like spin-1/2 particles where the Landre factors 
are $g_L=1$ and $g_s=2$. 

If we consider a Dirac particle moving in a central potential, the stationary state is the 
eigenstate of $\hat{H}, \hat{\vec J}^{~2}, \hat{J}_z$ and the parity $\hat{\cal P}$ 
with eigenvalues $(\varepsilon,j,m,{\cal P})$, i.e.,
\begin{equation}
\psi_{\varepsilon jm{\cal P}}(r,\theta,\phi,s)=
\left(\begin{array}{c} f_{\varepsilon l}(r)\Omega_{jm}^l(\theta,\phi)\\(-1)^{\frac{1}{2}(l-l'+1)} g_{\varepsilon l'}(r)\Omega_{jm}^{l'}(\theta,\phi)\end{array}\right), 
\label{eqpsispheric}
\end{equation}
where $\Omega_{jm}^{l}(\theta,\phi)$ is the $2\times1$ spheric harmonic wave function in the non-relativistic case, 
$f_{\varepsilon l}(r)$ and $g_{\varepsilon l'}(r)$ are the radial parts, $j=l\pm1/2=l'\mp1/2$ and ${\cal P}=(-1)^l$.
In the ground state $\varepsilon=\varepsilon_0$, $j=1/2$, ${\cal P}=+$,  
the magnetic moment is given by~\cite{Liang:1992hw},
\begin{equation}
\langle \varepsilon_0,1/2,m,+|\hat{\vec M}|\varepsilon_0,1/2,m,+\rangle=\mu_q\langle\xi(m)|\vec\sigma|\xi(m)\rangle, \label{eqMspheric}
\end{equation}
where $\mu_q=-2e\int r^3dr f_{00}(r)g_{01}(r)/3$ is a constant determined by ground state radial wave functions, 
$\xi(m)$ is the eigenstate of $\sigma_z$ and is a Pauli spinor.  
Eq.~\ref{eqMspheric} has exactly the same form as that for a quark at rest. 
This explains why the static quark model works well in describing the magnetic moment of baryon although we know that 
the quark mass is small and the relativistic treatment has to be used. 

 Third, we consider a Dirac particle moving in a magnetic field with potential $A=(\phi,\vec A)$. 
 By replacing $\hat p$ with $\hat p-eA$ in the Dirac equation and taking the non-relativistic approximation, 
 we obtain immediately,
 \begin{equation}
 \hat H_{\rm nr}=\frac{1}{2m}(\hat{\vec p}-e\vec A)^2-e\phi-\frac{1}{4m^2}\frac{d\phi}{rdr}\hat{\vec L}\cdot\vec\sigma,
\label{eqHeff}
 \end{equation}
 where the spin-orbit coupling is obtained automatically.
 
 \subsection{Spin-orbit coupling in systems under electromagnetic interactions}
 
Intuitively, the spin-orbit coupling in systems under electromagnetic interactions has a very clear physical picture 
and also leads to many well known effects.  
The most famous textbook example might be the fine structure of atomic light spectra. 
Here, we consider the electron moving in the electromagnetic field induced by the hydrogen atom, 
we take the extra $1/2$ factor due to Thomas precession into account and obtain immediately,
\begin{equation}
V_{ls}(\vec r)=-\frac{1}{2}\vec\mu\cdot \vec B=\frac{e}{4m}\vec\sigma\cdot \vec v\times\vec E=\frac{e}{4m^2}\frac{d\phi}{rdr}\vec\sigma\cdot \vec L.
\end{equation}
This is exactly the same as that in Eq.~(\ref{eqHeff}) derived from Dirac equation. 

The spin-orbit coupling plays also a very important in modern spintronics in condensed matter physics where spin transport in 
the electromagnetically interacting system is studied. 
There are also examples in electromagnetically interacting systems where 
spin polarization (magnetization) and orbital angular momentum (rotation) are transferred from one to the other.  
Earlier examples may even be traced back to 
Einstein and deHaas~\cite{EinsteindeHass} and Barnet~\cite{barnett}. 
It was known as the Einstein-deHaas effect where the rotation is caused by magnetization 
and the Barnett effect that is the gyromagnetic effect where magnetization is caused by rotation. 

 \subsection{Spin-orbit coupling in systems under strong interactions}

In systems under strong interactions, the spin-orbit coupling also leads to many distinguished effects.  
One of such famous examples is the nuclear shell model developed 
by Mayer and Jensen~\cite{shellmodel,Mayer:1949pd,Haxel:1949fjd}
where the spin-orbit coupling plays a crucial role to produce the magic numbers of atomic nuclei.  

There is no such a clear intuitive picture for the spin-orbit interaction in systems under strong interactions 
as that for electromagnetic interactions so the strength can not be derived explicitly. 
Usually in the covariant relativistic formalism, the spin-orbit coupling does appears explicitly.  
However, the role that it plays can be seen whenever one separates spin and orbital angular momentum from each other. 
Besides the famous example in the nuclear shell model, 
another explicit example is the heavy quarkonium spectra where spin-orbit 
coupling has to be taken into account~\cite{Brambilla:2004jw}. 

Even more interesting is that, 
in the frontier of high energy spin physics, it seems that spin-orbit coupling plays a key role in understanding 
all the four classes of striking spin effects mentioned in Sec.~\ref{secint} observed in experiments since 1970s.  
The simplest argument that orbital angular momentum contributes significantly to proton spin is that 
discussed in the first point in Sec.~\ref{seclsDirac} where it has been shown that 
the orbital angular momentum for a Dirac particle is not a good quantum number. 
Hence even if a quark is in the ground states in a central potential as given by Eq.~(\ref{eqpsispheric}) 
the average value of the orbital angular momentum is not zero. 
If we e.g. consider a quark in the ground state in a spheric potential well with infinite depth such as in the MIT bag model,  
the orbital angular momentum contributes $\sim35\%$ to the total angular momentum. 

Both phenomenological model~\cite{Liang:1992hw,Boros:1993ps} and pQCD calculations~\cite{Brodsky:2002cx} 
indicate that orbital angular momentum of quarks in a polarized nucleon and the initial or final state interactions 
are responsible for SSA observed~\cite{Klem:1976ui,Dragoset:1978gg,Adams:1991cs,Liang:2000gz} in inclusive hadron-hadron collisions. 
It has also been shown that transverse hyperon polarization 
observed~\cite{Lesnik:1975my,Bunce:1976yb,Bensinger:1983vc,Gourlay:1986mf,Krisch:2007zza} 
in unpolarized hadron-hadron collisions are closely related 
to SSA thus has the same physical origins~\cite{Liang:1997rt}.
The spin analyzing power observed~\cite{O'Fallon:1977cp,Crabb:1978km,Cameron:1985jy,Krisch:2007zza}  
in elastic $pp$ scattering is due to color magnetic interaction 
during the scattering~\cite{Liang:1989mb} thus originates also from the orbital angular momentum 
of the constituents in the polarized proton.  
The study of the role played by the orbital angular momentum is one of the core issues currently 
in high energy spin physics. See recent reviews such as~\cite{Bass:2004xa,DAlesio:2007bjf,Aidala:2012mv,Liang:2015nia,Chen:2015tca}.

\section{Theoretical predictions on the global polarization effect of QGP in HIC}
\label{secgp}

It has been shown~\cite{Liang:2004ph} that due to spin-orbit interactions in a strongly interacting system such as QGP, 
the orbital angular momentum can be transferred to the polarization of the constituents in the system 
such as the quarks and anti-quarks.

\subsection{Global quark polarization in QGP in HIC}
\label{secqpol}

In Sec.~\ref{secoam}, we have seen that in a non-central $AA$ collision, 
there is a huge global orbital angular momentum for the colliding system. 
Such a global angular momentum leads to the longitudinal fluid shear in the produced system of partons. 
A pair of interacting partons will have a finite value of relative
orbital angular momentum along the direction opposite to the normal of the reaction plane. 
We have also seen in Sec.~\ref{secls} that spin-orbit coupling is an intrinsic property of a relativistic system. 
It is thus natural to ask whether the orbital angular momentum or momentum shear lead to the polarization of partons in the system. 

There is no field theoretical calculation that can be applied directly to answer this question 
because usually the calculations are in the momentum space where the momentum shear with respect to $x$ coordinate 
can not be taken into account. 
To achieve this, Ref.~\cite{Liang:2004ph} took the approach by considering 
parton scattering with impact parameter in the preferred direction and reach the positive conclusion. 
We summarize the studies of Refs.~\cite{Liang:2004ph} and~\cite{Gao:2007bc} in this section.

\subsubsection{Quark scattering at fixed impact parameter}

To be explicit, we consider the scattering 
$q_1(p_1)+q_2(p_2)\to q_1(p_3)+q_2(p_4)$ of two quarks with different flavors.
The scattering matrix element in momentum space is given by,
\begin{equation}
S_{fi}=\langle f|\hat S|i\rangle={\cal M}_{fi}(q) (2\pi)^4\delta^4(p_1+p_2-p_3-p_4),
\label{eqsfi}
\end{equation}
where $p_i=(E_i,\vec p_i)$ is the four momentum of the quark, 
$q=p_1-p_3=p_4-p_2$ is the four momentum transfer 
and ${\cal{M}}_{fi}(q)$ is the scattering amplitude in momentum space. 
The incident momenta are taken as in $z$ or $-z$ direction 
and the transverse momentum is denoted as $\vec p_T=\vec p_{3T}=-\vec p_{4T}$.
The differential cross section in the momentum space is given by,
\begin{eqnarray}
&&d\sigma=\frac{c_{qq}}{F}
\frac{|S_{fi}(q)|^2}{TV}\frac{d^3{{p}}_3}{(2\pi)^{3}2E_3} \frac{d^3{{p}}_4}{(2\pi)^{3}2E_4},~~~~~~
\label{eqqqcs}
\end{eqnarray}
where $T$ and $V$ are interaction time and volume of the space, 
$c_{qq}=2/9$ is the color factor, and
$F=4\sqrt{(p_1\cdot p_2)^2-m_1^2m_2^2}$ is the flux factor. 
Here, just for clarity of equations, we omit the spin indices and will pick them up later in the following.

It can easily be verified that, 
\begin{equation}
S_{fi}=\int d^2x_T\int\frac{d^2q_\perp}{(2\pi)^2} {\mathcal{M}}_{fi}(q) e^{-i(\vec q_T+\vec p_T)\cdot\vec x_T}(2\pi)^4\delta^4(p_1+p_2-p_3-p_4),
\label{eqsfixT}
\end{equation}
where we use $\vec x_T$ to denote the impact parameter of the two scattering quarks to distinguish it from 
the impact parameter $\vec b=b~\vec e_x$ of the two nuclei. 
By inserting Eq.~(\ref{eqsfixT}) into (\ref{eqqqcs}), we obtain,
\begin{equation}
d\sigma=\frac{c_{qq}}{F}\int d^2x_T\int\frac{d^2q_\perp}{(2\pi)^2}\frac{d^2k_\perp}{(2\pi)^2} e^{-i(\vec q_T-\vec k_T)\cdot\vec x_T}
\frac{{\mathcal{M}}_{fi}(q)}{\Lambda(q)}\frac{{\mathcal{M}}_{fi}^{*}(k)}{\Lambda(k)} ,~~~
\label{eqcsxT}
\end{equation}
where ${\mathcal{M}}_{fi}(q)$ and ${\mathcal{M}}_{fi}(k)$
are scattering amplitudes in momentum space
with four momentum transfer $q=(q_0,\vec{q}_T,q_z)$ and $k=(k_0,\vec{k}_T,k_z)$ respectively;
$\Lambda(q)$ is a kinematic factor obtained in carrying out the integration and is given by,  
\begin{equation}
{\Lambda^{-2}(q)}=\int \delta^4(p_1+p_2-p_3-p_4)\delta^2(\vec q_T+\vec p_T)
\frac{d^3{{p}}_3}{2E_3} \frac{d^3{{p}}_4}{2E_4} 
=\frac{1}{(E_1+E_2)p_{3z}}~,
\label{eqLambdaqdef}
\end{equation}
where $p_{3z}$ is the positive solution of $\sqrt{\vec q_T^2+p_{3z}^2+m_3^2}+\sqrt{\vec q_T^2+p_{3z}^2+m_4^2}=E_1+E_2$. 
Here, in obtaining Eq.~(\ref{eqcsxT}), we have taken the symmetric form with exchange of $q$ and $k$ 
to guarantee the integrand of $d^2x_T$ to be positive definite. 

We pick up the spin indices and suppose that we are interested in the polarization of quark $q_1$ after the scattering.
We therefore average over the spins of initial quarks and sum over the spin of quark $q_2$ in the final state.
In this case, we have, 
\begin{equation}
\frac{d^2\sigma_{\lambda_3}}{d^2x_T}
=\frac{c_{qq}}{16F}\sum_{\lambda_1,\lambda_2,\lambda_4} \int\frac{d^2q_T}{(2\pi)^2}\frac{d^2k_T}{(2\pi)^{2}}
e^{i({\vec{k}}_{T}-{\vec{q}}_{T})\cdot{\vec{x}}_{T}}
\frac{\mathcal{M}(q)}{\Lambda(q)} \frac{{\mathcal{M}}^{*}(k)}{{\Lambda}(k)} .
\label{eqdcslambdadxT}
\end{equation}
We define, 
\begin{eqnarray}
\frac{d^2\Delta\sigma}{d^2x_T}&=& \frac{d^2\sigma_{+}}{d^2x_T}-\frac{d^2\sigma_{-}}{d^2x_T}, \label{eqdDcsdxT}\\
\frac{d^2\sigma}{d^2x_T}&=&\frac{d^2\sigma_{+}}{d^2x_T}+\frac{d^2\sigma_{-}}{d^2x_T}, \label{eqdcsunpoldxT}
\end{eqnarray}
where $\lambda_3=+$ or $-$ denotes that the spin of $q_1$ after the scattering is in the 
positive or negative direction of the normal $\vec n$ of the reaction plane; 
${d^2\sigma}/{d^2x_T}$ is just the unpolarized cross section at the fixed impact parameter. 

Suppose that the impact parameter $\vec x_T$ has a given distribution $f_{qq}(\vec x_T,b,Y,\sqrt{s})$, 
we can calculate the polarization in the following way, 
\begin{eqnarray}
\langle\Delta\sigma\rangle&=&\int d^2x_T f_{qq}(\vec x_T,b,Y,\sqrt{s}) \frac{d^2\Delta\sigma}{d^2x_T}, \label{eqDcsint} \\
\langle\sigma\rangle&=&\int d^2x_T f_{qq}(\vec x_T,b,Y,\sqrt{s}) \frac{d^2\sigma}{d^2x_T}, 
\label{eqcsint}
\end{eqnarray}
and the polarization of the quark $q_1$ after the scattering is given by,
\begin{equation}
P_q=\langle\Delta\sigma\rangle/\langle\sigma\rangle. \label{eqPqDcsovercs}
\end{equation}

As discussed in Sec.~\ref{secoam}, the average relative 
orbital angular momentum $\vec {\it l}$ of two scattering quarks is in the opposite direction
of the normal of the reaction plane in non-central $AA$ collisions.
Since a given direction of $\vec {\it l}$ corresponds to a given direction of $\vec x_T$, 
there should be a preferred direction of $\vec x_T$ at a given direction of the nucleus-nucleus
impact parameter $\vec b$. 
The distribution $f_{qq}(\vec x_T,b,Y,\sqrt{s})$ of $\vec x_T$ at given $\vec b$ 
depends on the collective longitudinal momentum distribution shown in Sec.~\ref{secoam}.
Clearly, it depends on the dynamics of QGP and that of $AA$ collisions. 

To see the qualitative features of the physical consequences explicitly, 
Refs.~\cite{Liang:2004ph,Gao:2007bc} took a simplified $f_{qq}(\vec x_T,b,Y,\sqrt{s})$ as an example, i.e.,  
a uniform distribution of $\vec x_T$ in the upper half $xy$-plane with $x>0$, i.e., 
\begin{equation}
f_{qq}(\vec x_T,b,Y,\sqrt{s})\propto \theta(x),
\label{eqfqq}
\end{equation}
so that 
\begin{eqnarray}
\langle\Delta\sigma\rangle&\approx &\int_0^\infty dx \int_{-\infty}^\infty dy \frac{d^2\Delta\sigma}{d^2x_T}, \label{eqDcsint2}\\
\langle\sigma\rangle&\approx&\int_0^\infty dx \int_{-\infty}^\infty dy \frac{d^2\sigma}{d^2x_T}. 
\label{eqcsint2}
\end{eqnarray}

\subsubsection{Quark scattering by a static potential}
\label{csSPM}

To see the characteristics of the physical consequences clearly, in~\cite{Liang:2004ph}, 
we considered first a quark scattering by a static potential. 
Here, it is envisaged that a quark incident in $z$-direction and is scattered by 
an effective static potential induced by other constituents of QGP. 
In this case, we obtain,
\begin{equation}
{\cal M}_{fi}(q)=\bar{u}_{\lambda}(p+q)~ A\hspace{-5pt}\slash (q) ~ u(p), ~
\label{eqmSPM}
\end{equation}
where $A(q)=(A_0(q),\vec 0)$ and $A_0(q)=g/(q^2+\mu_D^2)$ is the screened static potential
with Debye screen mass $\mu_D$~\cite{gw93}.
It follows that, 
\begin{equation}
{\cal M}_{fi}(q){\cal M}_{fi}^*(k)=A_0(q)A_0(k)\bar{u}_{\lambda}(p+q)
(\tilde p\hspace{-4.5pt}\slash+m_q) u_{\lambda}(p+k), 
\label{eqmmSPM}
\end{equation}
where $\tilde p\equiv(E,-\vec p)$.  
We choose $\vec n$ as the quantization axis of spin and denote the eigenvalue by $\lambda=\pm 1$.  
For small angle scattering, $q_T,k_T \sim \mu_D \ll E$, we obtain,
\begin{equation}
{\cal M}_{fi}(q){\cal M}_{fi}^*(k)\approx 4E^2A_0(q)A_0(k)
\left[1-i \lambda\frac{(\vec{q}_T-\vec{k}_T)\cdot (\vec{n}\times\vec{p})} {2E(E+m_q)} \right] , 
\label{eqmmresSPM}
\end{equation}
and the cross sections are given by,
\begin{eqnarray}
\frac{d^2\sigma}{d^2 x_T}&=&\frac{g^4c_{T}}{4} \int\frac{d^2q_T}{(2\pi)^{2}}\frac{d^2k_T}{(2\pi)^{2}}  
\frac{e^{i({\vec{k}}_{T}-{\vec{q}}_{T})\cdot\vec{x}_T}}{(q_T^{2}+\mu_D^2)
(k_T^{2}+\mu_D^2)},~~~ \label{eqcsSPM} \\
\frac{d^2\Delta{\sigma}}{d^2 x_T}&=&i\frac{g^4c_T}{8\vec p^2} \int\frac{d^2 q_T}{(2\pi)^2}\frac{d^2 k_T}{(2\pi)^2}  
\frac{~(\vec{n}\times\vec{p})\cdot (\vec{k}_T-\vec{q}_T)~e^{i(\vec{k}_T-\vec{q}_T)\cdot\vec{x}_T}}
{(q_T^{2}+\mu_D^2)(k_T^{2}+\mu_D^2)}.~~~ \label{eqDcsSPM}
\end{eqnarray}
where $c_T$ is the color factor. 
It is interesting to note that, under such approximation, these two parts of the cross section are related to each other,
\begin{equation}
\frac{d^2\Delta{\sigma}}{d^2 x_T}=\frac{1}{2\vec p^2}(\vec{n}\times\vec{p})\cdot\nabla\frac{d^2\sigma}{d^2 x_T}.
\label{eqcsDcsSPM}
\end{equation}
Completing the integrations over $d^2q_T$ and $d^2k_T$ by using the integration formulae,  
\begin{equation}
\int\frac{d^2 q_T}{(2\pi)^2} \frac{e^{i\vec q_T\cdot \vec x_T}}{q_T^2+\mu_D^2}
=\int\frac{q_Tdq_T}{2\pi}\frac{J_0(q_Tx_T)}{q_T^2+\mu_D^2}
=\frac{1}{2\pi}K_{0}(\mu_Dx_T),
\label{eqintform}
\end{equation}
we obtain from Eqs.~(\ref{eqcsSPM}) and (\ref{eqDcsSPM}) that~\cite{Liang:2004ph},
\begin{eqnarray}
\frac{d^2\sigma}{d^2x_T}&=&\alpha_s^2c_T K_0^2(\mu_{D}x_T), \label{eqcsresspm} \\
\frac{d^2\Delta\sigma}{d^2 x_T}&=&\alpha_s^2c_T
\left[(\vec{p}\times\vec{n})\cdot\hat{\vec{x}}_T/\vec p^2\right] \mu_D K_0(\mu_{D}x_T)K_1(\mu_{D}x_T). \label{eqDcsresspm}
\end{eqnarray}
where $J_0$ and $K_0$ are the Bessel and modified Bessel functions respectively and $x_T=|\vec x_T|$.
The unpolarized cross section just corresponds to 
$d^2\sigma/d^2q_T=4\pi\alpha_s^2c_T/(q_T^2+\mu_D^2)^2$ in the momentum space.

It is evident from Eq.~(\ref{eqDcsresspm}) that parton scattering polarizes quarks 
along the direction opposite to the normal of the parton reaction plane determined by the impact parameter $\vec{x}_T$, 
i.e., along the direction of the relative orbital angular momentum.  
This is essentially the manifest of spin-orbit coupling in QCD. 
Ordinarily, the polarized cross section along a fixed direction $\vec{n}$ vanishes when averaged
over all possible direction of the parton impact parameter $\vec{x}_T$. 
However, in non-central HIC the local relative orbital angular momentum $\langle l_y\rangle$ provides 
a preferred average reaction plane for parton collisions. 
This leads to a quark polarization opposite to the normal of the reaction plane of HIC. 
This conclusion should not depend on our perturbative treatment of parton scattering
as far as the effective interaction is mediated by the vector coupling in QCD.

Averaging over the relative angle between parton $\vec{x}_T$ and nuclear impact parameter $\vec{b}$
from $-\pi/2$ to $\pi/2$ and over $x_T$, one can obtain the global quark polarization,
\begin{equation}
P_q = - \pi\mu_D|\vec p|/2E(E+m_q)
\label{eqPqSPM}
\end{equation}
via a single scattering for given $E$. 

If one takes the non-relativistic limit, $E\sim m_q \gg |\vec p|, \mu_D$, one obtains, 
\begin{equation}
P_q\approx - \pi \mu_D |\vec p| /4m_q^2. \label{eqPqSPMnr}
\end{equation}

One of the advantages in this limit is that one can check effects due to spin-orbit coupling explicitly. 
Here, the spin-orbit coupling is given by Eq.~(\ref{eqHeff}). 
The corresponding energy is roughly given by 
$\langle E_{ls}\rangle \sim \langle \vec{\it l}\cdot \vec s~dV/rdr/m^2\rangle$. 
Given the interaction range is $r \sim 1/\mu_D$, $\langle dV/rdr\rangle \sim -\langle V\rangle \mu_D^2$; 
$\langle \vec{\it l}\cdot \vec s\rangle\sim \langle l\rangle/2\sim |\vec p|/2\mu_D$. 
The quark polarization is $P_q\sim \langle E_{ls}\rangle / \langle V\rangle$. 
We obtain $P_q\sim -\mu_D |\vec p| /m^2$ 
that is just the result given by Eq.~(\ref{eqPqSPMnr}). 

If one takes the ultra-relativistic limit $m_q=0$ and $|\vec p|\gg \mu_D$,
one expects from Eq.~(\ref{eqPqSPM}) that $P_q\sim -\pi\mu_D/2E$. 
However, given $dp_0/dx=0.34$ GeV/fm for semi-peripheral ($b=R_A$) collisions at RHIC,  
and an average range of interaction $\Delta x^{-1} \sim \mu_D \sim 0.5$ GeV, 
$\Delta p_z\sim 0.1$ GeV is smaller than the typical transverse momentum transfer $\mu_D$. 
In this case, one has to go beyond small angle approximation.

We also note that the cross sections can be written in a general form as,
\begin{eqnarray}
\label{eqcsgform}
\frac{d^2\sigma}{d^2 x_{T}}&=&F(x_T,E), \\
\label{eqdcsgform}
\frac{d^2\Delta\sigma}{d^2x_{T}}
&=&\vec{n}\cdot({\vec{x}}_T\times{\vec{p}}\ )~\Delta F(x_T,E),
\end{eqnarray}
where 
$F(x_T,E)$ and $\Delta F(x_T,E)$ are scalar functions of both
$x_T\equiv|\vec{x}_T|$ and the c.m. energy $E$ of the two quarks.
We would like to emphasize that Eqs. (\ref{eqcsgform}) and (\ref{eqdcsgform}) are in fact the 
most general forms of the two parts of the cross sections under parity conservation in the scattering process.
The unpolarized part of the cross section should be independent of any transverse direction thus can 
only take the form as given by Eq.~(\ref{eqcsgform}),
 i.e. it depends only on the magnitude of $x_T$ but not on the direction.
For the spin-dependent part, the only scalar that we can construct from
the available vectors is $\vec n\cdot(\vec p\times\vec x_T)$. 
Hence ${d^2\Delta\sigma}/{d^2x_{T}}$ can only take the form given by Eq.~(\ref{eqdcsgform}). 

We also note that, $\vec{x}_T\times\vec{p}$ is nothing but the
relative orbital angular momentum of the two-quark system,
$\vec{\it l}=\vec{x}_T\times\vec{p}$. Therefore, the polarized cross
section takes its maximum when
$\vec n$ is parallel or antiparallel to the relative
orbital angular momentum, depending on whether $\Delta F$ is
positive or negative. This corresponds to quark polarization in the direction $\vec{\it l}$ or $-\vec{\it l}$.

\subsubsection{Quark-quark scattering in a thermal medium}
\label{csqqhtl}

The quark-quark scattering amplitude in a thermal medium can be calculated 
by using the Hard Thermal Loop (HTL) resummed gluon propagator~\cite{WELD82,hw96},
\begin{equation}
\Delta^{\mu \nu}(q)=\frac{P_T^{\mu\nu}}{-q^2+\Pi_T(\xi)}
+\frac{P_L^{\mu\nu}}{-q^2+\Pi_L(\xi)}+(\alpha-1)\frac{q^{\mu}q^{\nu}}{q^4},
\label{eqhtl}
\end{equation}
where $q$ denotes the gluon four momentum and $\alpha$ is the gauge fixing parameter, 
$x=\omega/\sqrt{-\tilde q^2}$ and  $\omega=q\!\cdot\!u$, $\tilde{q}=q-\omega u$, 
$u$ is the fluid velocity of the local medium.
The longitudinal and transverse projectors $P_{T,L}^{\mu\nu}$
are defined by
\begin{eqnarray}
P_L^{\mu\nu}&=&\frac{1}{q^2\tilde q^2} (\omega q^{\mu}-q^2u^{\mu}) (\omega q^{\nu}-q^2u^{\nu})\, , \label{eqhtlPL} \\
P_T^{\mu\nu}&=&\tilde{g}^{\mu\nu}-\frac{\tilde{q}^{\mu}\tilde{q}^{\nu}}{\tilde q^2}\, , \label{eqhtlPT}
\end{eqnarray}
where $\tilde{g}_{\mu\nu}=g_{\mu\nu}-u_{\mu}u_{\nu}$.
$\Pi_L$ and $\Pi_T$ are the transverse and longitudinal self-energies and are given by~\cite{WELD82}
\begin{eqnarray}
\Pi_L(\xi)&=&\mu_D^2\left[1-\frac{\xi}{2}
\ln\left(\frac{1+\xi}{1-\xi}\right) + i \frac{\pi}{2}\xi \right](1-\xi^2)\,,
\label{eqhtlpiL} \\
\Pi_T(\xi)&=&\mu_D^2\left[\frac{\xi^2}{2}+\frac{\xi}{4}(1-\xi^2)
\ln\left(\frac{1+\xi}{1-\xi}\right) - i \frac{\pi}{4}\xi(1-\xi^2)\right], ~~~~~~
\label{eqhtlpiT}
\end{eqnarray}
where the Debye screening mass is $\mu_D^2=g^2(N_c+N_f/2)T^2/3$.

With the above HTL gluon propagator, the quark-quark scattering amplitude $\mathcal{M}_{fi}(q)$ 
in the momentum space can be expressed as,
\begin{equation}
\mathcal{M}_{fi}(q)= {\bar{u}}_{{\lambda}_3}(p_3){\gamma}_{\mu}
u_{{\lambda}_1}(p_1) {\Delta}^{\mu\nu}(q)
{\bar{u}}_{{\lambda}_4}(p_4){\gamma}_{\nu}u_{{\lambda}_2}(p_2). \label{eqMfihtl}
\end{equation}
The product ${\cal{M}}_{fi}(q){\cal{M^*}}_{fi}(k)$ can be converted to the following trace form,
\begin{eqnarray}
\sum_{\lambda_1,\lambda_2}{\cal{M}}_{fi}(q){\cal{M^*}}_{fi}(k) &=&
\Delta^{\mu\nu}(q)\Delta^{\alpha\beta*}(k) 
{\rm Tr}[u_{\lambda_3}(p_1-k)\bar u_{\lambda_3}(p_1+q)
\gamma_\mu(p_1\hspace{-9pt}\slash~+m_1)\gamma_\alpha]\nonumber\\
&&\times{\rm Tr}[u_{\lambda_4}(p_2-k)\bar u_{\lambda_4}(p_2-q)
\gamma_\nu(p_2\hspace{-9pt}\slash~+m_2)\gamma_\beta].~~~~~~\label{eqMMhtl}
\end{eqnarray}

In calculations of transport coefficients such as jet energy loss parameter~\cite{screen} 
and thermalization time~\cite{hw96} that generally involve cross sections weighted with transverse momentum transfer, 
the imaginary part of the HTL propagator in the magnetic sector is enough to 
regularize the infrared behavior of the transport cross sections. 
However, in the calculation of quark polarization, the total parton scattering cross section is involved. 
The contribution from the magnetic part of the interaction has therefore
infrared divergence that can only be regularized through the introduction of non-perturbative 
magnetic screening mass $\mu_m\approx0.255\sqrt{N_c/2} g^2T$~\cite{TBBM93}.

Since we have neglected the thermal momentum perpendicular to the longitudinal flow, 
the energy transfer $\omega=0$ in the center of mass frame of the two colliding partons. 
This corresponds to setting $x=0$ in the HTL resummed gluon propagator in Eq.~(\ref{eqhtl}).
In this case, the center of mass frame of scattering quarks coincides with 
the local co-moving frame of QGP and the fluid velocity is $u=(1,0,0,0)$. 
The corresponding HTL effective gluon propagator in Feynman gauge that contributes to
the scattering amplitudes reduces to,
\begin{equation}
{\Delta}^{\mu\nu}(q)=\frac{g^{\mu\nu}-u^\mu u^\nu}{q^2+\mu_m^2}+\frac{u^\mu u^\nu}{q^2+{{\mu}_D}^2}.
\end{equation}

The spin-dependent part determines the polarization of the final state quark $q_1$ via the scattering. 
The calculation is much involved.
A detailed study is given in~\cite{Gao:2007bc}. 
We summarize part of the key results in the following.

(i) {\it Small angle approximation}

We only consider light quarks and neglect their masses. 
Carrying out the traces in Eq.(\ref{eqMMhtl}),
we can obtain the expression of the cross section with HTL gluon propagators.
The results are much more complicated than those
as obtained in Sec.~\ref{csqqhtl} using a static potential model~\cite{Liang:2004ph}.
However, if we consider small transverse momentum transfer 
and use the small angle approximation, the results are still very simple.
In this case, with $q_z\sim 0$ and $q_T\equiv |\vec q_T|\ll p$,
we obtain,
\begin{eqnarray}
\frac{d^2\sigma}{d^2x_T}&=&\frac{g^4c_{qq}}{8}
\int\frac{d^2q_T}{(2\pi)^{2}}\frac{d^2k_T}{(2\pi)^{2}}
e^{i({\vec{k}}_{T}-{\vec{q}}_{T})\cdot\vec{x}_T}\nonumber\\
&&\times \left(\frac{1}{q_T^2+\mu_m^2}+\frac{1}{q_T^2+\mu_D^2}\right)
\left(\frac{1}{k_T^2+\mu_m^2}+\frac{1}{k_T^2+\mu_D^2}\right),~~~~~~
\label{eqcshtlsap} \\
\frac{d^2\Delta{\sigma}}{d^2x_T}
&=&-i\frac{g^4c_{qq}}{16\vec p^2}
\int\frac{d^2 q_T}{(2\pi)^{2}}\frac{d^2 k_T}{(2\pi)^{2}}
e^{i(\vec{k}_T-\vec{q}_T)\cdot\vec{x}_T}
\left[(\vec{k}_T-\vec{q}_T)\cdot(\vec{p}\times\vec{n})\right]\nonumber\\
&& \times \left(\frac{1}{q_T^2+\mu_m^2}+\frac{1}{q_T^2+\mu_D^2}\right)
\left(\frac{1}{k_T^2+\mu_m^2}+\frac{1}{k_T^2+\mu_D^2}\right). \label{eqdcshtlsap}
\end{eqnarray}
We note that there exist the same relationship between the polarized 
and unpolarized cross section as that as that given by Eq.~(\ref{eqcsDcsSPM}) obtained 
in the case of static potential model under the same small angle approximation. 
Completing the integration over $d^2q_T$ and $d^2k_T$ by using the formulae given by 
Eqs.~(\ref{eqintform}), we obtain,
\begin{eqnarray}
\frac{d^2\sigma}{d^2 x_T}&=&
\frac{c_{qq}}{2}\alpha_s^2
\left[K_0(\mu_{m}x_T)+K_0(\mu_{D}x_T)\right]^{2}, \label{eqcsreshtl} \\
\frac{d^2\Delta\sigma}{d^2 x_T} &=&
\frac{c_{qq}\alpha_s^2}{2\vec p^2}
\left[(\vec{p}\times\vec{n})\cdot\hat{\vec x}_T\right] \left[K_0(\mu_{m}x_T)+K_0(\mu_{D}x_T)\right] \nonumber\\
&&\times \bigl[\mu_{m}K_1(\mu_{m}x_T)+\mu_{D}K_1(\mu_{D}x_T)\bigr], \label{eqDcsreshtl}
\end{eqnarray}
where $\hat{\vec x}_T=\vec x_T/x_T$ is the unit vector of $\vec x_T$.
We compare the above results with those given by Eqs.~(\ref{eqcsresspm}) and  (\ref{eqDcsresspm}) 
obtained in the screened static potential model where one also made the
small angle approximation.
We see that the only difference between the two results is
the additional contributions from magnetic gluons,
whose contributions are absent in the static potential model.

(ii) {\it Beyond small angle approximation}

Now we present the complete results for the cross-section in
impact parameter space using HTL gluon propagators without small angle approximation. 
The unpolarized and polarized cross section can be expressed as,
\begin{equation}
\frac{d\sigma}{d^2x_T}= \frac{g^4c_{qq}}{16\hat{s}}
\int\frac{d^2q_T}{(2\pi)^{2}}\frac{d^2k_T}{(2\pi)^{2}}
~e^{i({\vec{k}}_{T}-{\vec{q}}_{T})\cdot\vec{x}_T}~\frac{f(q,k)}{\Lambda(q)\Lambda(k)}, 
\label{eqcshtlres}
\end{equation}
\begin{equation}
\frac{d\Delta{\sigma}}{d^2 x_T}= i\frac{g^4c_{qq}}{8\hat{s}^2}
\int\frac{d^2 q_T}{(2\pi)^{2}}\frac{d^2 k_T}{(2\pi)^{2}}
~e^{i(\vec{k}_T-\vec{q}_T)\cdot\vec{x}_T}~
\frac{\Delta f(q,k)}{\Lambda(q)\Lambda(k)}, \label{eqdcshtlres}
\end{equation}
where $\hat{s}$ is the c.m. energy squared of the quark-quark system, 
$f(q,k)$ and $\Delta f(q,k)$ are given by,
\begin{eqnarray}
f(q,k) &=&\sum_{a,b} \frac{A_{ab}(q,k)}{(q^2+\mu_a^2)(k^2+\mu_b^2)}, \\
\Delta f(q,k) &=&(\vec{p}\times\vec n)\cdot\sum_{ab} \frac{\Delta {\vec A}_{ab}(q,k)}{(q^2+\mu_a^2)(k^2+\mu_b^2)}, 
\end{eqnarray}
where the subscript $a$ or $b$ denotes $m$ or $D$ representing the magnetic or electric part 
and the sum runs over all possibilities of $(a,b)$.  
$A_{ab}$ are Lorentz scalar functions of $(q,k)$ given by, 
\begin{eqnarray}
A_{mm}(k,q)&=& \hat{s}[\hat{s}-(q+k)^2]+(q\cdot k)^2, \\ 
A_{DD}(q,k)&=&(\hat{s}-q^2-k^2)[\hat{s}-(q+k)^2]+(q\cdot k)^2,\\
A_{mD}(q,k)&=& A_{Dm}(k,q)=\hat{s}[\hat{s}-k^2-(q+k)^2]+(k^2-k\cdot q)^2+\frac{k^2q^2}{\hat{s}}(q+k)^2~, ~~~~~~~
\end{eqnarray}
$\Delta \vec A_{ab}(q,k)$ is a vector in the momentum space and can be written as,
\begin{equation}
\Delta \vec A_{ab}(q,k)=\Delta g_{ab}^{(q)}(q,k)~\vec q_T-\Delta g_{ab}^{(k)}(q,k)~\vec k_T, 
\end{equation}
where $\Delta g_{ab}^{(q)}(q,k)$ and $\Delta g_{ab}^{(k)}(q,k)$ are Lorentz scalar functions given by,  
\begin{eqnarray}
\Delta g^{(q)}_{mm}(q,k)&=&\Delta g^{(k)}_{mm}(k,q)=\hat{s}(\hat{s}-q\cdot k)-(\hat{s}+q^2+k^2-q\cdot k)k^2, ~~~~~~~~\\
\Delta g^{(q)}_{DD}(q,k)&=&\Delta g^{(k)}_{DD}(k,q)=(\hat{s}-q^2-k^2-q\cdot k)(\hat{s}-k^2)~, \\
\Delta g^{(q)}_{mD}(q,k)&=&\Delta g^{(k)}_{Dm}(k,q)=\hat{s}(\hat{s}-2k^2-q\cdot k)-(k^2-q\cdot k-\frac{q^2k^2}{\hat{s}})k^2~,~~~~~~\\
\Delta g^{(k)}_{mD}(q,k)&=&\Delta g^{(q)}_{Dm}(k,q)=\hat{s}(\hat{s}+q^2-k^2-q\cdot k)+(q^2-q\cdot k-\frac{q^2k^2}{\hat{s}})q^2, ~~~
\end{eqnarray}

We note that $A_{ab}(q,k)=A_{ab}(k,q)$,  $\Delta \vec A_{ab}(q,k)=-\Delta\vec A_{ab}(k,q)$ 
so that $f(q,k)=f(k,q)$ and $\Delta f(q,k)=-\Delta f(k,q)$, 
i.e., they are symmetric or anti-symmetric w.r.t. the two variables respectively. 
Hence, the integration result in Eq.~(\ref{eqcshtlres}) is real while that in Eq.~(\ref{eqdcshtlres}) is pure imaginary 
so that the cross section is real. 

We also note that $f(q,k)$ and $\Delta g_{\alpha\beta}^{(q/k)}(q,k)$ are all functions 
of Lorentz invariants ${\hat s},q^2,k^2$ and $q\cdot k$. 
Furthermore $A_{ab}(k,q)=\sum_{n=0-2} g_{ab}^{(n)}({\hat s},q^2,k^2)(\vec q_T\cdot\vec k_T)^n$, and 
$\Delta g_{ab}^{(q/k)}(k,q)=\sum_{n=0,1} \Delta g_{ab}^{(q/k,n)}({\hat s},q^2,k^2)(\vec q_T\cdot\vec k_T)^n$. 
The angular parts of the integrations in Eqs.~(\ref{eqcshtlres}) and (\ref{eqdcshtlres}) can be carried out. 
For this purpose, we note that, e.g., for any scalar function $f_s$ of $({\hat s},q^2,k^2)$, we have, 
\begin{eqnarray}
 &&\int\frac{d^2q_T}{(2\pi)^{2}}\frac{d^2k_T}{(2\pi)^{2}}
~e^{i({\vec{k}}_{T}-{\vec{q}}_{T})\cdot\vec{x}_T}~f_s({\hat s},q^2,k^2) \nonumber\\
&&~~~~~=\int \frac{dq_T^2}{4\pi}\frac{dk_T^2}{4\pi} J_0(q_Tx_T) J_0(k_Tx_T) f_s({\hat s},q^2,k^2)\equiv F^{(0)}(x_T,\hat{s}), \label{eqintfqk0}\\
&& \int\frac{d^2q_T}{(2\pi)^{2}}\frac{d^2k_T}{(2\pi)^{2}}
~e^{i({\vec{k}}_{T}-{\vec{q}}_{T})\cdot\vec{x}_T}(\vec q_T\cdot\vec k_T) f_s({\hat s},q^2,k^2)\nonumber\\
&&~~~~~=\int \frac{dq_T^2}{4\pi}\frac{dk_T^2}{4\pi} q_Tk_TJ'_0(q_Tx_T) J'_0(k_Tx_T) f_s({\hat s},q^2,k^2) \equiv F^{(1)}(x_T,\hat{s}) , ~~~\label{eqintfqk1} \\
&&\int\frac{d^2q_T}{(2\pi)^{2}}\frac{d^2k_T}{(2\pi)^{2}}
~e^{i({\vec{k}}_{T}-{\vec{q}}_{T})\cdot\vec{x}_T} g_s({\hat s},q^2,k^2) \vec q_T -i\hat{\vec x}_T G^{(0)}(x_T,\hat{s}), \nonumber\\
&&~~~G^{(0)}(x_T,\hat{s})= \int \frac{dq_T^2}{4\pi}\frac{dk_T^2}{4\pi} J'_0(q_Tx_T) J_0(k_Tx_T) g_s({\hat s},q^2,k^2), ~~~~~~~\label{eqintgqk0}\\
&&\int\frac{d^2q_T}{(2\pi)^{2}}\frac{d^2k_T}{(2\pi)^{2}}
~e^{i({\vec{k}}_{T}-{\vec{q}}_{T})\cdot\vec{x}_T} g_s({\hat s},q^2,k^2) (\vec q_T\cdot\vec k_T) \vec q_T =-i \hat{\vec x}_T G^{(1)}(x_T,\hat{s}), \nonumber\\
&&~~~G^{(1)}(x_T,\hat{s})= \int \frac{dq_T^2}{4\pi}\frac{dk_T^2}{4\pi} q_Tk_TJ''_0(q_Tx_T) J'_0(k_Tx_T) g_s({\hat s},q^2,k^2) . ~~~~~~~ \label{eqintgqk1}
 \end{eqnarray}
 Hence, we see clearly that,
\begin{eqnarray}
\frac{d\sigma}{d^2x_T} &=& \frac{g^4c_{qq}}{16\hat{s}} \sum_{a,b}  F_{ab}(x_T,\hat{s}), \label{eqcshtlfomal} \\
\frac{d\Delta{\sigma}}{d^2 x_T} &=&\frac{g^4c_{qq}}{8\hat{s}^2}
~(\vec{p}\times\vec n)\cdot\hat{\vec x}_T\sum_{a,b} \Delta F_{ab}(x_T,\hat{s}) . \label{eqdcshtlformal}
\end{eqnarray}

The scalar functions $F_{ab}(x_T,\hat{s})$ and $\Delta F_{ab}(x_T,\hat{s})$ are rather involved. 
However, if we take the simple form of $f_{qq}(\vec x_T,Y,b,\sqrt{s}~)$ given by Eq.~(\ref{eqfqq}) 
and calculate $\sigma$ and $\Delta\sigma$ using Eqs.~(\ref{eqDcsint2}) and (\ref{eqcsint2}), 
we may first carry out the integration over $\vec x_T$.
In this case we obtain,
\begin{eqnarray}
&& \langle\sigma\rangle=\frac{g^4 c_{qq}}{32\hat{s}} \int_{q_T\le p}\frac{d^{2}{{q}}_{T}}{(2\pi)^{2}}
\frac{f({q},{q})}{\Lambda^2(q)}, \\
&&\langle\Delta{\sigma}\rangle=-\frac{g^4c_{qq}}{8\hat{s}^2}
\int_{-E}^{E}\frac{dq_y}{2\pi}
\int_{-\sqrt{E^2-{q_y}^2}}^{\sqrt{E^2-{q_y}^2}}\frac{dq_x}{2\pi}
\int_{-\sqrt{E^2-{q_y}^2}}^{\sqrt{E^2-{q_y}^2}}\frac{dk_x}{2\pi} 
 \frac{\Delta f(q_x,q_y;k_x,q_y)}{({k_x-q_x})\Lambda(q)\Lambda(k)}.~~~~~~~~
\end{eqnarray}

\begin{figure}[htbp]
\begin{center}
\includegraphics[width=7.0cm]{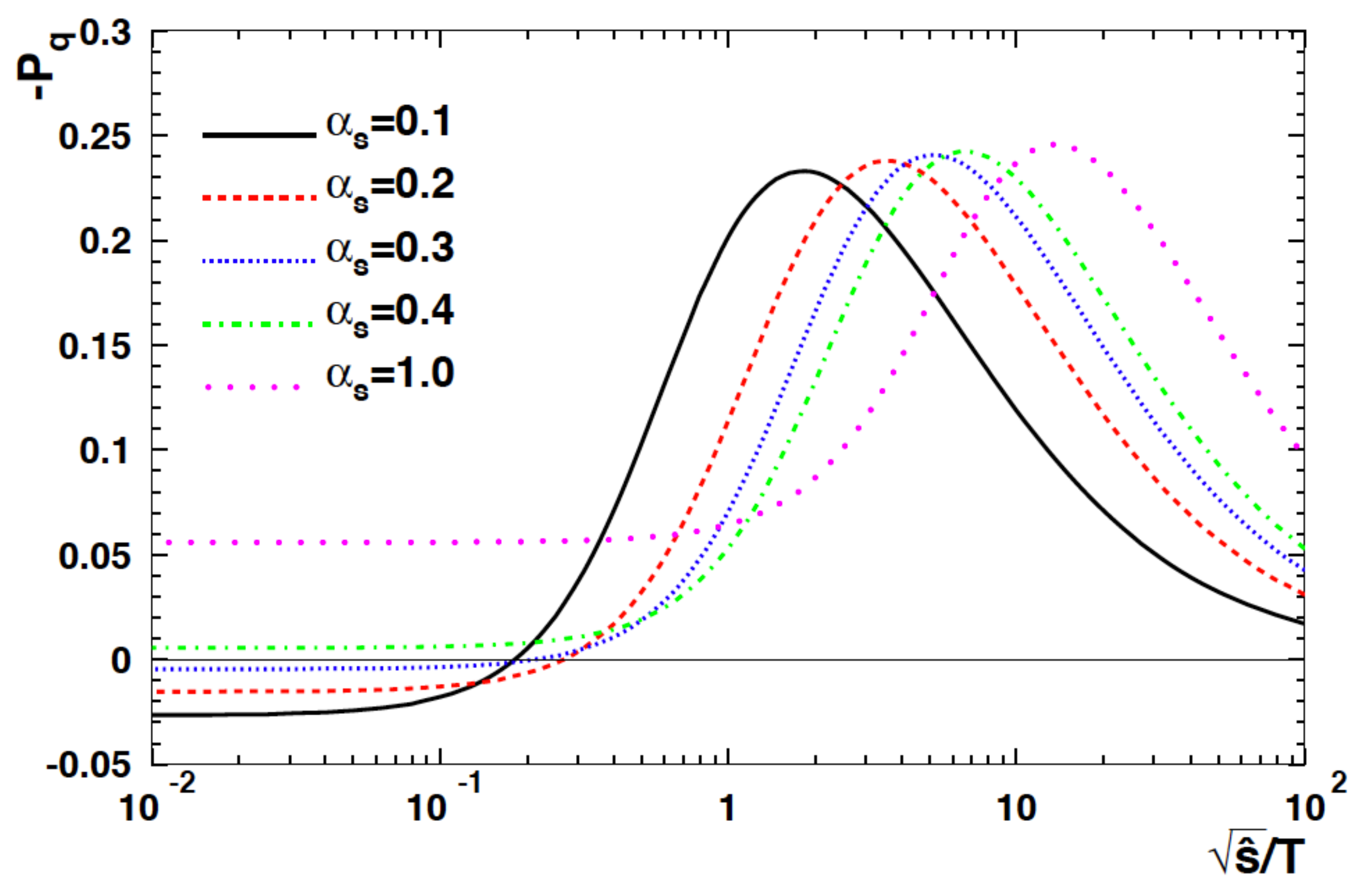}
\end{center}
\caption{Quark polarization -$P_q$ as a function of
$\sqrt{\hat{s}}/T$ for different $\alpha_s$'s obtained in quark-quark scattering with a hard thermal loop propagator. 
This figure is taken from~\cite{Gao:2007bc}. }
\label{figqpolhtl}
\end{figure}

These equations can be further simplified to the form suitable for carrying out numerical calculations.
Details are given in Ref.~\cite{Gao:2007bc} where cases are also studied. 
Here, we present only the result of the quark polarization $P_q$ 
as function of c,m, energy of the quark-quark system $\sqrt{\hat{s}}/T$ in Fig.~\ref{figqpolhtl}.

From Fig.~\ref{figqpolhtl}, we see that the quark polarization changes drastically with $\sqrt{\hat{s}}/T$.
It increases to some maximum values and then decreases with the growing energy, approaching
the result of small angle approximation in the high-energy limit.
This structure is caused by the interpolation between the high-energy and low-energy
behavior dominated by the magnetic part of the interaction in the weak coupling limit $\alpha_s<1$. 
Therefore, the position of the maxima in $\sqrt{\hat s}$ should
approximately scale with the magnetic mass $\mu_m$.

\subsubsection{Conclusions and discussions on global quark polarization}

\begin{figure}[htbp]
\begin{center}
\includegraphics[width=2.6in]{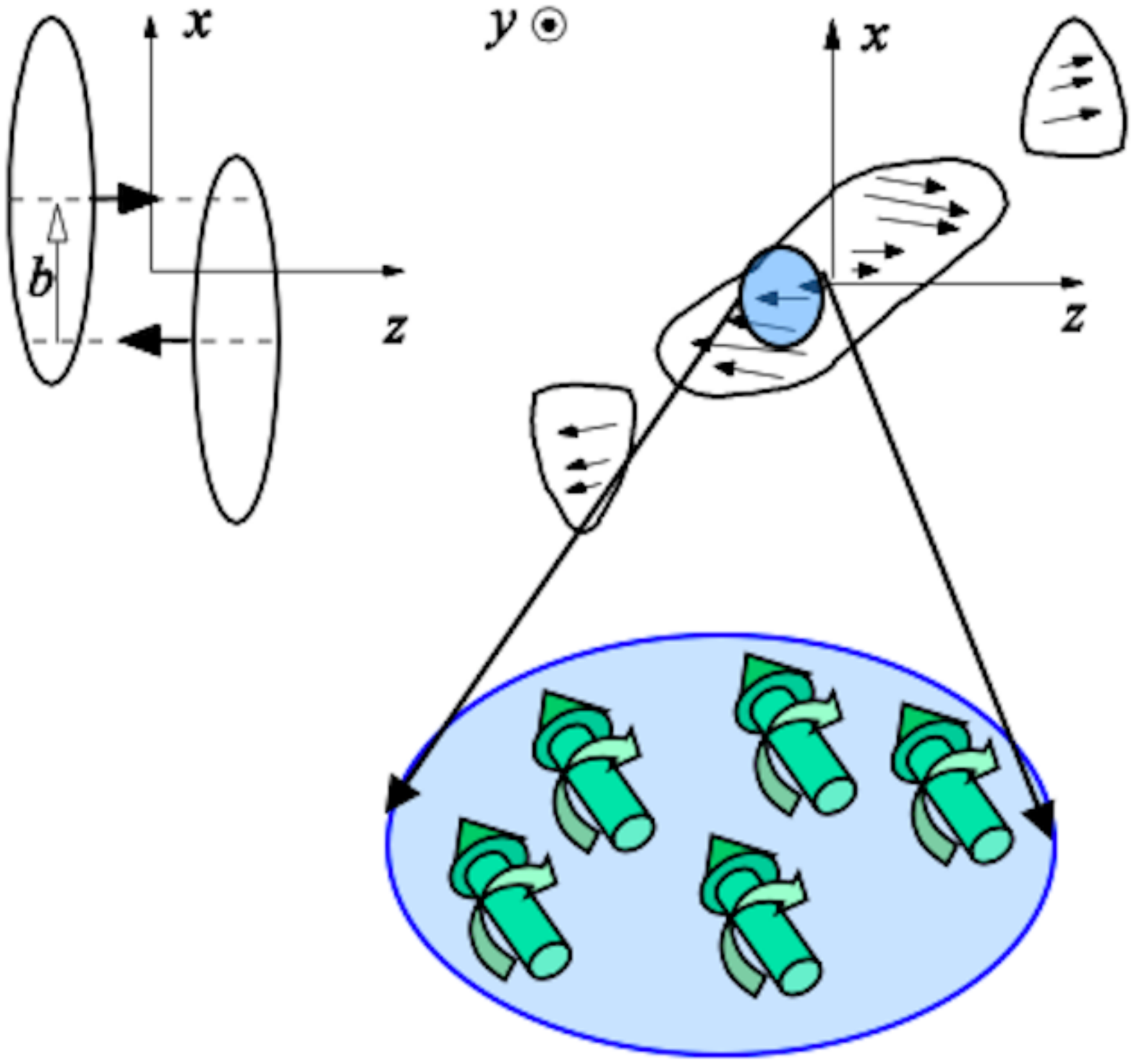}
\end{center}
\caption{Illustration of the global quark polarization effect in non-central heavy ion collisions.}
\label{figGP}
\end{figure}

Although approximations and/or models have to be used in the calculations presented above, 
the physical picture and consequence are very clear. 
It is confident that after the scattering of two constituents in QGP, 
the orbital angular momentum will be transferred partly to the polarization of quarks and anti-quarks 
in the system due to spin-orbit coupling in QCD. 
Such a polarization is very different from those that we meet usually in high energy physics
such as the longitudinal or the transverse polarization. 
The longitudinal polarization refers to the helicity or the polarization in the direction of the momentum,  
whereas the transverse polarization refers to directions perpendicular to the momentum, 
either in the production plane or along the normal of the production plane. 
These directions are all defined by the momentum of the individual particle 
and are in general different for different particles in the same collision event.  
In contrast, the polarization discussed here refers to the normal of the reaction plane. 
It is a fixed direction for one collision event and is independent of any particular hadron in the final state. 
Hence, in Ref.~\cite{Liang:2004ph}, this polarization was given a new name --- the global polarization, 
and the QGP was referred to the globally polarized QGP in non-central HIC. 
We illustrate this in Fig.~\ref{figGP}. 

The following three points should be addressed in this connection. 

(i) 
The results presented above are mainly a summary of those obtained in 
the original papers~\cite{Liang:2004ph,Gao:2007bc} where the global orbital angular momentum 
for the colliding system in HIC was first pointed out and the GPE were first predicted.  
These results are for a single quark-quark scattering. 
In a realistic HIC where QGP is created, such quark-quark scatterings may take place for a few times before they hadronize into hadrons. 
The calculations presented above or in~\cite{Liang:2004ph,Gao:2007bc} provide the theoretical basis for GPE. 
They do not provide final results of global quark polarizations.  

(ii) 
The numerical results on quark polarization presented above are based on the approximation by taking 
the simple form of $f_{qq}(\vec x_T,Y,b,\sqrt{s}~)$ given by Eq.~(\ref{eqfqq}). 
They provide a practical guidance for the magnitude of the quark polarization but can not give 
us the relationship between the polarization and the local orbital angular momentum. 
Further studies along this line are necessary. 
In practice, to describe the evolution of the global quark polarization, 
one can invoke a dynamical model of QGP evolution or 
effectively a dynamical model for $f_{qq}(\vec x_T,Y,b,\sqrt{s}~)$. 

(iii) 
If we consider QGP as a fluid, the momentum shear distribution discussed in Sec.~\ref{secoam} 
implies a non-vanishing vorticity $\vec\omega=(1/2) \nabla\times\vec v$. 
The spin-orbit coupling can be replaced by spin-vortical coupling. 
This provides a good opportunity to study spin-vortical effects in strongly interacting system 
and has  attracted much attention~\cite{Betz:2007kg,Becattini:2007sr,Deng:2016gyh,Fang:2016vpj,Pang:2016igs,Li:2017dan,Xia:2018tes,Florkowski:2018ahw,Wei:2018zfb}. 
See chapter on this topic in this series.

\subsection{A kinetic approach for quark polarization rate}


The global polarization in heavy ion collisions arises from scattering
processes of partons or hadrons with spin-orbit couplings. In a 2-to-2
particle scattering at a fixed impact parameter, one can calculate
the polarized cross section arising from the spin-orbit coupling.
In a thermal medium, however, momenta of incident particles are randomly
distributed and particles participating in the scattering are located
at different space-time points. In order to obtain observables we
have to take an ensemble average over random momenta of incident particles
and treat scatterings at different space-time points properly. To
this end, a microscopic model was proposed for the polarization from
the first principle through the spin-orbit coupling in particle scatterings
in a thermal medium with a shear flow \cite{Zhang:2019xya}. It is
based on scatterings of particles as wave packets, an effective method
to deal with particle scatterings at specified impact parameters.
The polarization is then the consequence of particle collisions in
a non-equilibrium state of spins. The spin-vorticity coupling naturally
emerges from the spin-orbit one encoded in polarized scattering amplitudes
of collisional integrals when one assumes local equilibrium in momentum
but not in spin.

As an illustrative example, we have calculated the quark polarization
rate per unit volume from all 2-to-2 parton (quark or gluon) scatterings
in a locally thermalized quark-gluon plasma. It can be shown that
the polarization rate for anti-quarks is the same as that for quarks
because they are connected by the charge conjugate transformation.
This is consistent with the fact that the rotation does not distinguish
particles and antiparticles. The spin-orbit coupling is hidden in
the polarized scattering amplitude at specified impact parameters.
We can show that the polarization rate per unit volume is proportional
to the vorticity as the result of particle scatterings. Thus we build
up a non-equilibrium model for the global polarization.

\subsubsection{Collision rate for spin-0 particles in a multi-particle system}

We aim to derive the spin polarization rate in a thermal medium with
a shear flow from particle scatterings through spin-orbit couplings.
Before we do it in the next section, let us first look at the collision
rate of spin-zero particles. It is easy to generalize it to the spin
polarization rate for spin-1/2 particles

\begin{figure}[h]
\centering{}\includegraphics[scale=0.3]{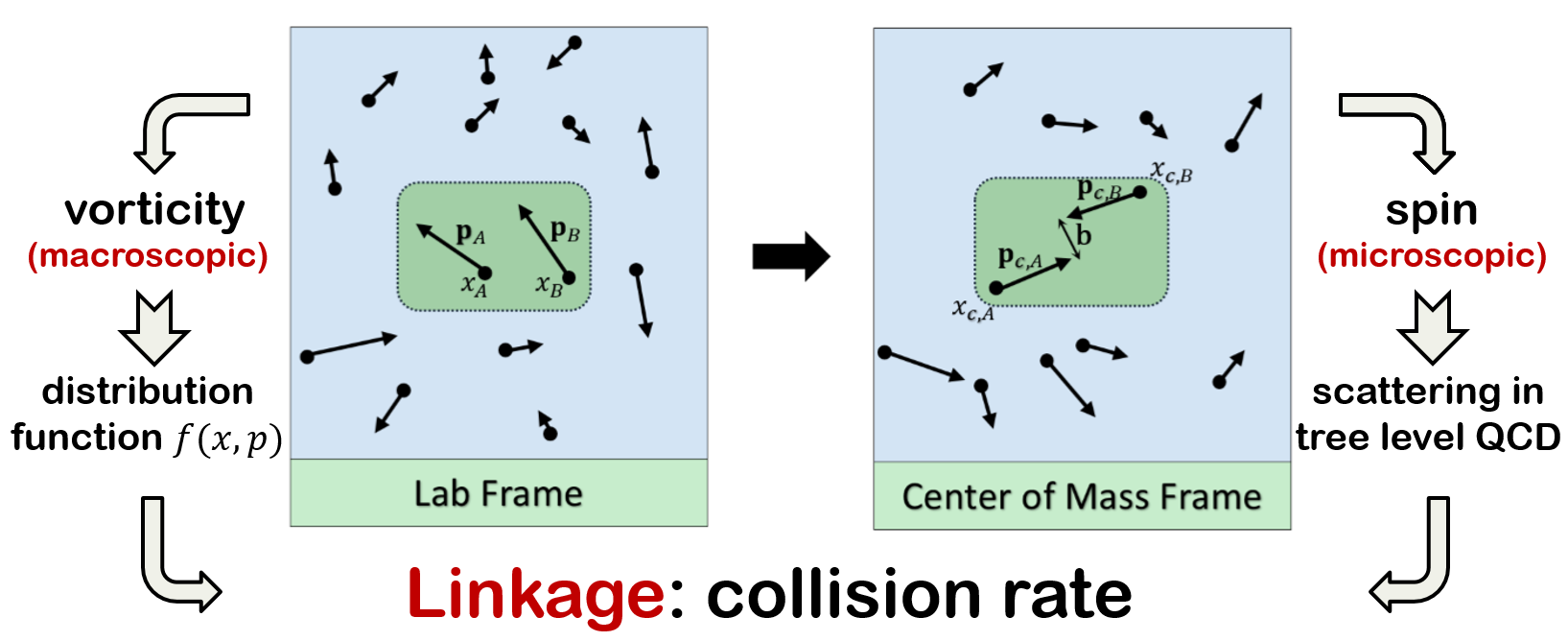}
\caption{A collision or scattering in the Lab frame
(left) and center-of-mass frame (right).} \label{fig:coll-heat-bath}
\end{figure}

In the center of mass frame (CMS) of the incident particle $A$ and
$B$, the collision rate (the number of collisions per unit time)
per unit volume is given by 
\begin{eqnarray}
R_{AB\rightarrow12} = n_{A}n_{B}|v_{A}-v_{B}|\sigma 
 \frac{d^{3}p_{A}}{(2\pi)^{3}}\frac{d^{3}p_{B}}{(2\pi)^{3}}f_{A}(x_{A},p_{A})f_{B}(x_{B},p_{B})|v_{A}-v_{B}|\Delta\sigma,~~\label{eq:collision rate}
\end{eqnarray}
where $v_{A}=|\mathbf{p}_{A}|/E_{A}$ and $v_{B}=-|\mathbf{p}_{B}|/E_{B}$
are the velocity of $A$ and $B$ respectively with $\mathbf{p}_{A}=-\mathbf{p}_{B}$,
$f_{A}$ and $f_{B}$ are the phase space distributions for $A$ and
$B$ respectively, and $\Delta\sigma$ denotes the infinitesimal element
of the cross section which is given by 
\begin{eqnarray}
\Delta\sigma  = \frac{1}{C_{AB}}d^{4}x_{A}d^{4}x_{B}\delta(\Delta t)\delta(\Delta x_{L}) 
 \frac{d^{3}p_{1}}{(2\pi)^{3}2E_{1}}\frac{d^{3}p_{2}}{(2\pi)^{3}2E_{2}}\frac{1}{(2E_{A})(2E_{B})}K,~~\label{eq:d-sigma}
\end{eqnarray}
where we assumed that the scattering takes place at the same time
and the same longitudinal position in the CMS (these conditions are
represented by two delta functions), the constant $C_{AB}$ makes
$\Delta\sigma$ have the correct dimension whose definition will be
given later, and $K$ is given by 
\begin{eqnarray}
K &=& (2E_{A})(2E_{B})|{}_{\text{out}}\langle p_{1}p_{2}|\phi_{A}(x_{A},p_{A})\phi_{B}(x_{B},p_{B})\rangle_{\text{in}}|^{2}, \label{eq:k-factor}
\end{eqnarray}
with ($i=A,B$)
\begin{eqnarray}
|\phi_{i}(x_{i},p_{i})\rangle_{\text{in}} & = & \int\frac{d^{3}k_{i}}{(2\pi)^{3}}\frac{1}{\sqrt{2E_{i,k}}}\phi_{i}(\mathbf{k}_{i}-\mathbf{p}_{i})e^{-i\mathbf{k}_{i}\cdot\mathbf{x}_{i}}|\mathbf{k}_{i}\rangle_{\text{in}},
\end{eqnarray}
being the wave packets for incident particles. If incoming particles
are described by two plane waves, there is no initial angular momentum.
This is why we should use wave packets for incoming particles. Normally
one can choose a Gaussian form for the wave packet amplitude, 
\begin{equation}
\phi_{i}(\mathbf{k}_{i}-\mathbf{p}_{i})=\frac{(8\pi)^{3/4}}{\alpha_{i}^{3/2}}\exp\left[-\frac{(\mathbf{k}_{i}-\mathbf{p}_{i})^{2}}{\alpha_{i}^{2}}\right],\label{eq:wave-packet-gs}
\end{equation}
where $\alpha_{i}$ denote the width of the wave packet. For simplicity,
we use plane waves to represent outgoing particles.

\begin{figure}
\begin{centering}
\includegraphics[scale=0.45]{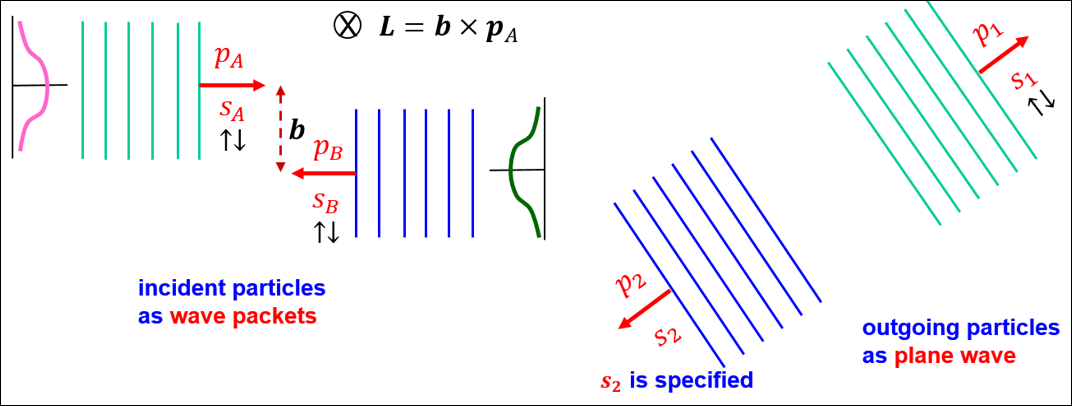}
\par\end{centering}
\caption{Scattering of two particles in the center of mass frame.} \label{fig:Scattering-of-two}
\end{figure}

Now we consider the scattering process in Fig.~\ref{fig:Scattering-of-two}.
The incoming particles are located at $x_{A}$ and $x_{B}$. We can
use new variables $X=(x_{A}+x_{B})/$ and $y = x_{A}-x_{B}$ to replace $x_{A}$ and $x_{B}$.
We then define $C_{AB}\equiv\int d^{4}X=t_{X}\Omega_{\mathrm{int}}$,
where $t_{X}$ and $\Omega_{\mathrm{int}}$ are the local time and
space volume for the interaction. The local collision rate from Eq.
(\ref{eq:collision rate}) can be written as 
\begin{eqnarray}
\frac{d^{4}N_{AB\rightarrow12}}{dX^{4}} &  &= \frac{1}{(2\pi)^{4}}\int\frac{d^{3}p_{A}}{(2\pi)^{3}2E_{A}}\frac{d^{3}p_{B}}{(2\pi)^{3}2E_{B}}\frac{d^{3}p_{1}}{(2\pi)^{3}2E_{1}}\frac{d^{3}p_{2}}{(2\pi)^{3}2E_{2}}\nonumber \\
 &  & \times|v_{A}-v_{B}|G_{1}G_{2}\int d^{3}k_{A}d^{3}k_{B}d^{3}k_{A}^{\prime}d^{3}k_{B}^{\prime}\nonumber \\
 &  & \times\phi_{A}(\mathbf{k}_{A}-\mathbf{p}_{A})\phi_{B}(\mathbf{k}_{B}-\mathbf{p}_{B})\phi_{A}^{*}(\mathbf{k}_{A}^{\prime}-\mathbf{p}_{A})\phi_{B}^{*}(\mathbf{k}_{B}^{\prime}-\mathbf{p}_{B})\nonumber \\
 &  & \times\delta^{(4)}(k_{A}^{\prime}+k_{B}^{\prime}-p_{1}-p_{2})\delta^{(4)}(k_{A}+k_{B}-p_{1}-p_{2})\nonumber \\
 &  & \times\mathcal{M}\left(\{k_{A},k_{B}\}\rightarrow\{p_{1},p_{2}\}\right)\mathcal{M}^{*}\left(\{k_{A}^{\prime},k_{B}^{\prime}\}\rightarrow\{p_{1},p_{2}\}\right)\nonumber \\
 &  & \times\int d^{2}\mathbf{b}f_{A}\left(X+\frac{y_{T}}{2},p_{A}\right)f_{B}\left(X-\frac{y_{T}}{2},p_{B}\right)\exp\left[i(\mathbf{k}_{A}^{\prime}-\mathbf{k}_{A})\cdot\mathbf{b}\right],~~~~~~~~~\label{eq:rate-ab12}
\end{eqnarray}
where $N_{AB\rightarrow12}$ is the number of collisions and $G_{i}$
$(i=1,2)$ denote the distribution factors which depends on the particle
types in the final state. We have $G_{i}=1$ for the Boltzmann particles
and $G_{i}=1\pm f_{i}(p_{i})$ for bosons (upper sign) and fermions
(lower sign).

\subsubsection{Polarization rate for spin-1/2 particles from collisions}

Based on the collision rate for spin-zero particles in the above section,
we now consider spin-1/2 particles. We assume that particle distributions
are independent of spin states, so the spin dependence comes only
from scatterings of particles carrying the spin degree of freedom.
In this section we will distinguish quantities in the CMS from those
in the lab frame, we will put an index $c$ for a CMS quantity.

If the system has reached local equilibrium in momentum, we can make
an expansion of $f_{A}f_{B}$ in $y_{c,T}=(0,\mathbf{b}_{c})$, and
thus,
\begin{eqnarray}
 &  & f_{A}\left(X_{c}+\frac{y_{c,T}}{2},p_{c,A}\right)f_{B}\left(X_{c}-\frac{y_{c,T}}{2},p_{c,B}\right)\nonumber \\
 & = & f_{A}\left(X,p_{A}\right)f_{B}\left(X,p_{B}\right)+\frac{1}{2}y_{c,T}^{\mu}[\Lambda^{-1}]_{\;\mu}^{\nu}\frac{\partial(\beta u_{\rho})}{\partial X^{\nu}}\nonumber \\
 &  & \times\left[p_{A}^{\rho}f_{B}\left(X,p_{B}\right)\frac{df_{A}\left(X,p_{A}\right)}{d(\beta u\cdot p_{A})}-p_{B}^{\rho}f_{A}\left(X,p_{A}\right)\frac{df_{B}\left(X,p_{B}\right)}{d(\beta u\cdot p_{B})}\right],\label{eq:expansion-ff}
\end{eqnarray}
where we have used the defination of the Lorentz transformation matrix
$\partial X^{\nu}/\partial X_{c}^{\mu}=[\Lambda^{-1}]_{\;\mu}^{\nu}=\Lambda_{\mu}^{\;\nu}$,
and the scalar invariance $f_{A}\left(X,p_{A}\right)=f_{A}\left(X_{c},p_{c,A}\right)$
and $f_{B}\left(X,p_{B}\right)=f_{B}\left(X_{c},p_{c,B}\right)$.
From Eq. (\ref{eq:expansion-ff}) we see that the local vorticity
$\partial(\beta u_{\rho})/\partial X^{\nu}$ shows up. We look closely
at the term $y_{c,T}^{\mu}[\partial(\beta u_{c,\rho})/\partial X_{c}^{\mu}]p_{c,A}^{\rho}$,
\begin{eqnarray}
y_{c,T}^{\mu}p_{c,A}^{\rho}\frac{\partial(\beta u_{\rho})}{\partial X_{c}^{\mu}} & = & \frac{1}{4}y_{c,T}^{[\mu}p_{c,A}^{\rho]}\left[\frac{\partial(\beta u_{c,\rho})}{\partial X_{c}^{\mu}}-\frac{\partial(\beta u_{c,\mu})}{\partial X_{c}^{\rho}}\right]\nonumber \\
 &  & +\frac{1}{4}y_{c,T}^{\{\mu}p_{c,A}^{\rho\}}\left[\frac{\partial(\beta u_{c,\rho})}{\partial X_{c}^{\mu}}+\frac{\partial(\beta u_{c,\mu})}{\partial X_{c}^{\rho}}\right]\nonumber \\
 & = & -\frac{1}{2}y_{c,T}^{[\mu}p_{c,A}^{\rho]}\varpi_{\mu\rho}^{(c)}+\frac{1}{4}y_{c,T}^{\{\mu}p_{c,A}^{\rho\}}\left[\frac{\partial(\beta u_{c,\rho})}{\partial X_{c}^{\mu}}+\frac{\partial(\beta u_{c,\mu})}{\partial X_{c}^{\rho}}\right]\nonumber \\
 & = & -\frac{1}{2}L_{(c)}^{\mu\rho}\varpi_{\mu\rho}^{(c)}+\frac{1}{4}y_{c,T}^{\{\mu}p_{c,A}^{\rho\}}\left[\frac{\partial(\beta u_{c,\rho})}{\partial X_{c}^{\mu}}+\frac{\partial(\beta u_{c,\mu})}{\partial X_{c}^{\rho}}\right],\label{eq:oam-vorticity}
\end{eqnarray}
where $[\mu\rho]$ and $\{\mu\rho\}$ denote the anti-symmetrization
and symmetrization of two indices respectively, $L_{(c)}^{\mu\rho}\equiv y_{c,T}^{[\mu}p_{c,A}^{\rho]}$
is the OAM tensor, and $\omega_{\mu\rho}^{(c)}\equiv-(1/2)[\partial_{\mu}^{X_{c}}(\beta u_{c,\rho})-\partial_{\rho}^{X_{c}}(\beta u_{c,\mu})]$
is the thermal vorticity. We see that the coupling term of the OAM
and vorticity appear in Eq. (\ref{eq:expansion-ff}). The second term
in last line of Eq. (\ref{eq:oam-vorticity}) is related to the Killing
condition required by the thermal equilibrium of the spin.

\begin{figure}
\begin{centering}
\includegraphics[width=3.0in,height=3.7in]{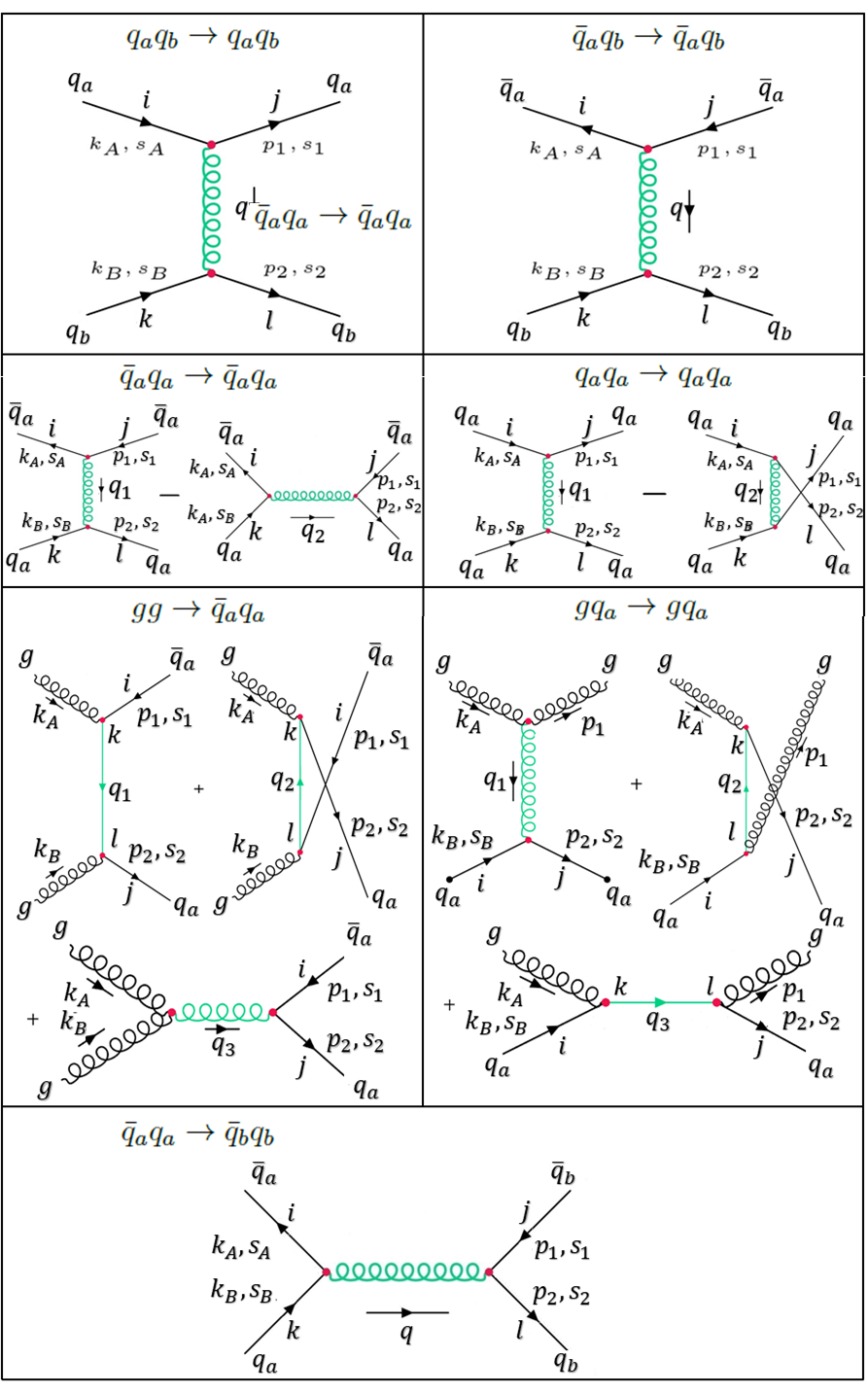}
\par\end{centering}
\caption{Tree level Feynman diagrams of all 2-to-2 parton scatterings. The
final states contain at least one quark. Here $a$ and $b$ denote
the quark flavor, $s_{i}=\pm1/2$ ($i=A,B,1,2$) denote the spin states,
$k_{i}$ ($i=A,B,1,2$) denote the momenta, $q,q_{1},q_{2},q_{3}$
denote the momenta in propagators. The processes for antiquark are
similar.} \label{fig:Tree-level-Feynman}
\end{figure}

Now we consider the scattering process $A+B\rightarrow1+2$ where
incoming and outgoing particles are in the spin state labeled by $s_{A}$,
$s_{B}$, $s_{1}$ and $s_{2}$ ($s_{i}=\pm1/2$, $i=A,B,1,2$) respectively.
For simplicity, we sum over $s_{A}$, $s_{B}$, $s_{1}$, and leave
$s_{2}$ open. Defining the direction of the reaction plane in the
CMS as $\mathbf{n}_{c}=\hat{\mathbf{b}}_{c}\times\hat{\mathbf{p}}_{c,A}$,
we have, from Eq. (\ref{eq:rate-ab12}), the polarization rate of
particle 2 per unit time and unit volume is 
\begin{eqnarray}
\frac{d^{4}\mathbf{P}_{AB\rightarrow12}(X)}{dX^{4}} & = & -\frac{1}{(2\pi)^{4}}\int\frac{d^{3}p_{A}}{(2\pi)^{3}2E_{A}}\frac{d^{3}p_{B}}{(2\pi)^{3}2E_{B}}\frac{d^{3}p_{c,1}}{(2\pi)^{3}2E_{c,1}}\frac{d^{3}p_{c,2}}{(2\pi)^{3}2E_{c,2}}\nonumber \\
 &  & \times|v_{c,A}-v_{c,B}|\int d^{3}k_{c,A}d^{3}k_{c,B}d^{3}k_{c,A}^{\prime}d^{3}k_{c,B}^{\prime}\nonumber \\
 &  & \times\phi_{A}(\mathbf{k}_{c,A}-\mathbf{p}_{c,A})\phi_{B}(\mathbf{k}_{c,B}-\mathbf{p}_{c,B})\phi_{A}^{*}(\mathbf{k}_{c,A}^{\prime}-\mathbf{p}_{c,A})\phi_{B}^{*}(\mathbf{k}_{c,B}^{\prime}-\mathbf{p}_{c,B})\nonumber \\
 &  & \times\delta^{(4)}(k_{c,A}^{\prime}+k_{c,B}^{\prime}-p_{c,1}-p_{c,2})\delta^{(4)}(k_{c,A}+k_{c,B}-p_{c,1}-p_{c,2})\nonumber \\
 &  & \times\frac{1}{2}\int d^{2}\mathbf{b}_{c}\exp\left[i(\mathbf{k}_{c,A}^{\prime}-\mathbf{k}_{c,A})\cdot\mathbf{b}_{c}\right]\mathbf{b}_{c,j}[\Lambda^{-1}]_{\;j}^{\nu}\frac{\partial(\beta u_{\rho})}{\partial X^{\nu}}\nonumber \\
 &  & \times\left[p_{A}^{\rho}-p_{B}^{\rho}\right]f_{A}\left(X,p_{A}\right)f_{B}\left(X,p_{B}\right)\Delta I_{M}^{AB\rightarrow12}\mathbf{n}_{c},\label{eq:polarization}
\end{eqnarray}
where $\mathbf{P}_{AB\rightarrow12}$ denotes the polarization vector.
In the derivation of Eq. (\ref{eq:polarization}), we have used Boltzmann
distributions for $f_{A}\left(X,p_{A}\right)f_{B}\left(X,p_{B}\right)$
with $G_{1}G_{2}=1$. The quantity $\Delta I_{M}^{AB\rightarrow12}$
is defined as
\begin{eqnarray}
\Delta I_{M}^{AB\rightarrow12} & = & \sum_{s_{A},s_{B},s_{1},s_{2}}\sum_{color}2s_{2}\mathcal{M}\left(\{s_{A},k_{c,A};s_{B},k_{c,B}\}\rightarrow\{s_{1},p_{c,1};s_{2},p_{c,2}\}\right)\nonumber \\
 &  & \times\mathcal{M}^{*}\left(\{s_{A},k_{c,A}^{\prime};s_{B},k_{c,B}^{\prime}\}\rightarrow\{s_{1},p_{c,1};s_{2},p_{c,2}\}\right).\label{eq:d-im-ab12}
\end{eqnarray}
Since we consider the polarization of quarks, there are seven processes
involved as shwon in Fig. \ref{fig:Tree-level-Feynman}. Evaluate
all these diagrams will give more than 5000 terms. However, all these
terms are spin-orbit coupling ones \cite{Liang:2004ph,Gao:2007bc}
that have four types of structures: $(\mathbf{n}\times\mathbf{p}_{1})\cdot\hat{\mathbf{k}}_{A}$,
$(\mathbf{n}\times\mathbf{p}_{1})\cdot\hat{\mathbf{k}}_{A}^{\prime}$,
$(\mathbf{n}\times\hat{\mathbf{k}}_{A})\cdot\hat{\mathbf{k}}_{A}^{\prime}$
and $(\mathbf{p}_{1}\times\hat{\mathbf{k}}_{A})\cdot\hat{\mathbf{k}}_{A}^{\prime}$.

\subsubsection{Numerical results for quark/antiquark polarization rate}

Finally the polarization rate of quarks per unit time and unit volume
in Eq. (\ref{eq:polarization}) can be put into a compact form 
\begin{eqnarray}
\frac{d^{4}\mathbf{P}_{q}(X)}{dX^{4}} & = & \frac{\pi}{(2\pi)^{4}}\frac{\partial(\beta u_{\rho})}{\partial X^{\nu}}
\sum_{A,B,1}\int\frac{d^{3}p_{A}}{(2\pi)^{3}2E_{A}}\frac{d^{3}p_{B}}{(2\pi)^{3}2E_{B}}|v_{c,A}-v_{c,B}|\nonumber \\
 &  & \times[\Lambda^{-1}]_{\;j}^{\nu}\mathbf{e}_{c,i}\epsilon_{ikh}\hat{\mathbf{p}}_{c,A}^{h} 
 f_{A}\left(X,p_{A}\right)f_{B}\left(X,p_{B}\right)\left(p_{A}^{\rho}-p_{B}^{\rho}\right)\Theta_{jk}(\mathbf{p}_{c,A})\nonumber \\
 & \equiv & \frac{\partial(\beta u_{\rho})}{\partial X^{\nu}}\mathbf{W}^{\rho\nu},\label{eq:diff-rate}
\end{eqnarray}
where the tensor $\mathbf{W}^{\rho\nu}$, defined in the last line,
contains 64 components, and each of its component a is 16 dimensional
integration. 

This is a major challenge in the numerical calculation. To handle
this high dimension integration, we split the integration into two
parts: a 10-dimension (10D) integration over $(\mathbf{p}_{c,1},\mathbf{p}_{c,2},\mathbf{k}_{c,A}^{T},\mathbf{k}_{c,A}^{\prime T})$
and a 6-dimension (6D) integration over $(\mathbf{p}_{A},\mathbf{p}_{B})$.
We first carry out the 10D integration by ZMCintegral-3.0, a Monte
Carlo integration package that we have newly developed and runs on
multi-GPUs \cite{Wu:2019tsf}. Then we save this 10D result $\Theta_{jk}(\mathbf{p}_{c,A})$
as a function of $\mathbf{p}_{c,A}$ (and $\mathbf{p}_{c,B}=-\mathbf{p}_{c,A}$).
Finally we perform the 6D integration using the pre-calculated 10D
integral. The main parameters are set to following values: the quark
mass $m_{q}=0.2$ GeV for all flavors ($u,d,s,\bar{u},\bar{d},\bar{s}$),
the gluon mass $m_{g}=0$ for the external gluon, the internal gluon
mass (Debye screening mass) $m_{g}=m_{D}=0.2$ GeV in gluon propagators
in the $t$ and $u$ channel to regulate the possible divergence,
the width $\alpha=0.28$ GeV of the Gaussian wave packet, and the
temperature $T=0.3$ GeV.

\begin{figure}
\includegraphics[scale=0.3]{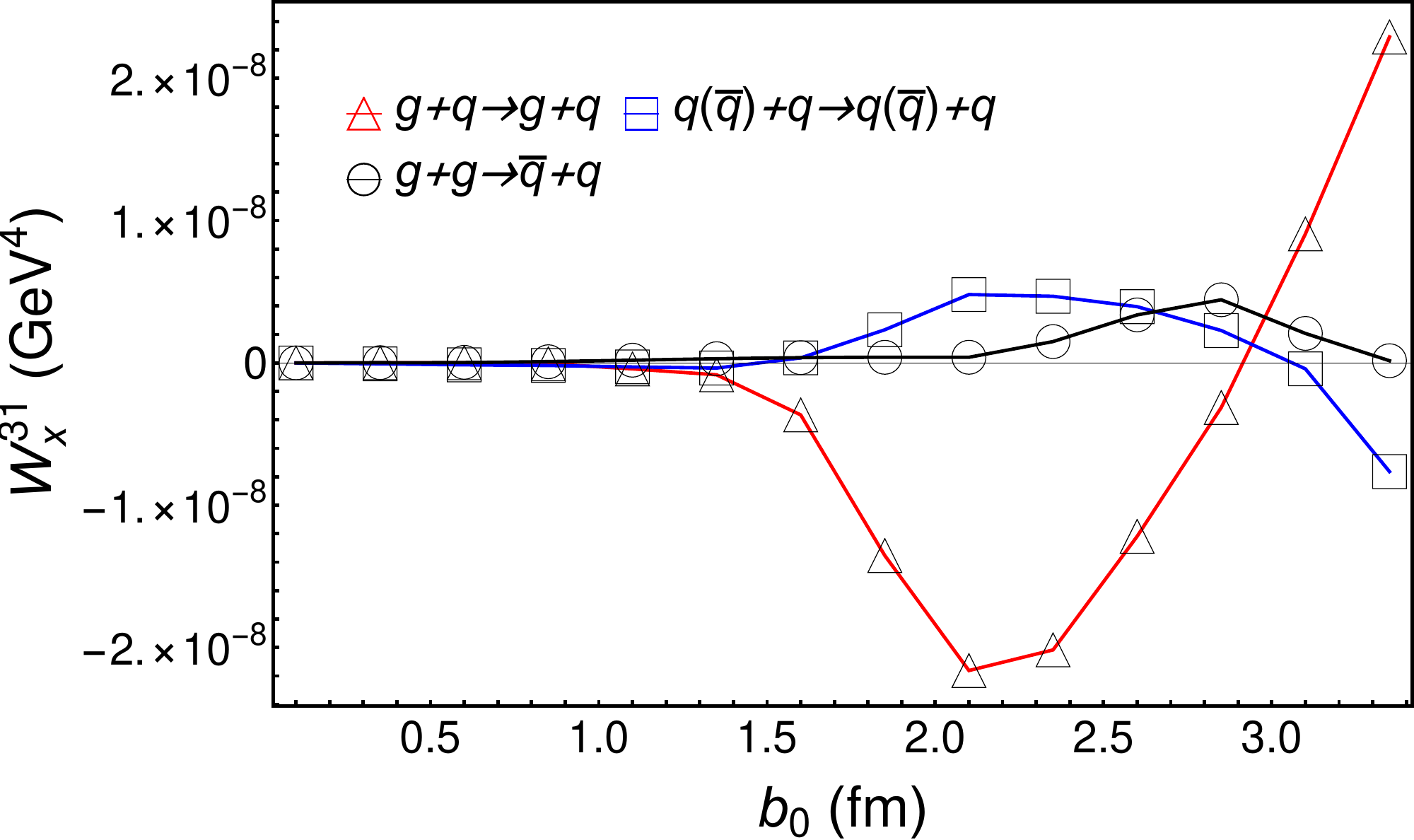}\includegraphics[scale=0.3]{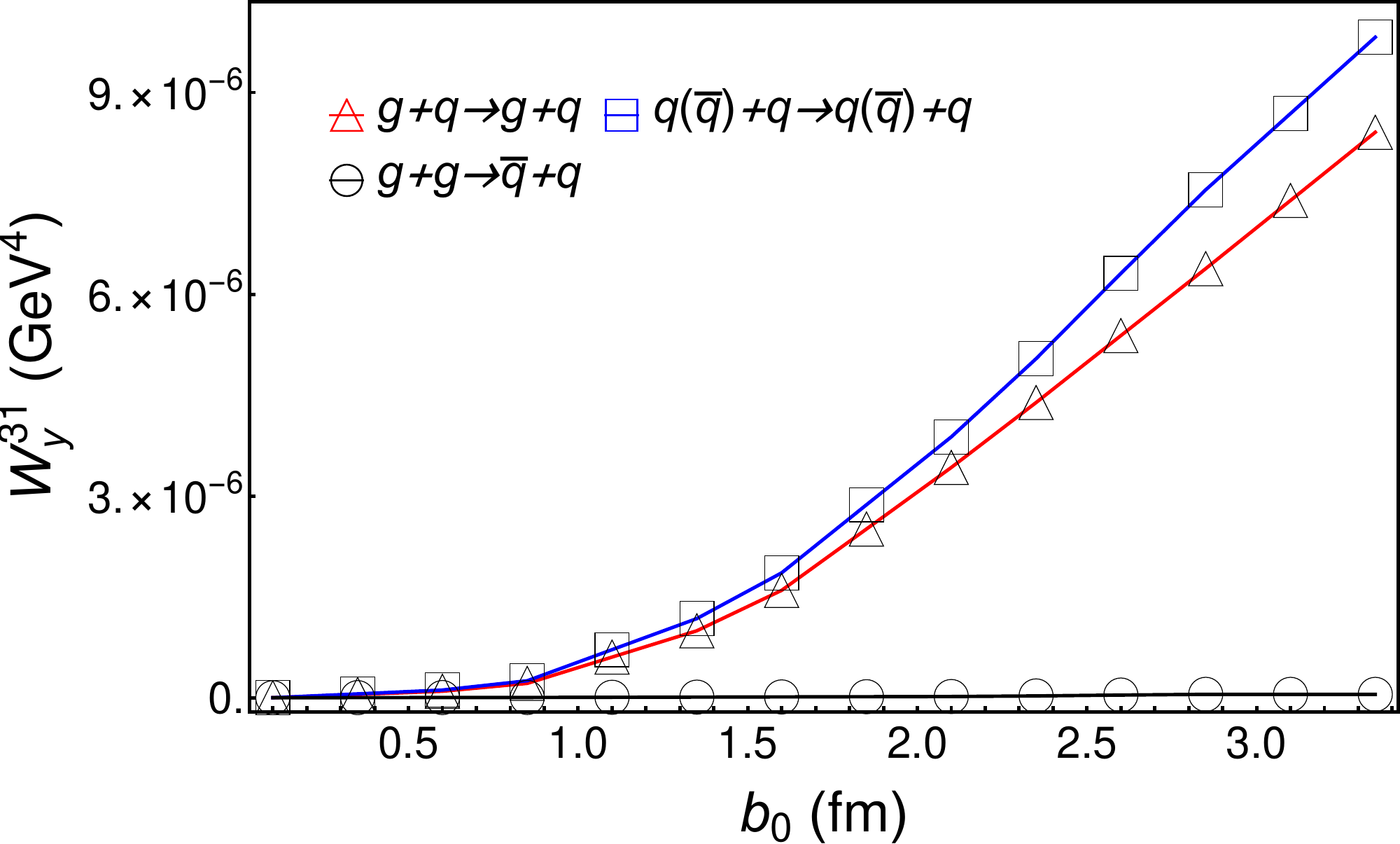}
\includegraphics[scale=0.3]{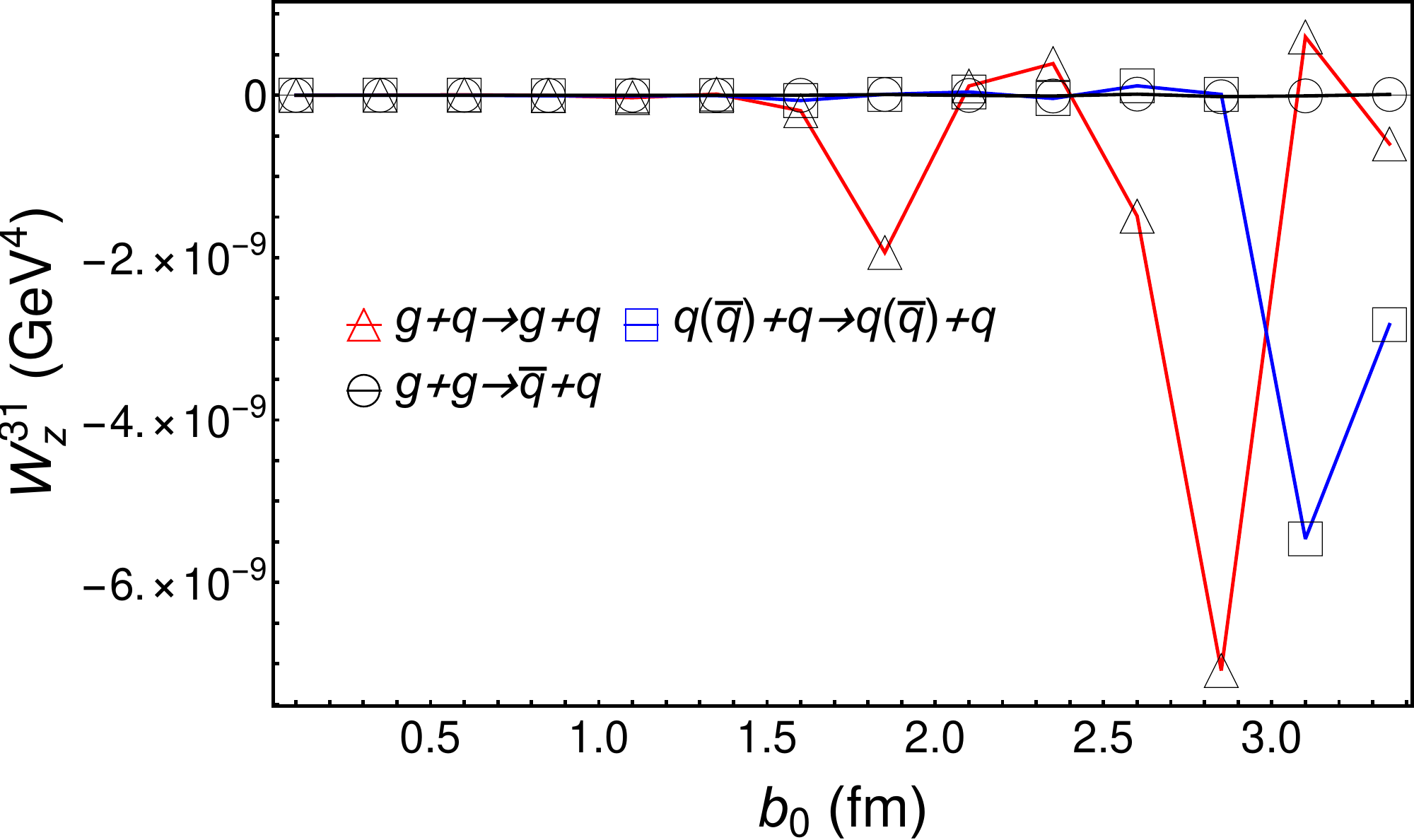}\includegraphics[scale=0.3]{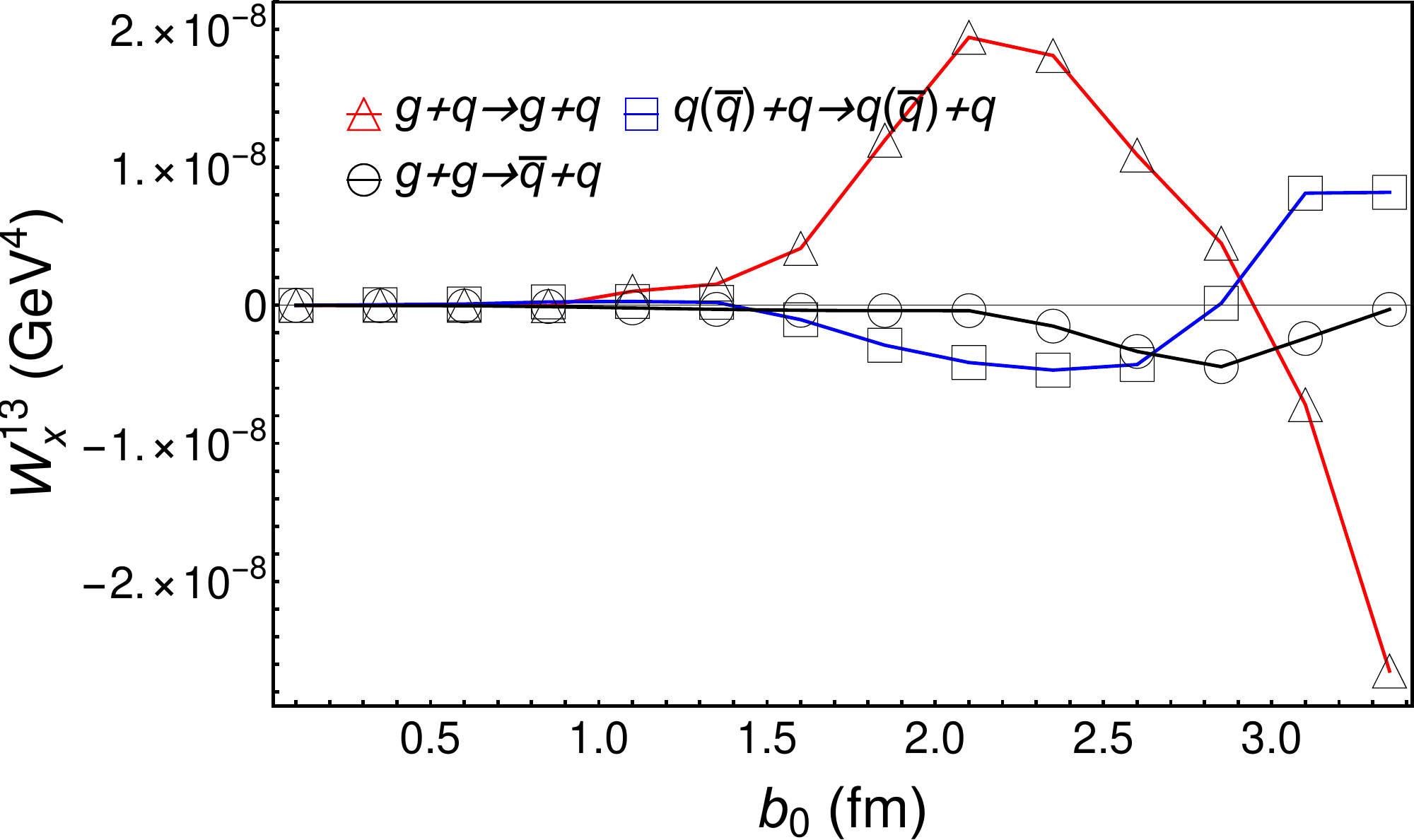}
\includegraphics[scale=0.3]{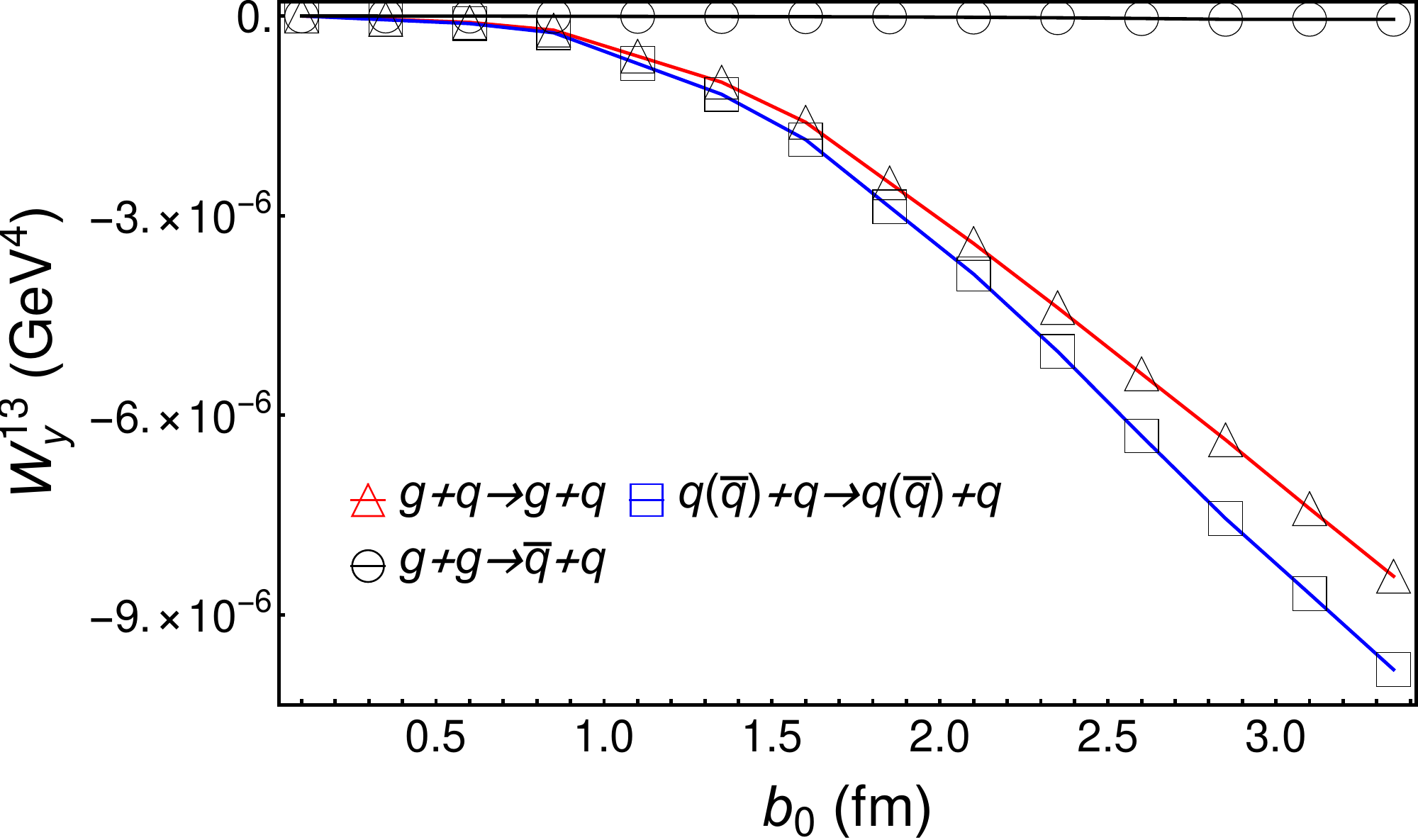}\includegraphics[scale=0.3]{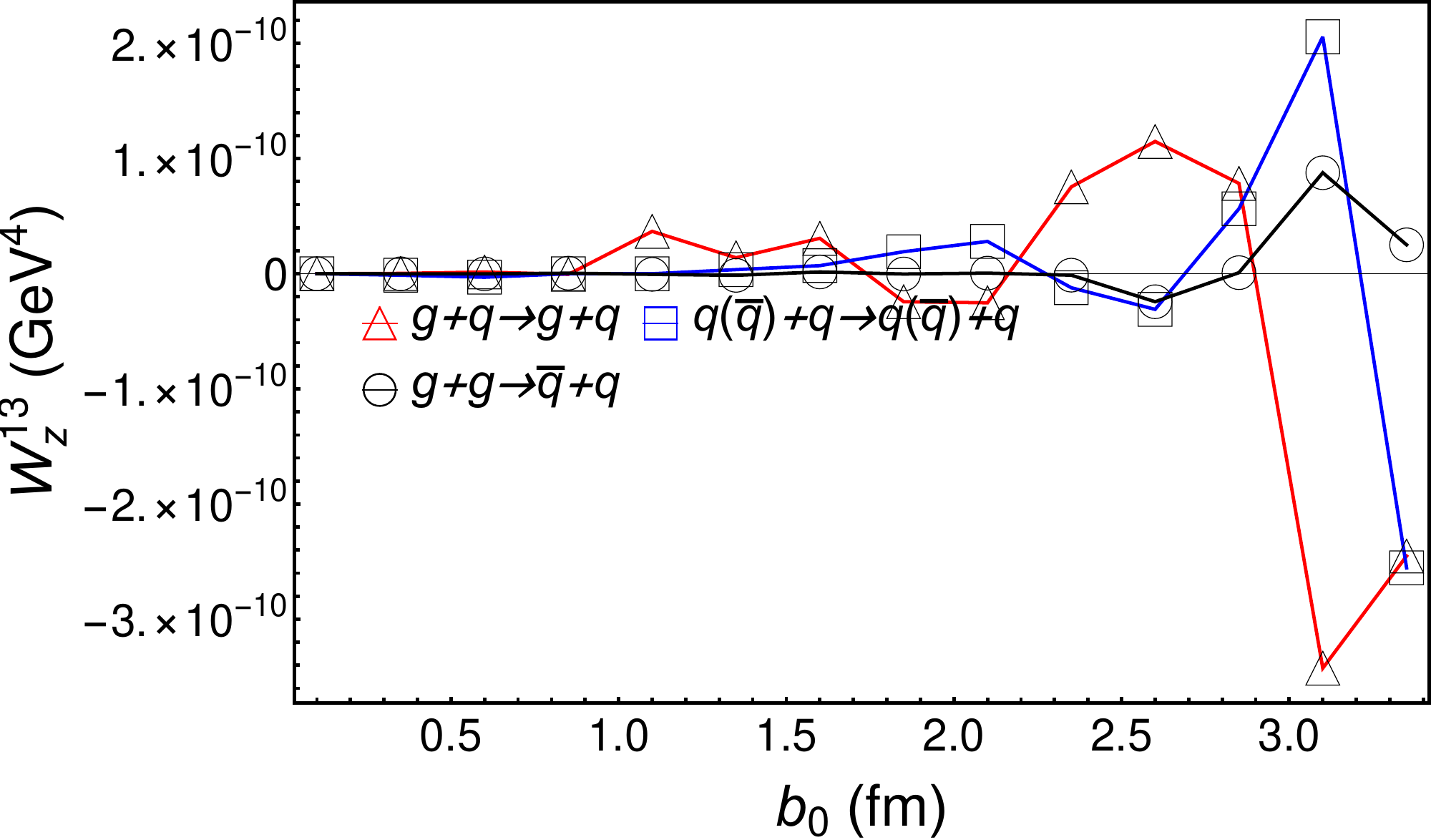}
\includegraphics[scale=0.3]{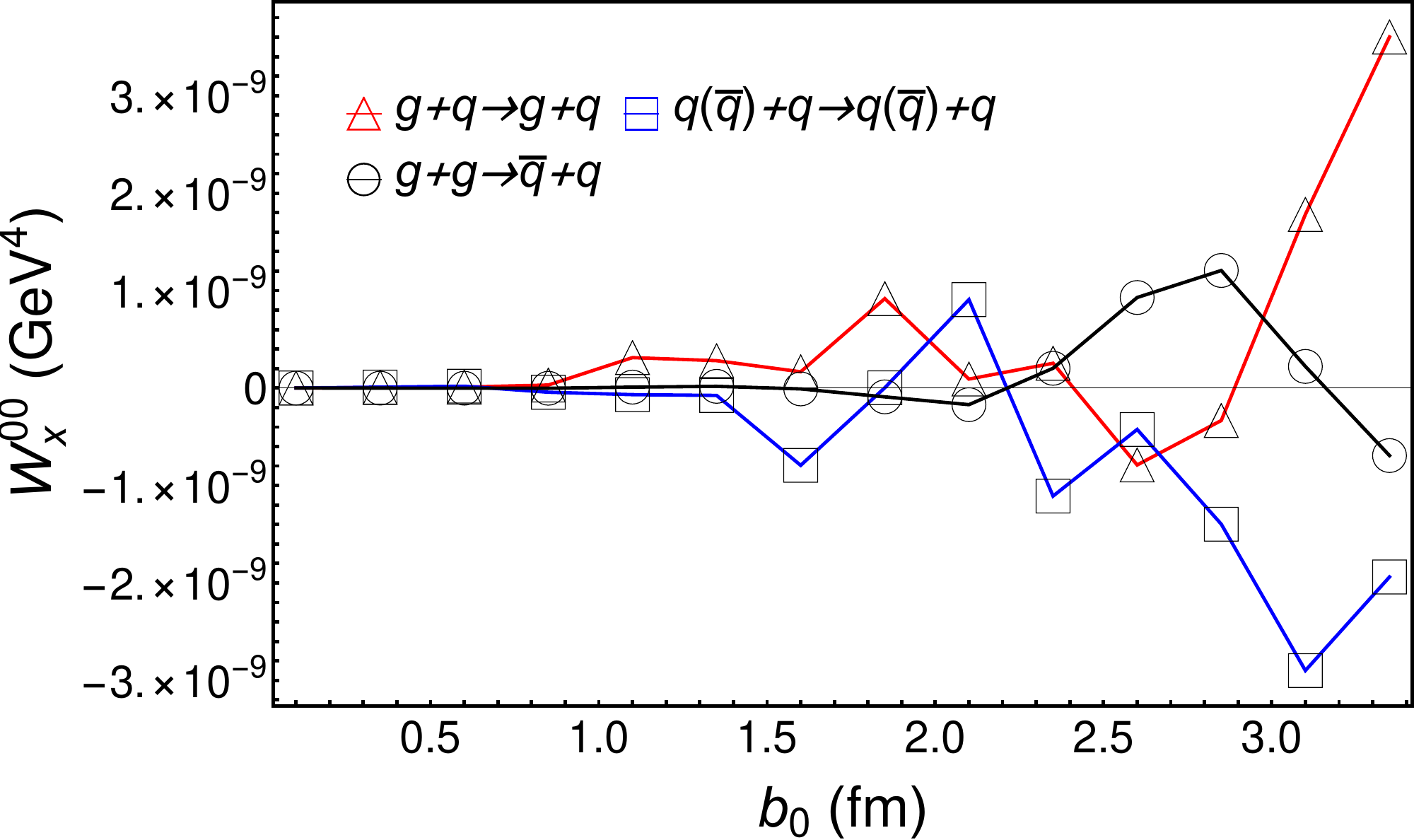}\includegraphics[scale=0.3]{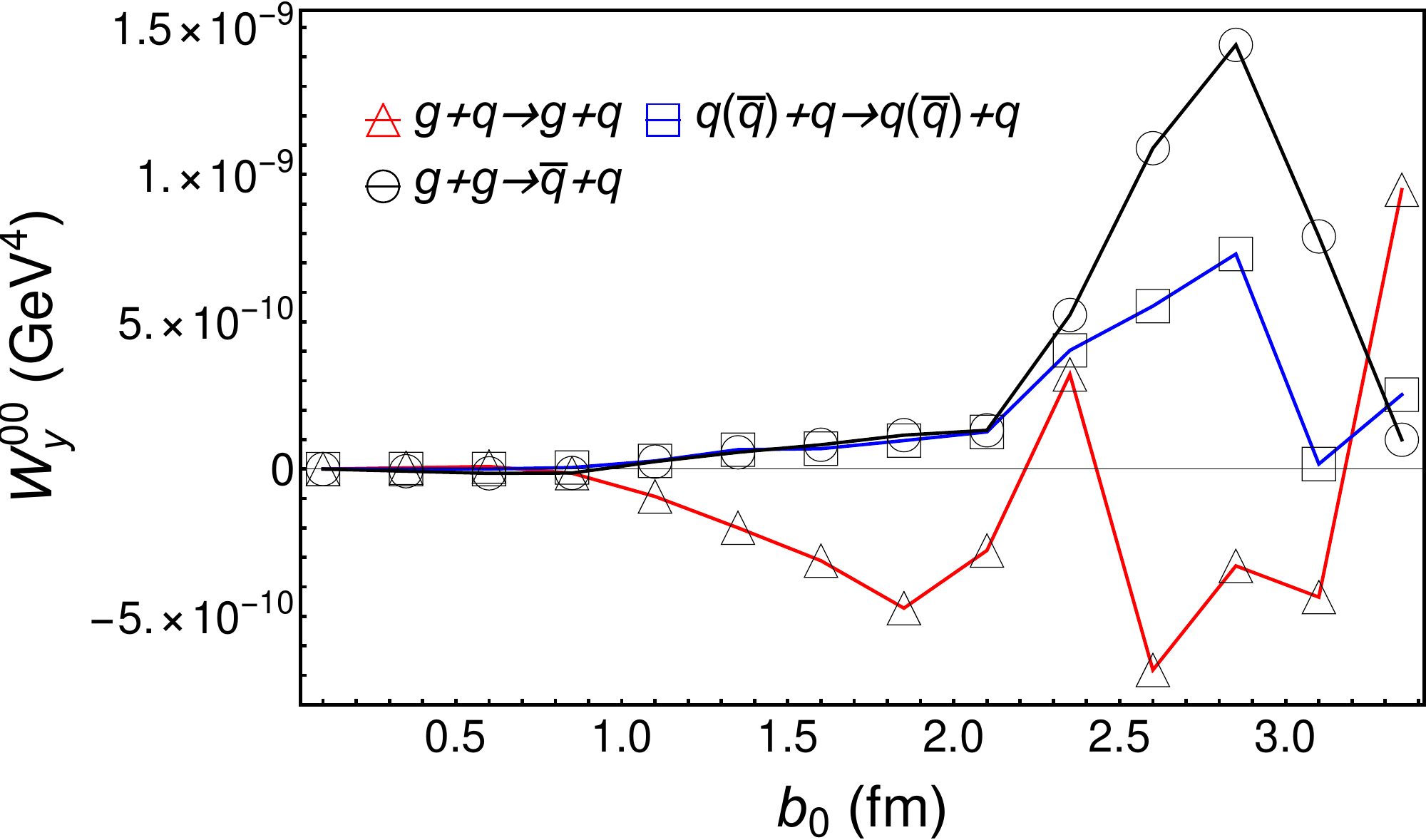}
\includegraphics[scale=0.3]{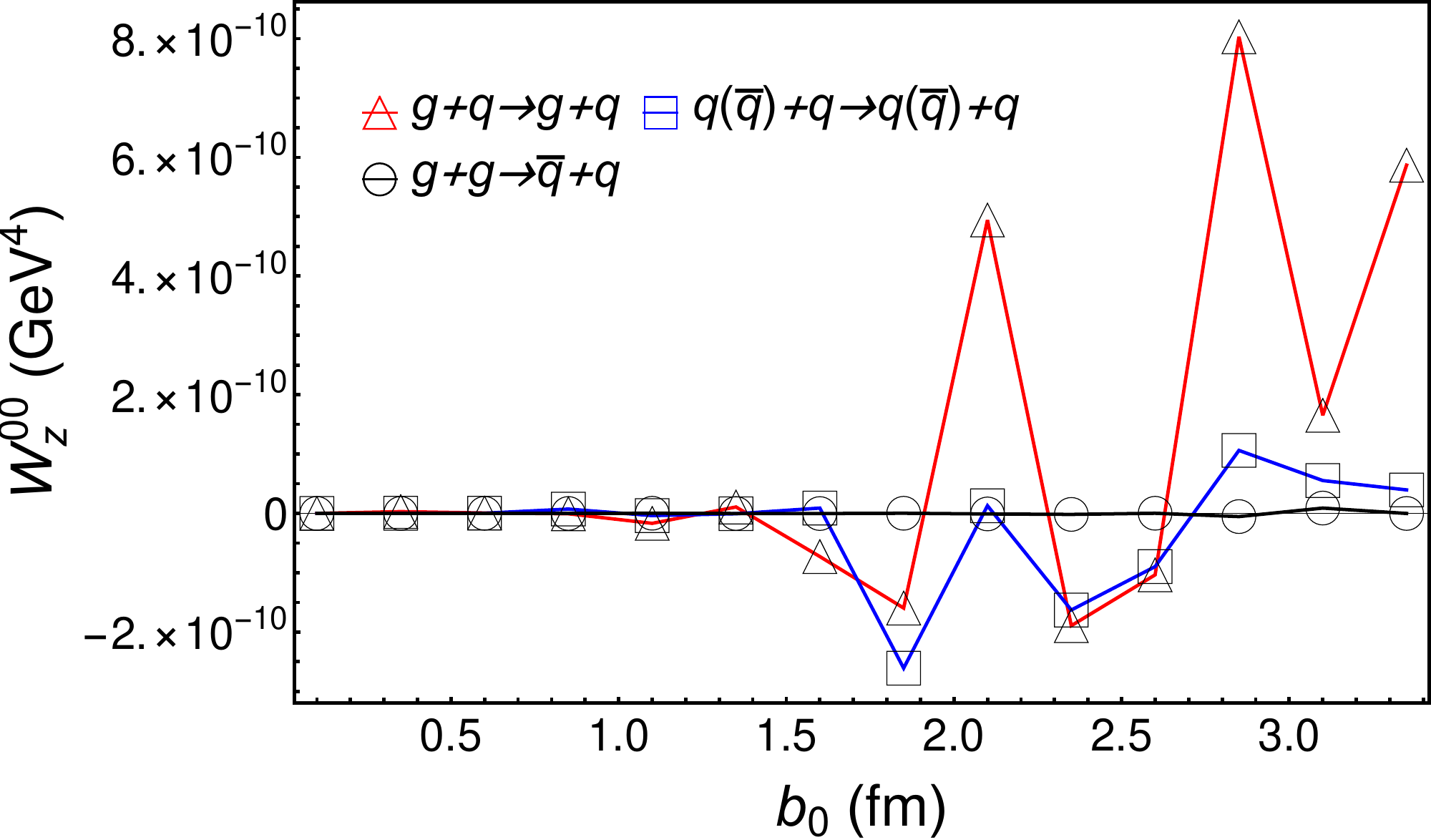}
\caption{Numerical results for components of $\mathbf{W}^{\rho\nu}$. Here
$b_{0}$ is the cut-off of the impact parameter in the CMS of the
scattering. }\label{fig:Numerical-results-for}
\end{figure}

The numerical results are shown in Fig. \ref{fig:Numerical-results-for},
from which we see an explicit form of $\mathbf{W}^{\rho\nu}$ as 
\begin{equation}
\mathbf{W}^{\rho\nu}=\left(\begin{array}{cccc}
0 & 0 & 0 & 0\\
0 & 0 & W\mathbf{e}_{z} & -W\mathbf{e}_{y}\\
0 & -W\mathbf{e}_{z} & 0 & W\mathbf{e}_{x}\\
0 & W\mathbf{e}_{y} & -W\mathbf{e}_{x} & 0
\end{array}\right),\label{eq:w-vector-exp}
\end{equation}
or in a compact form 
\begin{equation}
\mathbf{W}^{\rho\nu}=W\epsilon^{0\rho\nu j}\mathbf{e}_{j}.\label{eq:w-vector}
\end{equation}
Therefore Eq. (\ref{eq:diff-rate}) becomes 
\begin{eqnarray}
\frac{d^{4}\mathbf{P}_{q}(X)}{dX^{4}} & = & \epsilon^{0j\rho\nu}\frac{\partial(\beta u_{\rho})}{\partial X^{\nu}}W\mathbf{e}_{j}=2\epsilon_{jkl}\omega_{kl}W\mathbf{e}_{j}
=2W\nabla_{X}\times(\beta\mathbf{u}),\label{eq:w31}
\end{eqnarray}
where $\omega_{\rho\nu}=-(1/2)[\partial_{\rho}^{X}(\beta u_{\nu})-\partial_{\nu}^{X}(\beta u_{\rho})]$.

\subsubsection{Summary and discussions of this approach}

We have constructed a microscopic model for the global polarization
from particle scatterings in a many body system. The core of the idea
is the scattering of particles as wave packets so that the orbital
angular momentum is present in the initial state of the scattering
which can be converted to the spin polarization of final state particles.
As an illustrative example, we have calculated the quark/antiquark
polarization in a QGP. The quarks and gluons are assumed to obey the
Boltzmann distribution which simplifies the heavy numerical calculation.
There is no essential difficulty to treat quarks and gluons as fermions
and bosons respectively.

To simplify the calculation, we also assume that the quark distributions
are the same for all flavors and spin states. As a consequence, the
inverse process is absent that one polarized quark is scattered by
a parton to two final state partons as wave packets. So the relaxation
of the spin polarization cannot be described without inverse processes
and spin dependent distributions. We will extend our model by including
the inverse process in the future.
In Ref.~\cite{Weickgenannt:2020aaf}, local and nonlocal collision terms in the Boltzmann equation for massive spin-1/2 particles 
in the Wigner function approach~\cite{DeGroot:1980dk} have been derived for spin dependent distributions. 
The equilibration of spin degrees of freedom can be fully described by such a spin Boltzmann equation. 
Nonlocal collision terms are found to be responsible for the conversion of orbital into spin angular momentum. 
It can be shown that collision terms vanish in global equilibrium and that the spin potential is equal to the thermal vorticity. 
Such a Boltzmann equation can be applied to parton collisions in quark matter.

\subsection{Global hadron polarization in HIC}
\label{sechpol}

The global polarization of quarks and anti-quarks in QGP produced in non-central HIC has different direct consequences. 
The most obvious and measurable effects is the global polarization of hadrons produced 
after the hadronization of QGP.  
In~\cite{Liang:2004ph}, the global polarization of produced hyperons has been given. 
The spin alignment of vector mesons has been calculated in~\cite{Liang:2004xn}. 

It is clear that the global hadron polarization depends not only on the global quark polarization 
but also on the hadronization mechanism.  
In the following, we discuss the results obtained in quark combination and fragmentation respectively.

\subsubsection{Global hyperon polarization}

For all hyperons belong to the $J^P=(1/2)^+$ baryon octet except $\Sigma^0$, 
the polarization can be measured via the angular distribution of decay products 
in the corresponding weak decay. 
Such decay process is often called ``spin self analyzing parity violating weak decay". 
Because of this, hyperon polarizations are widely studied in the field of high energy spin physics.

(i) {\it Hyperon polarization in the quark combination} 

Different aspects of experimental data suggest that hadronization of QGP proceeds via 
combination of quarks and/or anti-quarks. 
This mechanism is phrased as ``quark re-combination'', or ``quark coalescence'' or simply as ``quark combination''. 
We simply refer it as ``the quark combination mechanism'' and use it to calculate 
the hyperon polarization in the following. 

In the quark combination mechanism, 
it is envisaged that quarks and anti-quarks evolve into constituent quarks and anti-quarks 
and combine with each other to form hadrons.  
We choose the minus direction of the normal of the reaction plane $-\vec n$ as the quantization axis. 
The spin density matrix of quark or anti-quark is given by,
\begin{equation}
\hat\rho_q=\frac{1}{2}\left(\begin{array}{cc} 1+P_q&0\\ 0&1-P_q \end{array}\right) .
\label{eqrhoq}
\end{equation}
We do not consider the correlation between the polarizations of different quarks and/or anti-quarks hence 
the spin density matrix for a $q_1q_2q_3$ is given by,
\begin{equation}
\hat\rho_{q_1q_2q_3}= \hat\rho_{q_1}\otimes\hat\rho_{q_2}\otimes\hat\rho_{q_3} .
\label{eqrhoq123}
\end{equation}
Suppose a hyperon $H$ is produced via the combination of $q_1q_2q_3$, 
we obtain, 
\begin{equation}
\rho_H(m',m)=\frac{\sum_{m_i,m_i'}\rho_{q_1q_2q_3}(m_i',m_i)
\langle j_H,m'|m'_1,m'_2,m'_3\rangle \langle m_1,m_2,m_3|j_H,m\rangle}
{\sum_{m,m_i,m_i'}\rho_{q_1q_2q_3}(m_i',m_i)
\langle j_H,m|m'_1,m'_2,m'_3\rangle \langle m_1,m_2,m_3|j_H,m\rangle}, \label{eqrhoH}
\end{equation}
where 
$|j_H,m\rangle$ is the spin wave function of $H$ in the constituent quark model, 
and $\langle j_H,m|m_1,m_2,m_3\rangle$ is the Clebsh-Gordon coefficient. 
The polarization of $H$ is, 
\begin{equation}
P_H=\rho_H(1/2,1/2)-\rho_H(-1/2,-1/2). 
\end{equation}

Since $\hat\rho_q$ is diagonal so is $\hat\rho_{q_1q_2q_3}$, i.e. 
$\rho_{q_1q_2q_3}(m_i',m_i)=\Pi_{i} (1+\tilde P_{q_i})\delta_{m_i,m'_i}/8$, 
where $\tilde P_{q_i}\equiv {\rm sign}(m_i) P_{q_i}$,  Eq.~(\ref{eqrhoH}) reduces to,
\begin{equation}
\rho_H(m',m)=\frac{\sum_{m_i} \Pi_{j} (1+\tilde P_{q_j}) 
\langle j_H,m'|m_1,m_2,m_3\rangle \langle m_1,m_2,m_3|j_H,m\rangle}
{\sum_{m,m_i}\Pi_{j} (1+\tilde P_{q_j}) 
|\langle j_H,m|m_1,m_2,m_3\rangle|^2}.  \label{eqrhoHdia}
\end{equation}
The remaining calculations are straight forward and we list the results in table~\ref{tabHpol}. 
It is also obvious that if $P_u=P_d=P_s\equiv P_q$, we obtain $P_H=P_q$ for all hyperons. 

\begin{table}
\caption{Polarization of hyperons directly produced in the quark combination or fragmentation mechanism. 
 The results for fragmentation are for the leading hadrons only where $n_s$ and $f_s$ in fragmentation 
 are the strange quark abundances relative to up or down quarks in QGP and quark fragmentation, respectively.
These results are taken from~\cite{Liang:2004ph}.} \label{tabHpol}
\begin{tabular}{ccccccc}\hline
hyperon & $\Lambda$ & $\Sigma^+$ & $\Sigma^0$ &$\Sigma^-$ &$\Xi^0$ &$\Xi^-$  \rule[-0.20cm]{0mm}{0.6cm} \\ \hline 
combination &~~~$P_s$~~~ & $\frac{4P_u-P_s}{3}$& $\frac{2(P_u+P_d)-P_s}{3}$ 
&$\frac{4P_d-P_s}{3}$&$\frac{4P_s-P_u}{3}$ & $\frac{4P_s-P_d}{3}$ \rule[-0.25cm]{0mm}{0.7cm} \\ \hline 
fragmentation &~~~$\frac{n_sP_s}{n_s+2f_s}$ &
~~$\frac{4f_sP_u-n_sP_s}{3(2f_s+n_s)}$ & ~~$\frac{2f_s(P_u+P_d)-n_sP_s}{3(2f_s+n_s)}$ & ~~$\frac{4f_sP_d-n_sP_s}{3(2f_s+n_s)}$ &
~~$\frac{4n_sP_s-f_sP_u}{3(2n_s+f_s)}$& ~~$\frac{4n_sP_s-f_sP_d}{3(2n_s+f_s)}$ 
\rule[-0.25cm]{0mm}{0.7cm} \\ \hline 
\end{tabular}
\end{table}

(ii) {\it Hyperon polarization in the quark fragmentation} 

In the high $p_T$ region, hadron production is dominated by the quark fragmentation mechanism,  
described by quark fragmentation functions defined via the quark-quark correlator such as,
\begin{eqnarray}
D_1(z)&=&\sum_{S_h}\int \frac{d\xi^-}{2\pi} e^{-i\xi^-p_h^+/z}~~{\rm Tr}~\gamma^+~\langle 0| {\cal L}(0,+\infty) \psi(0)|p_h,S_h,X\rangle \nonumber \\
&&~~~~~~\times \langle p_h,S_h,X| \bar\psi(\xi){\cal L}(\xi,+\infty)|0\rangle, \label{eqffD1}
\end{eqnarray}
which is the number density of hadron $h$ produced in the fragmentation process $q\to h+X$; 
$z=p_h^+/p^+$ is the momentum fraction of quark $q$ carried by hadron $h$, 
where $p$ and $p_h$ denote the momenta of $q$ and $h$ respectively. 
Here the light cone coordinate is used and the superscript $+$ denotes the $+$ component.
${\cal L}$ is the gauge link that originates from the multiple gluon scattering and guarantees the gauge invariance. 
The polarization transfer is described by 
\begin{eqnarray}
G_1(z)&=&\int \frac{d\xi^-}{2\pi} e^{-i\xi^-p_h^+/z}~{\rm Tr}~ \gamma_5\gamma^+ \langle 0| \psi(0)|p_h,+,X\rangle  \langle p_h,+,X| \bar\psi(\xi)|0\rangle, \label{eqffG1} \\
H_{1T}(z)&=&\int \frac{d\xi^-}{2\pi} e^{-i\xi^-p_h^+/z} ~{\rm Tr}~ \gamma_T\gamma^+ \langle 0| \psi(0)|p_h,+_T,X\rangle \langle p_h,+_T, X| \bar\psi(\xi)|0\rangle, ~~~~~~ \label{eqffH1}
\end{eqnarray}
for the longitudinal and transverse polarization respectively;  
the $+$ or $+_T$ in $|p_h,S_h,X\rangle$ represents that the spin of $h$ is in the $S_{hz}=+1/2$ or $S_{hT}=+1/2$ state
and gauge links are omitted for clarity of equations. 
The presence of $\gamma_5$ or $\gamma_T=\vec\gamma\cdot \vec n_T$ 
introduces the dependence on the spin of the fragmenting quark $q$. 

Fragmentation functions are best studied in $e^+e^-$ annihilations.  
They can not be calculated using pQCD so currently we have to rely on parameterizations or models. 
There are still not much data available yet. 
For longitudinal polarization, we have data from LEP at CERN for $\Lambda$ 
polarization~\cite{Buskulic:1996vb,Ackerstaff:1997nh}. 
A recent parameterization of $G_1$ can be found in~\cite{Chen:2016moq}. 
For the transversely polarized case, little data and no parameterization of $H_{1T}$ is available. 

\begin{figure}[htbp]
\begin{center}
\includegraphics[width=2.7in]{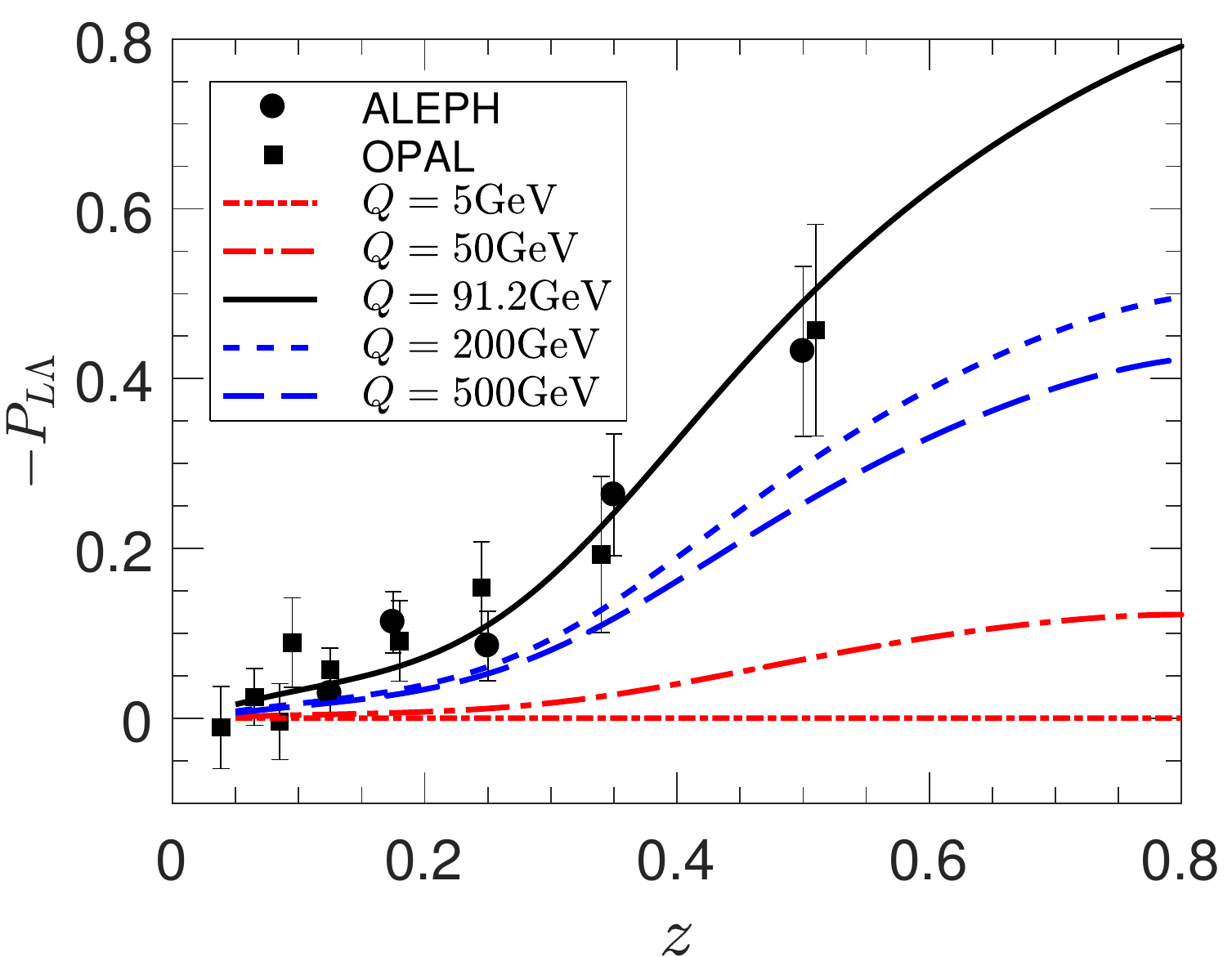}
\end{center}
\caption{Longitudinal polarization of $\Lambda$ in $e^+e^-\to \Lambda+X$ as described by using 
a parameterization of $G_{1L}(z)$. 
The data points are from experiments at LEP~\cite{Buskulic:1996vb,Ackerstaff:1997nh}. 
This figure is taken from \cite{Chen:2016moq}.}
\label{figlambdapolLEP}
\end{figure}

To get a feeling of the $z$-dependence of the spin transfer in quark fragmentations, 
we show the fit obtained in~\cite{Chen:2016moq} to the LEP data in Fig.~\ref{figlambdapolLEP}. 
We see that, although the accuracy is still need to be improved, 
it is definite that there is a strong $z$-dependence of  $G_1$ and 
the spin transfer $G_1/D_1$ is usually significantly smaller than unity. 
This implies that the hyperon polarization obtained in the fragmentation mechanism should be much smaller than that obtained in the combination case.

In \cite{Liang:2004ph}, a model estimation was made for the polarization of the leading hyperon 
produced in the fragmentation of a polarized quark. 
It was assumed that two unpolarized quarks are created in the fragmentation 
and they combine with the polarized $q$ to form the leading hyperon. 
In this case, we obtain the results as given in table~\ref{tabHpol}.
We see if $n_s=f_s$ the result from fragmentation 
is just $1/3$ of the corresponding result from combination, i.e., 
much smaller than the latter even for the leading hyperon.

\subsubsection{Global spin alignment of vector mesons}

Vector meson spin alignment can also be measured via angular distribution of decay products 
in the strong two body decay $V\to 1+2$ into two spinless mesons. 
Hence it is also frequently studied in high energy spin physics.

(i) {\it Vector meson alignment in the quark combination} 

Similar to $q_1q_2q_3$, we do not consider the correlation between polarizations of quarks and anti-quarks, 
and obtain the spin density matrix for a $q_1\bar q_2$-system as,
\begin{equation}
\hat\rho_{q_1\bar q_2}= \hat\rho_{q_1}\otimes\hat\rho_{\bar q_2} .
\label{eqrhoq1q2bar}
\end{equation}
The spin density matrix for a vector meson $V$ produced via the combination of $q_1\bar q_2$ is given by, 
\begin{equation}
\rho^V_{m'm}=\frac{\sum_{m_i,m_i'}\rho_{q_1\bar q_2}(m_i',m_i)
\langle j_V,m'|m'_1,m'_2\rangle \langle m_1,m_2|j_V,m\rangle}
{\sum_{m,m_i,m_i'}\rho_{q_1\bar q_2}(m_i',m_i)
\langle j_V,m|m'_1,m'_2\rangle \langle m_1,m_2|j_V,m\rangle}, \label{eqrhoV}
\end{equation}
where $|j_V,m\rangle$ is the spin wave function of $V$ in the constituent quark model. 
For diagonal $\hat\rho_q$ and $\hat\rho_{\bar q}$, we have,
 \begin{equation}
\rho^V_{m'm}=\frac{\sum_{m_i} (1+\tilde P_{q_1}) (1+\tilde P_{\bar q_2})
\langle j_V,m'|m_1,m_2\rangle \langle m_1,m_2|j_V,m\rangle}
{\sum_{m,m_i} (1+\tilde P_{q_1})(1+\tilde P_{\bar q_2}) 
|\langle j_V,m|m_1,m_2\rangle|^2} , \label{eqrhoVdia}
\end{equation}
The spin alignment is described by $\rho^V_{00}$ and is obtained as~\cite{Liang:2004xn},
\begin{equation}
\rho^V_{00}
=\frac{1-P_{q_1}P_{\bar q_2}}{3+P_{q_1}P_{\bar q_2}} . \label{eqrhoV00}
\end{equation}
From Eq.~(\ref{eqrhoV00}), we see clearly that the global vector meson spin alignment $\rho^V_{00}$ obtained in quark combination 
should be less than $1/3$. 
We also see that in contrast to the hyperon polarization $P_H$, $\rho^V_{00}$ is a quadratic effect of $P_q$.   
 
(ii) {\it Vector meson spin alignment in the quark fragmentation} 

To define the fragmentation functions for spin-1 hadrons in $q\to V+X$, 
one usually decomposes the $3\times3$ spin density matrix $\rho$ in terms of the $3\times3$ representation 
of the spin operator $\Sigma^i$ 
and $\Sigma^{ij}= \frac{1}{2} (\Sigma^i\Sigma^j + \Sigma^j \Sigma^i) - \frac{2}{3} \mathbf{1} \delta^{ij}$, i.e., 
\begin{align}
\rho = \frac{1}{3} (\mathbf{1} + \frac{3}{2}S^i \Sigma^i + 3 T^{ij} \Sigma^{ij}), \label{eqspin1rho}
\end{align}
where the spin polarization tensor $T^{ij}={\rm Tr}(\rho \Sigma^{ij})$ and is parameterized as,
\begin{align}
\mathbf{T}= \frac{1}{2} \left( \begin{array}{ccc}
-\frac{2}{3}S_{LL} + S_{TT}^{xx} & S_{TT}^{xy} & S_{LT}^x  \\
S_{TT}^{xy}  & -\frac{2}{3} S_{LL} - S_{TT}^{xx} & S_{LT}^{y} \\
S_{LT}^x & S_{LT}^{y} & \frac{4}{3} S_{LL}
\end{array} \right).
\label{eqspintensor}
\end{align}
The spin alignment $\rho_{00}$ is directly related to $S_{LL}$ by $\rho_{00}=(1-2S_{LL})/3$ and 
$S_{LL} = {3}\langle \Sigma_z^2 \rangle/2 - 1$ is a Lorentz scalar.  
The complete set of fragmentation functions for spin-1 hadrons can be found in~\cite{Chen:2016moq}.
The $S_{LL}$-dependence is given by, 
\begin{eqnarray}
D_{1LL}(z)&=&\sum_\lambda(-1)^{\lambda+1} \int \frac{d\xi^-}{4\pi} e^{-i\xi^-p^+} ~ {\rm Tr}~ \gamma^+ \langle 0|  \psi(0) 
|p_h\lambda X\rangle \langle p_h\lambda X| \bar\psi(\xi)|0\rangle,~~~~~~~\label{eqffD1LL}
\end{eqnarray}
where $\lambda=\pm 1, 0$ represents the spin of the vector meson.
It very interesting to see that $D_{1LL}(z)$ in fact does not depends on the spin of the fragmenting quark $q$. 

\begin{figure}[htbp]
\begin{center}
\includegraphics[width=2.7in]{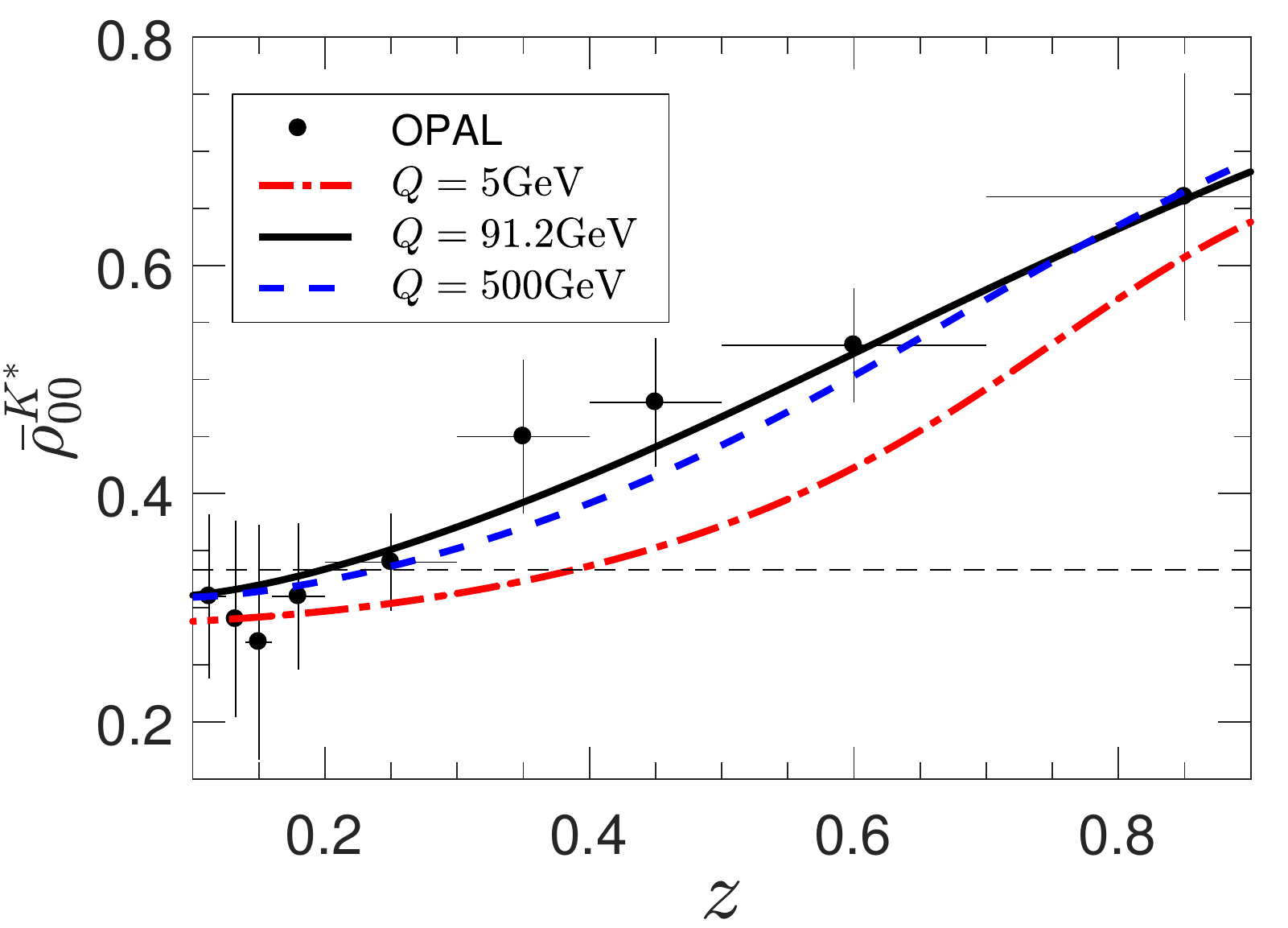}
\end{center}
\caption{Spin alignment of $K^*$ in $e^+e^-\to K^*+X$ as described by using 
a parameterization of $D_{1LL}(z)$. 
The data points are from experiments at LEP~\cite{Ackerstaff:1997kj,Abreu:1997wd}. 
This figure is taken from \cite{Chen:2016iey}.}
\label{figVpolLEP}
\end{figure}

There are data available on the vector meson spin alignment from experiments 
at LEP~\cite{Ackerstaff:1997kj,Abreu:1997wd,Ackerstaff:1997kd}. 
A parameterization of $D_{1LL}(z)$ is given in~\cite{Chen:2016iey,Chen:2020pty} 
and the fit to the data is shown in Fig.~\ref{figVpolLEP}.

From Fig.~\ref{figVpolLEP}, we see clearly that, in contrast to quark combination mechanism, 
$\rho_{00}$ obtained in fragmentation is larger than $1/3$. 
This indicates that the spin of $\bar q$ produced in the fragmentation $q\to h+X$ has larger probability 
to be in the opposite direction as $q$. 
For the leading meson, a parameterization of  $P_{\bar q}=-\beta P_q$ (where $\beta\sim 0.5$) 
for the anti-quark $\bar q$ produced in the fragmentation process and combine 
with the fragmenting quark to form the vector meson 
was obtained~\cite{Xu:2001hz} to fit the data~\cite{Ackerstaff:1997kj,Abreu:1997wd}.
Ref.~\cite{Liang:2004xn} also made an estimation for such leading vector mesons in fragmentation based on the 
this empirical relation and obtained that,  
\begin{equation}
\rho^V_{00}=(1+ \beta P_q^2)/(3-\beta P_q^2) . \label{eqrhoV00phe}
\end{equation}
We see that the spin alignment $\rho^V_{00}$ obtained this way is indeed larger than $1/3$. 

\subsubsection{Decay contributions}

It is clear that final state hadrons in a high energy reaction usually contain the contributions 
from decays of heavier resonances in particular those from strong and electromagnetic decays. 
To compare with the data, we need to take such decay contributions into account. 

The decay contributions have influences both on the momentum distribution and on the polarization of final hadrons.  
Such influences have been discussed repeatedly in literature calculating hyperon polarizations in high energy reactions  
(see e.g.~\cite{Gatto:1958qmn,Gustafson:1992iq,Boros:1998kc,Liu:2000fi} and recently in HIC~\cite{Becattini:2016gvu,Xia:2019fjf}). 
For hadrons consisting of light flavors 
of quarks, we usually consider only the production of $J^P=(1/2)^+$ octet and $J^P=(3/2)^+$ decuplet baryons, and $J^P=0^-$ pseudo-scalar and $J^P=1^-$ vector mesons. 
In this case, there is no decay contribution to vector mesons. 
We only need to consider those to hyperons and 
most of them are just two body decay $H_j\to H_i+M$ where $H_j$ and $H_i$ are two hyperons 
and $M$ is a pseudo-scalar meson.  We limit our discussions to this process in the following. 

To be explicit, we consider the fragmentation mechanism and study decay contributions to fragmentation functions.  
For quark combination, we need only to replace the fragmentation function by the
corresponding distribution function and $z$ by the corresponding variable. 
We start with the unpolarized case and the contribution from $H_j\to H_i+M$ 
to the unpolarized fragmentation function of $H_i$ is given by,
\begin{equation}
D_1^{ij}(z_i,\vec p_{Ti})={\rm Br}(H_i,H_j) \int dz_j d^2p_{Tj} K_{ji}(z_i,\vec p_{Ti};z_j,\vec p_{Tj}) D_1^{j}(z_j,\vec p_{Tj}),  
\end{equation}
where ${\rm Br}(H_i,H_j) $ is the decay branch ratio. 
$K_{ji}(z_i,\vec p_{Ti};z_j,\vec p_{Tj})$ is a kernel function representing the probability 
for a $H_j$ with $(z_j,\vec p_{Tj})$ to decay into a $H_i$ with $(z_i,\vec p_{Ti})$. 
It is just the normalized distribution of $H_i$ from $H_j\to H_i+M$ 
and should be determined by the dynamics of the decay process. 
However, in the unpolarized case, for two body decay, it is determined completely by the energy momentum conservation. 

From energy conservation, we obtain that, in the rest frame of $H_j$,  
\begin{eqnarray}
&&E_i^*=(M_j^2+M_i^2-M_m^2)/2M_j\equiv E_0^*, \\
&& |\vec p^*|
=\lambda^{1/2}(M_j^2,M_i^2,M_m^2)/2M_j\equiv p_0^*,
\end{eqnarray}
where the $\lambda$-function is $\lambda(x,y,z)=x^2+y^2+z^2-2xy-2yz-2zx$. 
We see that the magnitude of $\vec p^*$ is completely fixed. 
Furthermore, because there is no specified direction in the initial state, the decay product should be distributed isotropically.
Hence, in the Lorentz invariant form, the distribution of $H_i$ from $H_j\to H_i+M$ is given by,
\begin{equation}
E_i \frac{d^3N}{d^3p_i} 
= \frac{M_j^2}{\pi \lambda^{1/2}(M_j^2,M_i^2,M_m^2)} \delta\left((p_j-p_i)^2-M_m^2\right).
\end{equation}
By replacing variables $\vec p$ with $z$ and $\vec p_T$, 
we obtain the kernel function $K_{ji}$ as, 
\begin{eqnarray}
&& K_{ji}(z_i,\vec p_{Ti};z_j,\vec p_{Tj}) = \frac{d^3N}{dz_id^2p_{Ti}}  \nonumber\\
&&~~~~~~ =\frac{z_jM_j^2}{\pi \lambda^{1/2}(M_j^2,M_i^2,M_m^2)}  \delta\left((\frac{\vec p_{Tj}}{z_j}-\frac{\vec p_{Ti}}{z_i})^2
+(\frac{M_j}{z_j}-\frac{M_i}{z_i})^2-\frac{\Delta M^2-M_m^2}{z_iz_j}\right),~~~~~~~~~~~
\end{eqnarray}
where $\Delta M\equiv M_j-M_i$ is the mass difference between the two hyperons.

In practice, we often use the following approximation. 
We note that the Lorentz transformation of the four-momentum of $H_i$ from the rest frame of $H_j$ to the laboratory frame is given by, 
\begin{eqnarray}
&& E_i=(E_jE_i^*+\vec p_j\cdot \vec p_i^*)/M_j, \\
&& \vec p_i =  \vec p_i^*+ \frac{ \vec p_j \cdot\vec p_i^* +(E_j-M_j) E_i^*}{M_j(E_j-M_j)}~ \vec p_j,
\end{eqnarray} 
We take the average over the distribution of $\vec p_i^*$ at given $\vec p_j$, and obtain,
\begin{equation}
\langle p_i\rangle 
=p_j ~\xi_{ij},\phantom{XX} 
\xi_{ij}=
{(M_j^2+M_i^2-M_m^2)}/{2M_j^2}. ~~~
\end{equation} 
In the case that $\Delta M \ll M_j\sim M_i$ and $p_0^*\ll |\vec p_i|$, one can simply neglect the distribution and 
take,  $p_i\approx \langle p_i\rangle =p_j \xi_{ij}$ 
so that $z_i\approx z_j \xi_{ij}$, $\vec p_{Ti}\approx \vec p_{Tj} \xi_{ij}$ and,
\begin{eqnarray}
&&K_{ij}(z_i,\vec p_{Ti};z_j,\vec p_{Tj})\approx \delta(z_j- z_i/ \xi_{ij})~\delta^2(\vec p_{Tj}- \vec p_{Ti}/ \xi_{ij})~, \label{eqKjiapprox}\\
&&D_1^{ij}(z_i,\vec p_{Ti})\approx {\rm Br}(H_i,H_j) D^{j}_1(z_i/\xi_{ij},\vec p_{Ti}/\xi_{ij})~. \label{eqD1ijapprox}
\end{eqnarray}

In the polarized case, we need also to consider the polarization transfer $t_D^{ij}$. 
In general, in the rest frame of $H_j$, $t_D^{ij}$ may depend on the momentum $\vec p_i^*$ of $H_i$. 
By transforming it to the Lab frame, we should obtain a result depending 
on $(z_i,\vec p_{Ti})$ and $(z_j,\vec p_{Tj})$ and it is different for the longitudinal and transverse polarization.  
This is much involved. 
In practice, we often take the approximation by neglecting the momentum dependence 
and calculate $t_D^{ij}$ in the rest frame of $H_j$.  
In this case it is the same for the longitudinal and transverse polarization.  
E.g., for the longitudinal polarized case, we have,   
\begin{equation}
G_{1L}^{ij}(z_i,\vec p_{Ti})={\rm Br}(H_i,H_j)~ t_D^{ij} \int dz_j d^2p_{Tj} K_{ij}(z_i,\vec p_{Ti};z_j,\vec p_{Tj}) G_{1L}^{j}(z_j,\vec p_{Tj}).
\end{equation}
Under the approximation given by Eq.~(\ref{eqKjiapprox}), we have,
\begin{equation}
G_{1L}^{ij}(z_i,\vec p_{Ti})\approx {\rm Br}(H_i,H_j) ~ t_D^{ij} ~G_{1L}^{j}(z_i/\xi_{ij},\vec p_{Ti}/\xi_{ij}),  
\end{equation}

For parity conserving decays, 
the polarization transfer factor $t_D^{ij}$ can easily be calculated from angular momentum conservation. 
The results are given in table~\ref{tabdecay}.
For the weak decay $\Xi\to\Lambda\pi$,  $t_D=(1+\gamma)/2$ where  
$\gamma$ is a decay parameter that can be found in Review of Particle Properties (see e.g. \cite{Tanabashi:2018oca}).

\begin{table}
\caption{The decay spin transfer factor in parity conserving tow body decay $H_j\to H_i+M$. 
The first column specifies the spin and parity $J^P$ of hadrons.}\label{tabdecay} 
\begin{center}
\begin{tabular}{cccc}
\hline
$H_j\to H_i+M$~~  & ~~relative orbital angular momentum~~  & ~~$t_D^{ij}=P_{H_i}/P_{H_j}$~~ \rule[-0.20cm]{0mm}{0.55cm} \\ \hline
\phantom{X} $1/2^+\to 1/2^++0^-$ \phantom{X} & $l=1$ ~ (P-wave decay) & $-1/3$  \rule[-0.20cm]{0mm}{0.55cm} \\ \hline
$1/2^-\to 1/2^++0^-$ & $l=0$  ~ (S-wave decay) & $1$ \rule[-0.20cm]{0mm}{0.55cm} \\ \hline
$3/2^+\to 1/2^++0^-$ & $l=1$  ~ (P-wave decay) & $1$ \rule[-0.20cm]{0mm}{0.55cm} \\ \hline
$3/2^-\to 1/2^++0^-$ & $l=2$  ~ (D-wave decay)  & $-3/5$ \rule[-0.20cm]{0mm}{0.55cm} \\ \hline
\end{tabular}
\end{center}
\end{table}

If we taken only $J^P=(1/2)^+$ hyperons into account and use spin counting for relative production weights, we obtain
\begin{equation}
P_\Lambda^{final}=P_\Lambda^{direct} [2+3\lambda(1+\gamma)]/6(1+\lambda),
\end{equation}
where $\lambda$ is the strangeness suppression factor for $s$-quarks.
This leads to a reduction factor between $0.33$ and $0.44$ for $\lambda=0$ and $1$ respectively.  
In this sense, it is more sensitive to study polarization of $\Sigma^\pm$ or $\Xi$ where decay influences are negligible.

\subsection{Comparison with experiments}

The novel predictions~\cite{Liang:2004ph,Liang:2004xn} on GPE attracted immediate attention, both experimentally and theoretically. 
A new preprint~\cite{Voloshin:2004ha} only three days after the first prediction~\cite{Liang:2004ph} attempted to extend the idea to other reactions. 
Experimentalists in the STAR Collaboration had started measurements shortly after the publication of theoretical predictions~\cite{Liang:2004ph,Liang:2004xn},  
both on the global $\Lambda$ hyperon polarization and on spin alignments of $K^*$ and $\phi$ 
\cite{Selyuzhenkov:2005xa,Selyuzhenkov:2006fc,Selyuzhenkov:2006tj,Selyuzhenkov:2007ab,Chen:2007zzq,Abelev:2007zk,Abelev:2008ag}. 
Studies on both aspects have advantages and disadvantages. 
Hyperon polarization is a linear effect where the polarization for directly produced $\Lambda$ is equal to that of quarks.
The spin alignment of vector meson is a quadratic effect proportional to the square of the quark polarization. 
Hence the magnitude of the latter should be much smaller than that of the former.   
However, to measure the polarization of hyperon, one has to determine the direction of  
the normal of the reaction plane, which is not needed for measurements of vector meson spin alignments. 
Also the contamination effects due to decay contributions to vector mesons 
are negligible but not for $\Lambda$ hyperons. 
 
Although there were some promising indications, the results obtained 
in the early measurements~\cite{Abelev:2007zk,Abelev:2008ag} by the STAR Collaboration 
were consistent with zero within large errors.  
STAR measurements continued during the beam energy scan (BES) experiments   
and positive results were obtained in lower energy region with improved accuracies~\cite{STAR:2017ckg}. 
The obtained value averaged over energy is $1.08 \pm 0.15  \pm 0.11$  per cent 
and  $1.38 \pm 0.30 \pm  0.13$ per cent for $\Lambda $ and $\bar\Lambda$ respectively.
With much higher statistics, the STAR Collaboration has repeated 
measurements~\cite{Adam:2018ivw} in Au-Au collisions at 200AGeV 
and obtained positive result of $P_\Lambda\sim -0.003$ with much higher accuracies.

To compare with experiments at this stage, we start with the following rough estimations:
(i) From both Figs.~\ref{figdndydxHIJING} and \ref{figllyBGK} obtained using HIJING and BGK respectively, 
we obtain at $Y\sim 0$, $\Delta p\sim 0.002$GeV for $\Delta x\sim 1$fm. 
If we take $T\sim 140$ MeV, $\Delta p/T\sim 0.015$. 
From Fig.~\ref{figqpolhtl}, 
we see that the quark polarization $P_q$ is unfortunately in the small and rapidly changing region.  
Nevertheless, the order of magnitude is in the same range of STAR data~\cite{Adam:2018ivw}.  
(ii) If we take $\omega\sim \partial u_z/\partial x$, $u_z\sim \langle p_z\rangle /p_T$, we obtain,
$\omega\sim {\Delta^2_Y} \cosh Y~\xi_p/12$ from Eq.~(\ref{eqdapzdxxip}). 
By using the results for $\langle \xi_p\rangle$ shown in Fig.~\ref{figdndydxHIJING} 
or Fig.~\ref{figllyBGK} and $T\sim 140$ MeV, 
we obtain $P_q\sim -0.003$ at $\sqrt{s}=200$GeV that is consistent with STAR experimental results~\cite{Adam:2018ivw}.  
(iii) If we take the result at non-relativistic limit given by Eq.~(\ref{eqPqSPMnr}), 
and note that $\delta u\sim |\vec p|/m_q$, $\delta x\sim 1/\mu_D$, so that $\omega\sim \delta u/\delta x\sim \mu_D|\vec p|/m_q$, 
and quark polarization is $P_q\sim \pi \omega/4m_q$. 
If we take an effective quark mass $m_q\sim 200$ MeV at the hadronization, 
this is clearly also of the same order of magnitude  as $\omega/T$.  

Such rough estimations are rather encouraging. We  continue with more realistic estimations.  
We note that quark polarization is given by Eqs.~(\ref{eqDcsint}-\ref{eqPqDcsovercs}) 
and $d\sigma$ and $d\Delta\sigma$ take the general form given by Eqs.~(\ref{eqcsgform}) and (\ref{eqdcsgform}). 
Before we construct a dynamical model for $f_{qq}(\vec x_T, b,Y,\sqrt{s})$, we present the following qualitative discussion. 

\begin{figure}[htbp]
\begin{center}
\includegraphics[width=1.75in]{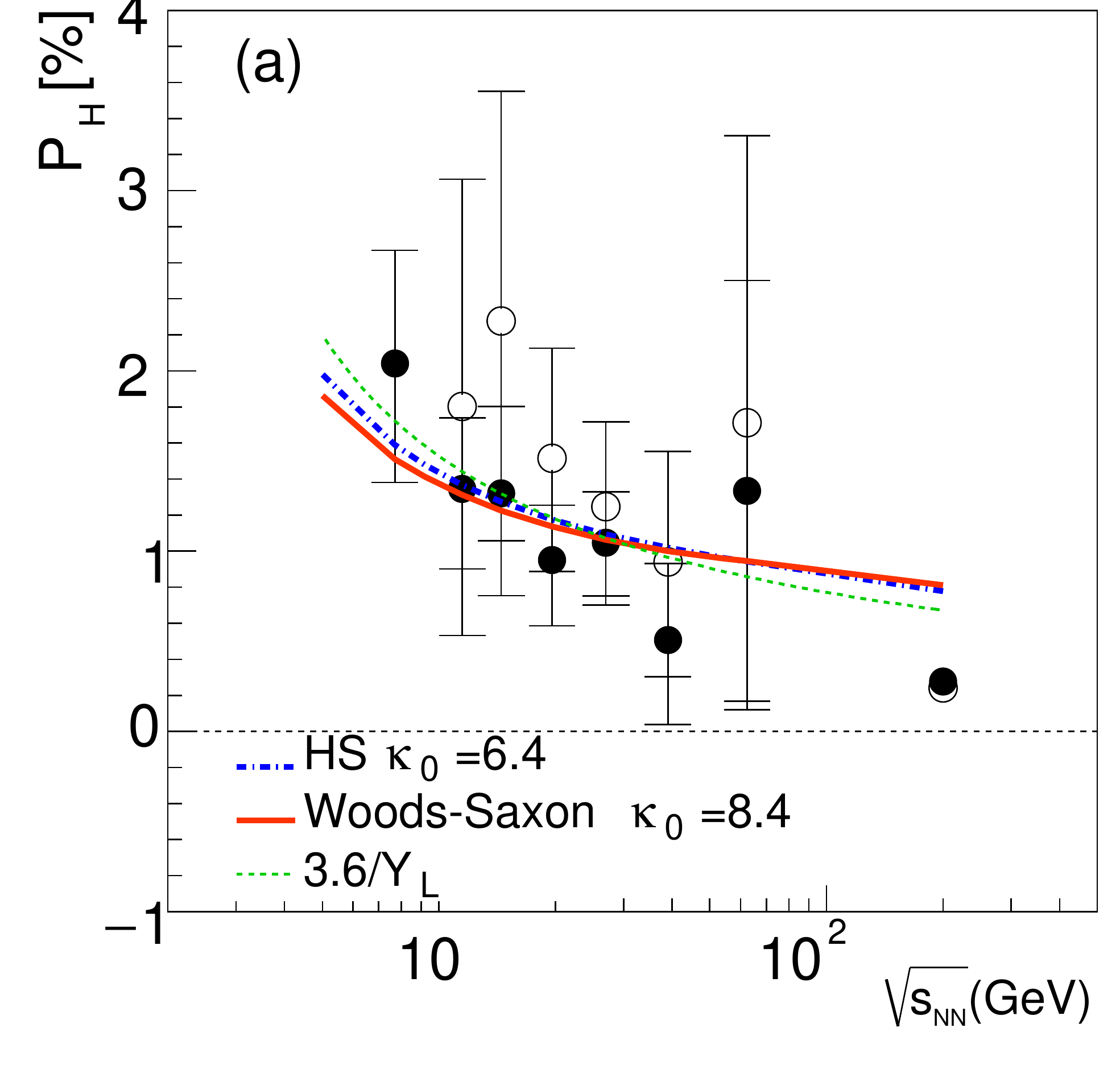}
\includegraphics[width=1.75in]{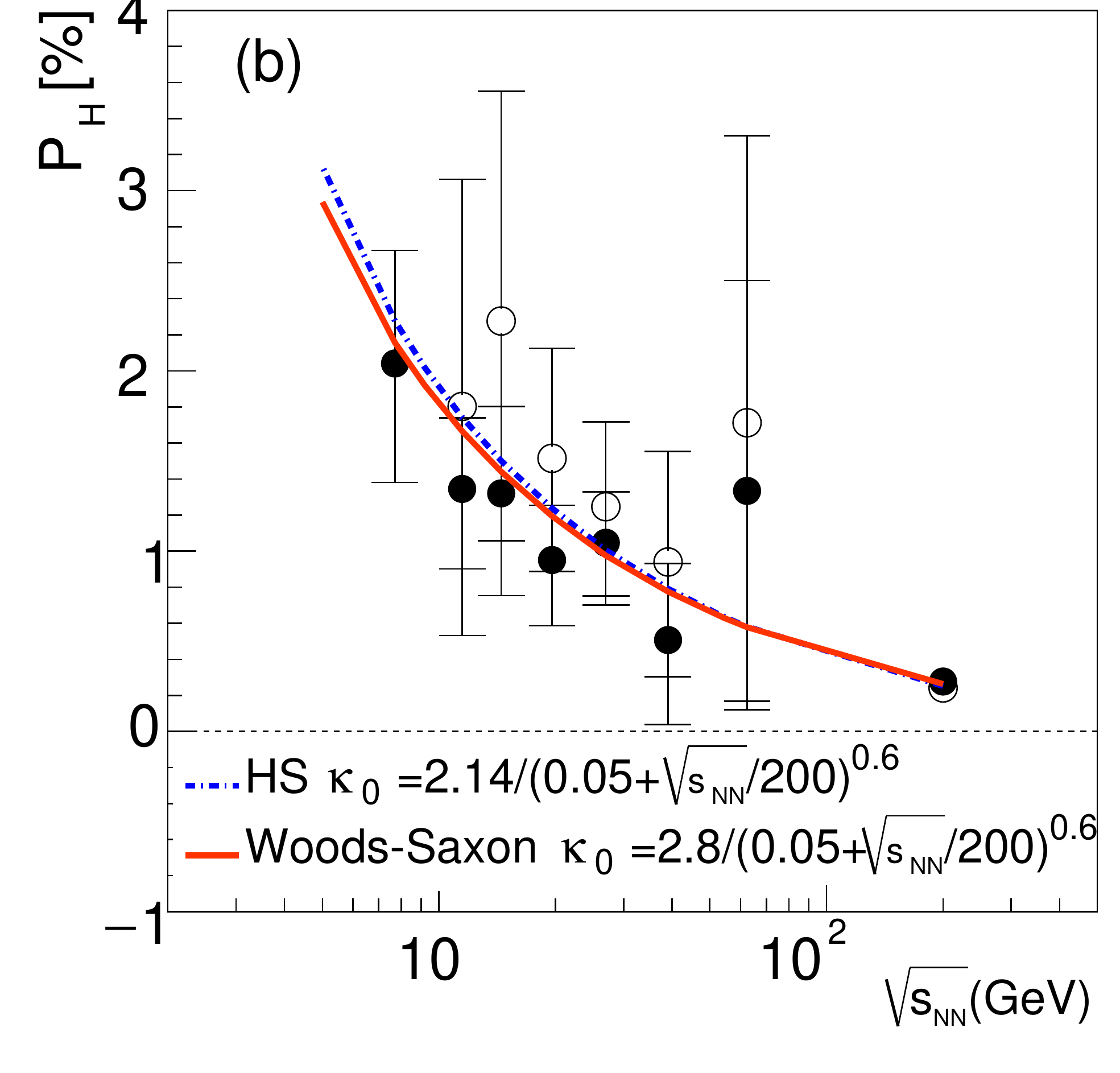}
\end{center}
\caption{Energy dependence of the global polarization of $\Lambda$ obtained by taking $\kappa$ as a constant or an energy dependent form. 
The data points are taken from \cite{STAR:2017ckg,Adam:2018ivw}.
This figure is taken from \cite{Liang2019}.}
\label{figgpolEnergy}
\end{figure}

It is clear that at $b=0$,  $f_{qq}(\vec x_T,0,Y,\sqrt{s})$ should be independent of the direction of $\vec x_T$. 
The $\hat{\vec x}_T$-dependent term should given by $\hat{\vec x}_T\cdot\vec b$. 
We take the linearly dependent term into account and have, 
\begin{equation}
f_{qq}(\vec x_T,b,Y,\sqrt{s})=f_{qq}(x_T,0,Y,\sqrt{s})+f_{qq}(x_T,b,Y,\sqrt{s})~\hat{\vec x}_T\cdot\vec b,  \label{eqfqqxTb}
\end{equation}
We insert Eq.~(\ref{eqfqqxTb}) into (\ref{eqcsgform}) and (\ref{eqdcsgform}) and obtain immediately that $P_q\propto \langle l_y^*\rangle$, i.e., 
\begin{equation}
P_q=\alpha(b,Y,\sqrt{s}) \langle l_y^*(b,Y,\sqrt{s}~)\rangle.  \label{eqPqlly}
\end{equation}
We insert the result of $\langle l_y^*\rangle$ given by Eq.~(\ref{eqllyYxip}) into (\ref{eqPqlly}), average over the impact parameter $b$ and obtain, 
\begin{equation}
P_q=-\kappa(Y,\sqrt{s}) \langle p_T\rangle \langle \xi_p\rangle.  \label{eqPqxip}
\end{equation}
where $\kappa=\alpha(\Delta x)^2\Delta_Y^2/24$. 
The proportional coefficient $\alpha$ in Eq.~(\ref{eqPqlly}) hence also $\varkappa$ in Eq.~(\ref{eqPqxip}) are very involved. 
They are determined by the dynamics in QGP formation and evolution. 
Averaged over $b$, $\kappa$ can still be dependent of $Y$ and $\sqrt{s}$. 
In \cite{Liang2019}, the simplest choice, i.e.,  $\kappa$ is taken as a constant independent of $\sqrt{s}$ at $Y=0$, was first considered and obtained 
the energy dependence of $P_q$ shown in Fig.~\ref{figgpolEnergy}(a).  
Taking an energy dependent $\kappa$, Ref.~\cite{Liang2019} made a better fit to the data~\cite{STAR:2017ckg,Adam:2018ivw} 
available as shown in Fig.~\ref{figgpolEnergy}(b).

\begin{figure}[htbp]
\begin{center}
\includegraphics[width=1.75in]{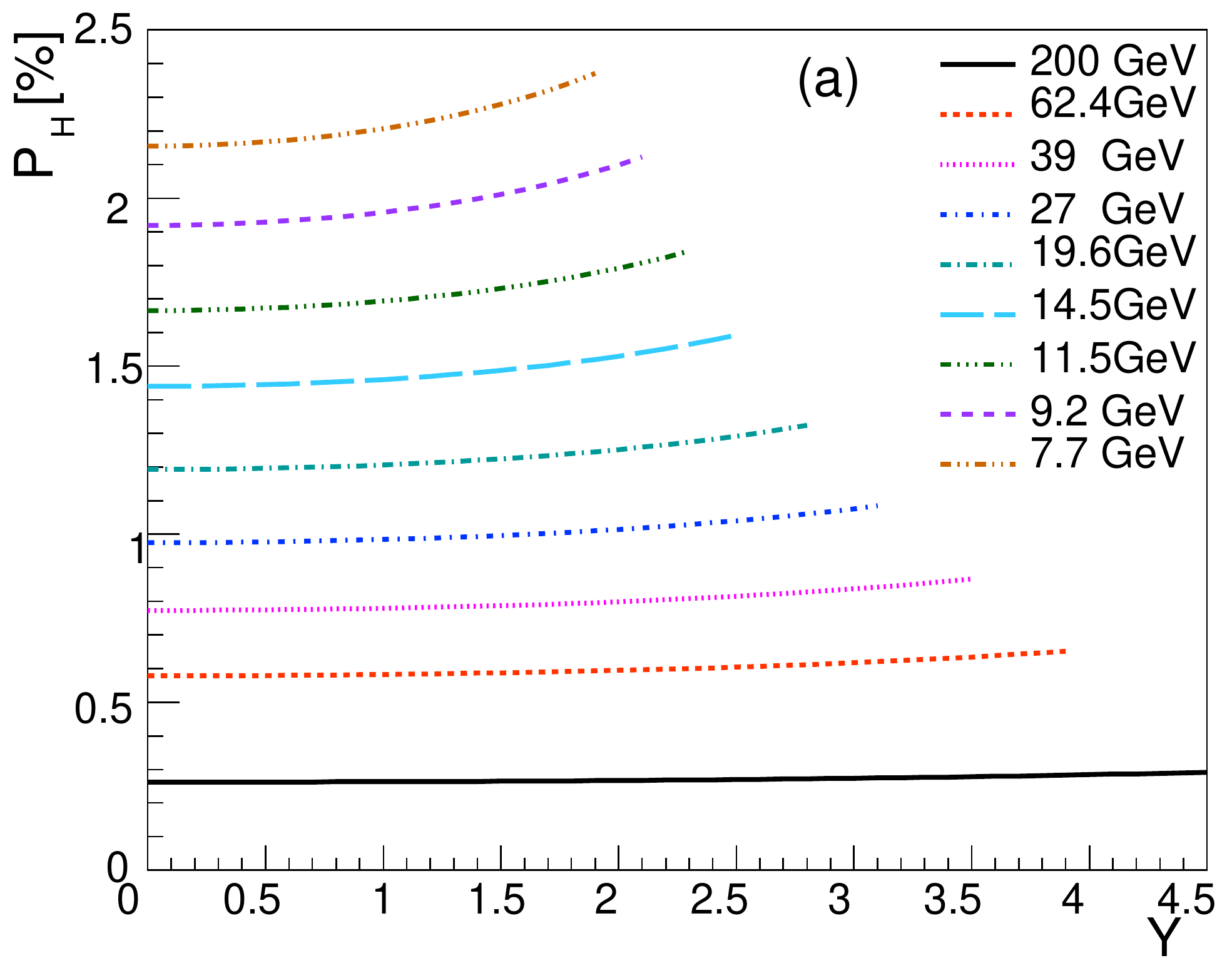}
\includegraphics[width=1.75in]{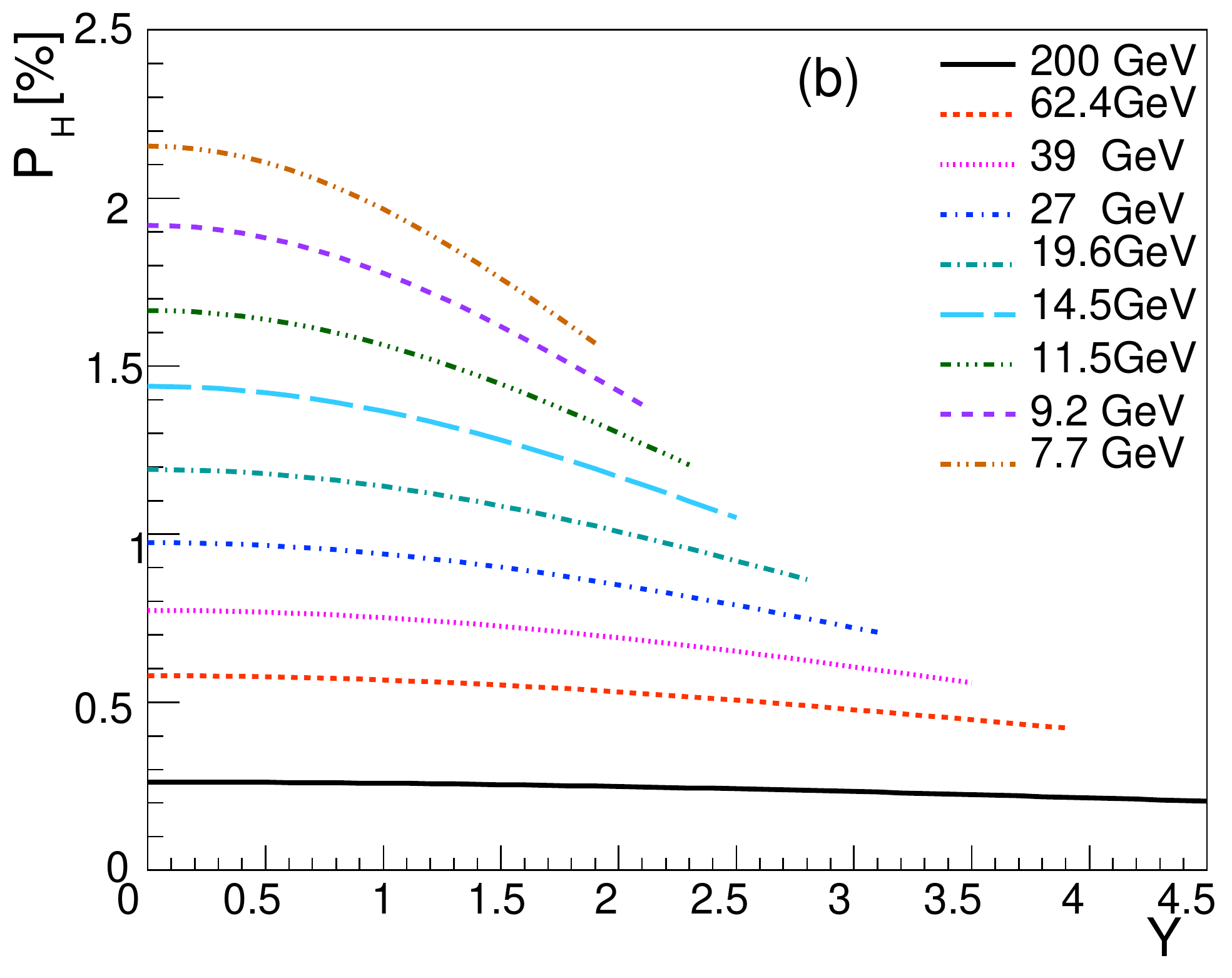}
\includegraphics[width=1.75in]{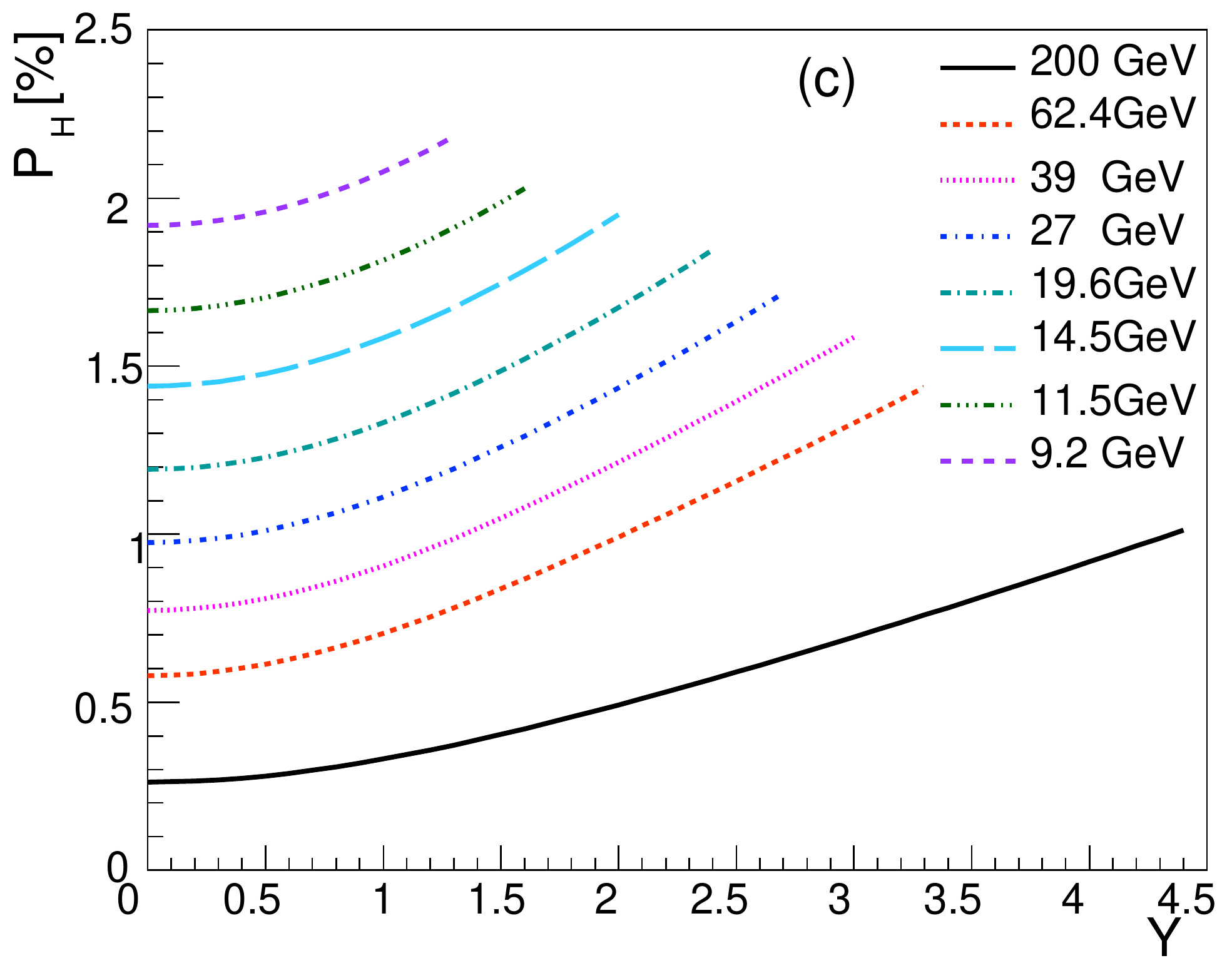}
\includegraphics[width=1.75in]{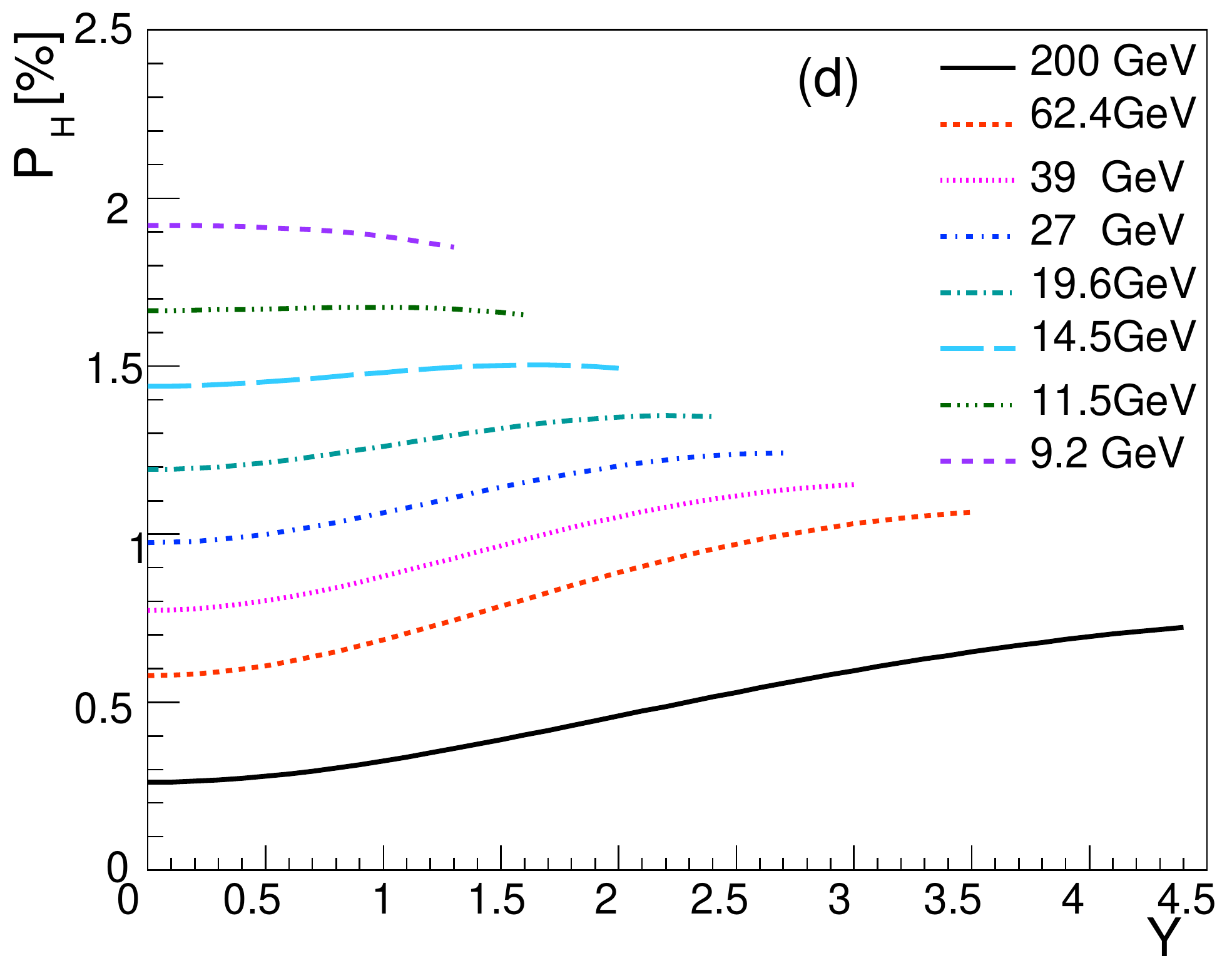}
\end{center}
\caption{Rapidity dependence of the global polarization of $\Lambda$ 
obtained in four different cases (a) neither $\kappa$ nor $\langle p_T\rangle$ depends on $Y$; 
(b) $\kappa$ depends on $Y$ but  $\langle p_T\rangle$ does not; 
(c) $\kappa$ does not but  $\langle p_T\rangle$ depends on $Y$; 
(d) both $\kappa$ and $\langle p_T\rangle$ depend on $Y$. 
This figure is taken from \cite{Liang2019}.}
\label{figgpolRapidity}
\end{figure}

Fig.~\ref{figgpolRapidity} shows the rapidity dependence of the polarization at different energies obtained in \cite{Liang2019} in the different cases. 
The $Y$-dependence of $\kappa$ was obtained by assuming that the dependence is mediated by the chemical potential. 
The $Y$-dependence of $\langle p_T\rangle$ was taken empirically~\cite{Liang2019}. 
See \cite{Liang2019} for details.

\section{Summary and Outlook}

To summarize, 
high energy HIC is usually non-central thus the colliding system and the produced partonic system 
QGP carries a huge global orbital angular momentum as large as $10^5\hbar$ in Au-Au collisions at RHIC energies.
Due to the spin-orbit coupling in QCD, such huge orbital angular momentum can be transferred 
to quarks and anti-quarks thus leads to a globally polarized QGP.  
The global polarization of quarks and anti-quarks manifest itself as the global polarization of 
hadrons such as hyperons and vector mesons produced in HIC. 

The early theoretical prediction~\cite{Liang:2004ph} 
and discovery by the STAR Collaboration~\cite{STAR:2017ckg}  
open a new window to study properties of QGP and a new direction in high energy heavy ion physics. 
Similar measurements have been carried in other experiments such as those 
by ALICE Collaboration at the Large Hadron Collider (LHC) in Pb-Pb collisions~\cite{Acharya:2019ryw}.
Other efforts have also been made on measurements of vector meson spin alignments~\cite{Zhou:2019lun,Acharya:2019vpe}.
The STAR Collaboration has just finished major detector upgrades and started 
the beam energy scan at phase II (BES II). 
The successful detector upgrade with improved 
inner time projection chamber (iTPC) and event plane detector (EPD) 
will be crucial to the measurements of global hadron polarizations. 
The STAR BES II will provide an excellent opportunity to study GPE in HIC and 
we expect new results with higher accuracies in next years.  

The experimental efforts in turns further inspire theoretical studies. 
The rapid progresses and continuous studies along this line lead to a very active research direction 
-- the Spin Physics in HIC in the field of high energy nuclear physics. 
Among the most active aspects, we have in particular the following. 

(i) {\it GPE phenomenology}

This includes different model 
approaches~\cite{Betz:2007kg,Becattini:2007sr,Ipp:2007ng,Barros:2007pt,Xie:2015xpa,Jiang:2016woz,
Montenegro:2017rbu,Xie:2017upb,Li:2017slc,Sun:2017xhx,Yang:2017sdk,Pang:2018zzo,Hirono:2018bwo,Sun:2018bjl,Liang2019} 
to numerical calculations of GPE in HIC and 
its dependences on different kinematic variables. 
The model approaches are basically divided into two categories, i.e.,  
microscopic approaches based on the spin-orbit (or spin-vorticity) coupling 
and hydrodynamic approaches based on equilibrium assumptions.
The various dependences of GPE are studied on kinematic variables describing 
(a) the initial state such as energy, centrality (impact parameter), different incident nuclei even $pA$ collisions etc;  
(b) the produced hadron such as transverse momentum, rapidity, azimuthal angle, different types of hyperons and/or vector mesons; 
(c) other related measurable effects such as longitudinal polarization, the interplay with other effects and so on.  
Short summaries can e.g. be found in plenary talks given at recent Quark Matter conferences~\cite{Liang:2007ma,Wang:2017jpl}.  

(ii) {\it Spin-vortical effects in strong interacting system}

If we can treat QGP as a vortical ideal fluid consisting of quarks and anti-quarks, 
the global polarization of hadrons is directly related to the vorticity of the system~\cite{Becattini:2016gvu}.
The fluid vorticity may be estimated from the data~\cite{STAR:2017ckg} on GPE of $\Lambda$ hyperon using the relation given in 
the hydro-dynamic model, and it leads to a vorticity $\omega\approx(9\pm1)\times 10^{21} s^{-1}$. 
This far surpasses the vorticity of all other known fluids.  
It was therefore concluded that QGP created in HIC is the most vortical fluid in nature observed yet.
GPE in HIC therefore provides a very special place to study spin-vortical effects in strong interaction 
and attracts many studies~\cite{Betz:2007kg,Becattini:2007sr,Deng:2016gyh,Fang:2016vpj,Pang:2016igs,Li:2017dan,Xia:2018tes,Florkowski:2018ahw,Wei:2018zfb}. 
See chapter on this topic for discussions in this aspect.

(iii) {\it Spin-magnetic effects in HIC}

Because of the huge orbital angular momentum, 
there exists also a very strong magnetic field for the colliding system in HIC. 
In Au-Au collision at RHIC, it can reach at least instantaneously the order $10^{14}-10^{16}$~Tesla.  
Such a strong magnetic field can manifest itself in different aspects and lead to different measurable effects. 
The most frequently discussed currently are the following three aspects.

(a) The fine structure of GPE of different hadrons. 
The spin-orbit coupling in QCD predicts e.g. the same polarization of quarks and anti-quarks 
thus also the same for hyperons and anti-hyperons. 
The strong magnetic field can lead to differences between the polarization of quarks and 
that of anti-quarks thus lead to difference in the polarization of hyperons and anti-hyperons. 
Indeed, the STAR data in Ref.~\cite{STAR:2017ckg} suggests such a fine-structure pattern, 
and if errorbars are ignored, would indicate $B\sim10^{14}$~T.  
However, much smaller uncertainties-- available with the new BES-II data-- will be needed to resolve the issue. 
Also the magnetic field may lead to different behavior of vector meson spin alignment~\cite{Sheng:2019kmk}. 

(b) Chiral magnetic effect. 
In Ref.~\cite{Kharzeev:2007jp,Fukushima:2008xe}, 
a novel electromagnetic spin effect -- the chiral magnetic effect was proposed. 
It was argued that such effects have deep connection to $P$ and $CP$ violation.
Clearly, if they exist, strong magnetic filed in HIC provides good opportunity to detect such effects~\cite{Kharzeev:2010gr}.
This has attracted much attention both experimentally and theoretically. 
See a number of reviews such as~\cite{Huang:2015oca,Kharzeev:2015znc,Florkowski:2018fap,Zhao:2019hta,Liu:2020ymh,Wang:2018ygc},    
plenary talks at QM2019 by Xu-guang Huang and Jin-feng Liao~\cite{HuangQM2019,LiaoQM2019} 
and related chapter in this series.  

(c) Spin-electromagnetic effects in ultra-peripheral collisions (UPC) in HIC. 
From a field theoretical point of view, the electromagnetic coupling for a HIC is enhanced 
by a factor $Z$ (number of protons in the nucleus). 
Hence, many electromagnetic effects become visible in UPC with nuclei of large $A$. 
This provides a good place to study the spin-electromagnetic effects and develop 
the theoretical methodology in particular those developed in studying nucleon structure in the small-$x$ region. 
See e.g. the plenary talk at QM2019 by Zhangbu Xu~\cite{XuQM2019} for a brief summary.  

(iv) {\it Spin transport theory in relativistic quantum system}

Theoretically, a very challenging task is to derive GPE, describe the spin transport, calculate the polarization 
and other related spin effects directly from QCD. 
This is rather involved since, to describe orbital angular momentum or vorticity of the system, 
not only momentum but also space coordinate are needed.  
It seems that quantum kinetic theory based on the Wigner function 
formalism~\cite{Heinz:1983nx,Elze:1986qd,Vasak:1987um,Zhuang:1995pd} is very promising 
thus has attracted much attention recently. 
Many progresses have been made. 
Besides others, the local polarization effect has been first derived~\cite{Gao:2012ix} 
and a disentanglement theorem~\cite{Gao:2018wmr} in the massless case has proposed.   
It has now extended to massive case~\cite{Gao:2019znl,Weickgenannt:2019dks,Li:2019qkf,Gao:2020vbh} 
and has been shown that different spin effects can indeed be derived. 
See chapter on this topic for more discussions in this aspect.

\section*{Acknowledgements}
We thank in particular many collaborators for excellent collaborations on this subject. 
This work was supported in part by 
the National Natural Science Foundation of China (Nos. 11890713, 11675092, 11535012, 11935007, 11861131009 and 11890714), 
and by the Director, Office of Energy Research, Office of High Energy and Nuclear Physics, Division of Nuclear Physics, 
of the U.S. Department of Energy under grant No. DE-AC02-05CH11231, the National Science Foundation (NSF) 
under grant No. ACI-1550228 within the framework of the JETSCAPE Collaboration.

%
%
%


\end{document}